\pdfoutput=1
\documentclass[reprint,prb,nofootinbib]{revtex4-2}
\usepackage{tikz}
\usetikzlibrary{patterns.meta}
\usetikzlibrary{fadings}
\tikzfading[name=fade out, inner color=black!40, outer color=black] % radial fade for label light background. this needs to be done before "append style=alignmid" from network_drawing!?!

\usepackage{tikz}
\usetikzlibrary{decorations,decorations.pathmorphing,decorations.markings,calc,snakes,positioning,fixedpointarithmetic,shapes,shapes.geometric,patterns}

\usepackage{etoolbox}
\AtBeginEnvironment{tikzpicture}{\catcode`$=3 } % $ % was important for auctex preview??

\tikzset{alignmid/.style={baseline={([yshift=-.5ex]current bounding box.center)}}} % adjust pictures vertically
\tikzset{every picture/.append style=alignmid}

\tikzset{
bottomzigzag/.style={postaction={draw,decorate, decoration={zigzag,amplitude=1pt,segment length=3pt,raise=1pt}}},
zigzag/.style={draw,decorate, decoration={zigzag,amplitude=1pt,segment length=3pt}},
rc/.style=rounded corners,
}

\tikzset{
    -|/.style={to path={-| (\tikztotarget)}},
    |-/.style={to path={|- (\tikztotarget)}},
}

\usepackage{ifthen}
\usepackage{xparse,pgffor}

\tikzset{
mark/.code={
\tikzset{postaction={/network/mark/.cd,#1,/tikz/.cd,decorate,decoration={name=markings,mark=at position \netmarkpos with{%+\netmarkposoff} with{
\begin{scope}[netmarktrafo]
\netmarkcode
\end{scope}
}}}}
\def\netmarkpos{0.5}%\pgfdecoratedpathlength}
},
}

\def\netmarkpos{0.5}%\pgfdecoratedpathlength}
\def\netmarkposoff{0cm}
\def\netmarkcode{}

\tikzset{
netmarktrafo/.style={},
netmarkstyle/.style={solid,semithick,sharp corners},
}

\pgfkeys{
/network/mark/.cd,
sty/.code={
\tikzset{netmarkstyle/.style={#1}}
},
asty/.code={
\tikzset{netmarkstyle/.append style={#1}}
},
f/.style={asty=fill},
p/.code={ % relative positioning on decorated line
\def\netmarkpos{#1}%\pgfdecoratedpathlength}
},
poff/.code={ % absolute shift on decorated line in cm
\def\netmarkposoff{#1cm}
},
e/.code={ % mark at the End of line
\def\netmarkpos{\pgfdecoratedpathlength-0.005cm-\netmarkposoff}

\tikzset{netmarktrafo/.append style={shift={(-\netmarkwidth,0)}}}
},
s/.code={ % mark at the Start of line
\def\netmarkpos{0.005cm+\netmarkposoff}

\tikzset{netmarktrafo/.append style={shift={(\netmarkwidth,0)},xscale=-1,yscale=-1}}
},
a/.code={ % mark After end of line
\def\netmarkpos{\pgfdecoratedpathlength-0.005cm}

\tikzset{netmarktrafo/.append style={xscale=-1,shift={(-\netmarkwidth,0)}}}
},
b/.code={ % mark Before start of line
\def\netmarkpos{0.005cm}

\tikzset{netmarktrafo/.append style={xscale=-1,shift={(\netmarkwidth,0),yscale=-1}}}
},
-/.code={ % flip horizontally (mark backwards)
\tikzset{netmarktrafo/.append style={xscale=-1}}
},
r/.code={ % flip vertically (mark on the Right instead of left)
\tikzset{netmarktrafo/.append style={yscale=-1}}
},
sideoff/.code={ % shift sideways
\tikzset{netmarktrafo/.append style={shift={(0,#1)}}}
},
slab/.code={ % label next to the line
\def\netmarkwidth{0}
\def\netmarkcode{
\node[inner sep=0.04cm,netmarkstyle,draw=none] (mylabelwidthtest) at (0,0){\phantom{#1}};
\path let \p1=(mylabelwidthtest.north east), \p2=(mylabelwidthtest.south east), \n1 = {max(abs(\y1),abs(\y2))} in node[inner sep=0.04cm,netmarkstyle] at (0,\n1) {#1};
}
},
lab/.code={ % label on the line
\def\netmarkwidth{0}
\def\netmarkcode{
\node[inner sep=0.04cm,anchor=\netmarkanchor] (mylabelwidthtest) at (0,0) {\phantom{#1}};
\draw[white] (mylabelwidthtest.\pgfdecoratedangle)--(mylabelwidthtest.\pgfdecoratedangle+180);
\node[inner sep=0.04cm,anchor=\netmarkanchor,netmarkstyle] at (0,0) {#1};
}
},
rotlab/.code={ % label on the line rotated with the line
\def\netmarkwidth{0}
\def\netmarkcode{
\node[inner sep=0.04cm,fill=white,transform shape,rotate=90,anchor=\netmarkrotanchor,netmarkstyle] (mydecorationnodename) at (0,0) {#1};
}
},
mrotlab/.style={ % small math label on the line rotated with the line
rotlab={$\scriptstyle{#1}$}
},
arr/.code={ % arrow
\def\netmarkwidth{0.04}
\def\netmarkcode{\draw[netmarkstyle] (-0.04,0.08)--(0.04,0)--(-0.04,-0.08);}
},
arrbar/.code={ % arrow
\def\netmarkwidth{0.08}
\def\netmarkcode{\draw[netmarkstyle] (-0.08,0.08)--(0,0)--(-0.08,-0.08) (0.04,0.08)--(0.04,-0.08);}
},
rarr/.code={ % round arrow
\def\netmarkwidth{0.04}
\def\netmarkcode{\draw[netmarkstyle] (-0.04,-0.08)arc(90-180:90:0.08);}
},
dot/.code={ % dot
\def\netmarkwidth{0.08}
\def\netmarkcode{\draw[netmarkstyle] (0,0)circle(0.08);}
},
flag/.code={ % flag to the left
\def\netmarkwidth{0.06}
\def\netmarkcode{\draw[netmarkstyle] (-0.06,0)--(0,0.09)--(0.06,0)--cycle;}
},
darr/.code={ % "double arrow": half-arrow on the left
\def\netmarkwidth{0.08}
\def\netmarkcode{\draw[netmarkstyle] (-0.04,0)--(0.04,0)--(-0.04,0.08)--cycle;}
},
circ/.code={
\def\netmarkwidth{0.1}
\def\netmarkcode{\draw[netmarkstyle] (0,0) circle (0.1);}
},
hcirc/.code={ % half-circle on the left
\def\netmarkwidth{0.1}
\def\netmarkcode{\draw[netmarkstyle] (-0.1,0) arc (180:0:0.1);}
},
diamond/.code={
\def\netmarkwidth{0.1}
\def\netmarkcode{\draw[netmarkstyle] (-0.1,0)--(0,-0.1)--(0.1,0)--(0,0.1)--cycle;}
},
bar/.code={ % tick, short line perpendicular to line
\def\netmarkwidth{0.05}
\def\netmarkcode{
\draw[netmarkstyle] (0,-0.08cm-0.5*\pgflinewidth)--(0,0.08cm+0.5*\pgflinewidth);
}
},
dbar/.code={ % double tick
\def\netmarkwidth{0.13}
\def\netmarkcode{
\draw[netmarkstyle] (-0.04cm,-0.08cm-0.5*\pgflinewidth)--(-0.04cm,0.08cm+0.5*\pgflinewidth) (0.04cm,-0.08cm-0.5*\pgflinewidth)--(0.04cm,0.08cm+0.5*\pgflinewidth);
}
},
hbar/.code={ % tick, short perpendicular line towards the left
\def\netmarkwidth{0.05}
\def\netmarkcode{
\draw[netmarkstyle] (0, 0.5*\pgflinewidth)--++(0,0.12);
}
},
mill/.code={
\def\netmarkwidth{0.16}
\def\netmarkcode{
\draw[netmarkstyle] (0,-0.5*\pgflinewidth)--++(-0.08,-0.08)--++(0,0.08);
\draw[netmarkstyle] (0,0.5*\pgflinewidth)--++(0.08,0.08)--++(0,-0.08);
}
},
three dots/.code={
\def\netmarkwidth{0.2}
\def\netmarkcode{
\fill (-0.12,0) circle (0.5*0.05) (0,0) circle (0.5*0.05) (0.12,0) circle (0.5*0.05);
}
},
}

\tikzset{wid/.style={minimum width=#1cm}}
\tikzset{hei/.style={minimum height=#1cm}}

\tikzset{sx/.style={xshift=#1cm}}
\tikzset{sy/.style={yshift=#1cm}}

\tikzset{box/.style={draw,rectangle}}
\tikzset{fbox/.style={draw,rectangle, line width=1.1}}
\tikzset{roundbox/.style={draw,rectangle,rounded corners}}
\tikzset{froundbox/.style={draw,rectangle, rounded corners, line width=1.1}}
\tikzset{rounddiamond/.style={draw,diamond,rounded corners}}
\tikzset{dot/.style={draw, shape=circle, fill=black, scale=0.5}}

\tikzset{
netbox/.code={
\node[draw,netbdstyle] (\atomname) at (0,0) {#1};
\coordinate (\atomname-r) at (\atomname.east);
\coordinate (\atomname-l) at (\atomname.west);
\coordinate (\atomname-t) at (\atomname.north);
\coordinate (\atomname-b) at (\atomname.south);
\coordinate (\atomname-tr) at (\atomname.north east);
\coordinate (\atomname-br) at (\atomname.south east);
\coordinate (\atomname-tl) at (\atomname.north west);
\coordinate (\atomname-bl) at (\atomname.south west);
},
}

\tikzset{bdlw/.code={\tikzset{mybdstyle/.style={draw, line width=#1}}}}
\tikzset{bdcol/.code={\tikzset{mybdstyle/.append style={#1}}}}

\newcommand\setelements[1]{
\pgfkeys{/network/atom/.cd,#1}
}

\newcommand\atoms[2]{
\foreach \name/\keys in {#2}{
\expandafter\atom\expandafter{\keys,#1}{\name}
}
}

\newcommand\atom[2]{
\def\atomname{#2}
\tikzset{
nettrafo/.style={},
netatompos/.style={},
netdeco/.style={},
netpostdeco/.style={},
}

\pgfkeys{/network/atom/.cd,#1}

\begin{scope}[netatompos] % shift to atom position
\begin{scope}[nettrafo] % rotate, flip and scale
\netshapecoords % set the anchor coordinates
\fill[netbackstyle] \netshapepath;
\clip \netshapepath;
\tikzset{netdeco}
\draw[netbdstyle] \netshapepath;
\end{scope}
\tikzset{netpostdeco} % draw post-decorations, not rotated, flipped, or scaled
\end{scope}

}

\pgfkeys{
/network/atom/.cd,
square/.code={

\def\netshapepath{(-\tempsize,-\tempsize)rectangle (\tempsize,\tempsize)}
\def\netshapecoords{
\node[rectangle,wid=2*\tempsize,hei=2*\tempsize,inner sep=0,transform shape](\atomname)at(0,0){};
\coordinate(\atomname-c) at (0,0);
\coordinate(\atomname-r) at (\tempsize,0);
\coordinate(\atomname-l) at (-\tempsize,0);
\coordinate(\atomname-t) at (0,\tempsize);
\coordinate(\atomname-b) at (0,-\tempsize);
\coordinate(\atomname-br) at (\tempsize,-\tempsize);
\coordinate(\atomname-tr) at (\tempsize,\tempsize);
\coordinate(\atomname-bl) at (-\tempsize,-\tempsize);
\coordinate(\atomname-tl) at (-\tempsize,\tempsize);
}},
circ/.code={

\def\netshapepath{(0,0)circle(\tempsize)}
\def\netshapecoords{
\node[circle,wid=2*\tempsize,hei=2*\tempsize,inner sep=0,transform shape](\atomname)at(0,0){};
\coordinate(\atomname-c) at (0,0);
\coordinate(\atomname-r) at (\tempsize,0);
\coordinate(\atomname-l) at (-\tempsize,0);
\coordinate(\atomname-t) at (0,\tempsize);
\coordinate(\atomname-b) at (0,-\tempsize);
}},
triang/.code={

\def\netshapepath{(-30:\tempsize)--(90:\tempsize)--(-150:\tempsize)--cycle}
\def\netshapecoords{
\node[regular polygon,regular polygon sides=3,wid=2*\tempsize,inner sep=0,transform shape](\atomname)at(0,0){};
\coordinate(\atomname-c) at (0,0);
\coordinate(\atomname-cr) at (-30:\tempsize);
\coordinate(\atomname-cl) at (-150:\tempsize);
\coordinate(\atomname-ct) at (90:\tempsize);
\coordinate(\atomname-mb) at (-90:0.5*\tempsize);
\coordinate(\atomname-mr) at (30:0.5*\tempsize);
\coordinate(\atomname-ml) at (150:0.5*\tempsize);
}},
diamond/.code={

\def\netshapepath{(0,-\tempsize)--(\tempsize,0)--(0,\tempsize)--(-\tempsize,0)--cycle}
\def\netshapecoords{
\node[rotate=45,rectangle,wid=sqrt(2)*\tempsize,hei=sqrt(2)*\tempsize,inner sep=0,transform shape](\atomname)at(0,0){};
\coordinate(\atomname-c) at (0,0);
\coordinate(\atomname-r) at (\tempsize,0);
\coordinate(\atomname-l) at (-\tempsize,0);
\coordinate(\atomname-t) at (0,\tempsize);
\coordinate(\atomname-b) at (0,-\tempsize);
}},
pentagon/.code={

\def\netshapepath{(-126:\tempsize)--(-54:\tempsize)--(18:\tempsize)--(90:\tempsize)--(162:\tempsize)--cycle}
\def\netshapecoords{
\node[regular polygon,regular polygon sides=5,wid=2*\tempsize,inner sep=0,transform shape](\atomname)at(0,0){};
\coordinate(\atomname-c) at (0,0);
\coordinate (\atomname-mb)at(-90:{\tempsize*cos(36)});
\coordinate (\atomname-mbr)at(-18:{\tempsize*cos(36)});
\coordinate (\atomname-mtr)at(54:{\tempsize*cos(36)});
\coordinate (\atomname-mtl)at(126:{\tempsize*cos(36)});
\coordinate (\atomname-mbl)at(-162:{\tempsize*cos(36)});
\coordinate (\atomname-cbr)at(-54:\tempsize);
\coordinate (\atomname-cr)at(18:\tempsize);
\coordinate (\atomname-ct)at(90:\tempsize);
\coordinate (\atomname-cl)at(162:\tempsize);
\coordinate (\atomname-cbl)at(-126:\tempsize);
}},
halfcirc/.code={

\def\netshapepath{(\tempsize,0)arc(0:180:\tempsize)--++(0,-0.04)-|cycle}
\def\netshapecoords{
\node[circle,wid=2*\tempsize,hei=2*\tempsize,inner sep=0,transform shape](\atomname)at(0,0){};
\coordinate(\atomname-c) at (0,0);
\coordinate(\atomname-r) at (\tempsize,0);
\coordinate(\atomname-l) at (-\tempsize,0);
\coordinate(\atomname-t) at (0,\tempsize);
\coordinate(\atomname-b) at (0,0);
}},
void/.code={
\def\netshapepath{}
\def\netshapecoords{
\coordinate(\atomname) at (0,0);
\coordinate(\atomname-c) at (0,0);
}},
labbox/.code={
\def\netshapepath{(0,0)}
\def\netshapecoords{}
\tikzset{netpostdeco/.append style={netbox=#1}}
},
}

\pgfkeys{
/network/atom/.cd,
p/.code={\tikzset{netatompos/.append style={shift={(#1)}}}},
xscale/.code={\tikzset{nettrafo/.append style={xscale=#1}}},
yscale/.code={\tikzset{nettrafo/.append style={yscale=#1}}},
scale/.code={\tikzset{nettrafo/.append style={scale=#1}}},
rot/.code={\tikzset{nettrafo/.append style={rotate=#1}}},
hflip/.style={xscale=-1},
vflip/.style={yscale=-1},
big/.style={scale=3/2},
small/.style={scale=2/3},
huge/.style={scale=9/4},
tiny/.style={scale=4/9},
flat/.style={yscale=2/3},
fflat/.style={yscale=4/9},
}

\tikzset{
netbdstyle/.style={line width=0.15em}, % changed from pt(default)
netdecstyle/.style={},
netpostdecstyle/.style={},
netbackstyle/.style={white},
}
\pgfkeys{
/network/atom/.cd,
bdstyle/.code={\tikzset{netbdstyle/.style={#1}}},
bdastyle/.code={\tikzset{netbdstyle/.append style={#1}}},
decstyle/.code={\tikzset{netdecstyle/.style={#1}}},
decastyle/.code={\tikzset{netdecstyle/.append style={#1}}},
postdecstyle/.code={\tikzset{netpostdecstyle/.style={#1}}},
postdecastyle/.code={\tikzset{netpostdecstyle/.append style={#1}}},
backstyle/.code={\tikzset{netbackstyle/.style={#1}}},
backastyle/.code={\tikzset{netbackstyle/.append style={#1}}},
style/.style={bdstyle=#1, decstyle=#1, postdecstyle=#1},
astyle/.style={bdastyle=#1, decastyle=#1, postdecastyle=#1},
whiteblack/.style={decstyle=white,backstyle=black},
nobd/.style={bdstyle={line width=0}},
}

\tikzset{
netbscope/.code={\begin{scope}[#1]},
netescope/.code={\end{scope}},
}

\pgfkeys{
/network/atom/.cd,
bscope/.code={\tikzset{netdeco/.append style={netbscope={#1}}}},
escope/.code={\tikzset{netdeco/.append style={netescope}}},
dec/.style 2 args={bscope={#1},#2,escope}, % for applying transformation keys to certain decorations only
vbar/.style={dec={rotate=90}{hbar}},
dcross/.style={dec={rotate=45}{hbar},dec={rotate=-45}{hbar}},
cross/.style={hbar,vbar},
sect6/.style={sect=60},
sect8/.style={sect=45},
sect10/.style={sect=36},
}

\def\regdec#1{\pgfkeys{/network/atom/.cd,#1/.code={\tikzset{netdeco/.append style={net#1}}}}}
\regdec{all}\regdec{rhalf}\regdec{rquart}\regdec{brquart}\regdec{dot}\regdec{spiral}\regdec{swirl}\regdec{hstripe}\regdec{hbar}\regdec{rrey}
\pgfkeys{/network/atom/.cd,sect/.code={\tikzset{netdeco/.append style={netsect=#1}}}}

\tikzset{
netall/.code={\fill[netdecstyle] (-0.3,-0.3)rectangle (0.3,0.3);}, % fill all
netrhalf/.code={\fill[netdecstyle] (0,-0.3)rectangle (0.3,0.3);}, % right half
netrquart/.code={\fill[netdecstyle] (0.075,-0.3)rectangle (0.3,0.3);}, % right quarter
netbrquart/.code={\fill[netdecstyle] (0,0)rectangle (0.3,-0.3);}, % bottom right quarter
netsect/.code={\fill[netdecstyle] (0,0)--(0,-0.3)arc(-90:-90+#1:0.3)--cycle;}, % section of angle #1 starting from -90
netdot/.code={\fill[netdecstyle] (0,0)circle(0.07);}, % dot in the middle
netspiral/.code={\draw[netdecstyle] plot [variable=\t,domain=0:4] ({0.075*\t*cos(pi*(\t-0.5) r)},{0.075*\t*sin(pi*(\t-0.5) r)});}, % spiral
netswirl/.code={\fill[netdecstyle] plot [variable=\t,domain=0:2] ({0.15*\t*cos(pi*(\t-0.5) r)},{0.15*\t*sin(pi*(\t-0.5) r)}) arc(-90:-450:0.3)--cycle;}, % filled swirl
nethstripe/.code={\fill[netdecstyle] (-0.3,-0.05)rectangle(0.3,0.05);}, % horizontal stripe
nethbar/.code={\draw[netdecstyle] (-0.3,0)--(0.3,0);}, % horizontal line
netrrey/.code={\draw[netdecstyle] (0,0)--(0.3,0);} % line from the middle to the right
}

\pgfkeys{
/network/atom/.cd,
postbscope/.code={\tikzset{netpostdeco/.append style={netbscope={#1}}}},
postescope/.code={\tikzset{netpostdeco/.append style={netescope}}},
postdec/.style 2 args={postbscope={#1},#2,postescope}, % for applying transformation keys to certain decorations only
arc/.code={\tikzset{netpostdeco/.append style={netarc=#1}}},
lab/.code={\tikzset{netpostdeco/.append style={netlab={#1}}}},
shadecirc/.code={\tikzset{netpostdeco/.append style=netshadecirc}},
shaderect/.code={\tikzset{netpostdeco/.append style={netshaderect={#1}}}},
debug/.code={\tikzset{netpostdeco/.append style=netdebug}},
markline/.code 2 args={\tikzset{netpostdeco/.append style={netmarkline={#1}{#2}}}},
postcirc/.code={\tikzset{netpostdeco/.append style=netpostcirc}},
}

\tikzset{
netlab/.code={
\pgfkeys{/network/atom/lab/.cd,#1}
\node[netpostdecstyle] at (\ifdefined\netlabpos\netlabpos\else\netlabang:\netlabdist\fi) {\netlabwrap{\netlabtext}};
},
netarc/.code args={#1:#2:#3}{
\draw[netpostdecstyle] (#1:#3) arc (#1:#2:#3);
},
netshadecirc/.code= {
\fill[opacity=0.4,netpostdecstyle] (0,0)circle(0.4);
},
netpostcirc/.code= {
\draw[netpostdecstyle] (0,0)circle(0.15);
},
netshaderect/.code= {
\fill[rc,opacity=0.4,netpostdecstyle] ($-1*(#1)$) rectangle (#1);
},
netdebug/.code= {
\node[red] at (0,0){\atomname};
},
netmarkline/.code 2 args= {
\draw (\atomname)edge[mark={#2}]++(#1);
},
}

\def\netlabwrap#1{#1}
\pgfkeys{
/network/atom/lab/.cd,
t/.code={\def\netlabtext{#1}}, % label text
p/.code={\def\netlabpos{#1}}, % position of the label
ang/.code={\def\netlabang{#1}},
dist/.code={\def\netlabdist{#1}},
wrap/.code={\def\netlabwrap##1{#1}}, % store in doesn't work
}

\usetikzlibrary{patterns.meta}

\usepackage{xcolor}
\usepackage{amsmath}
\usepackage{amssymb}

\usepackage{bbold}
\usepackage{mathtools}
\usepackage{hyperref}
\usepackage{makecell}
\usepackage{braket}

\def\zz{\mathbb{Z}}
\def\cc{\mathbb{C}}
\def\idop{\mathbb{1}}
\def\mmod{\operatorname{mod}}

\definecolor{mypurple}{rgb}{0.7,0,0.5}
\definecolor{darkgreen}{rgb}{0,0.6,0.2}
\newcommand{\qubita}{mypurple}
\newcommand{\qubitb}{darkgreen}

\tikzset{
ind/.style={mark={lab=#1,a}}, % normal open index label
startind/.style={mark={lab=#1,b}}, % normal open index label
worldline/.style={red,line width=5,opacity=0.3,line join=round},
smoothbd/.style={orange, line width=0.05cm},
roughbd/.style={orange, line width=0.08cm,densely dotted},
corner/.style={orange, line width=0.1cm},
back/.style={opacity=0.5},
lightback/.style = {circle, fill=white, inner sep=0.5,path fading=fade out}, % lighten up background for label nodes
configcol/.style = {black},
membranecol/.style = {blue},
roughbdfill/.style={pattern={Dots[radius=0.045cm]},opacity=0.6},
roughbdfillback/.style={pattern={Dots[radius=0.045cm,xshift=0.04cm,yshift=0.04cm]},opacity=0.6},
}

\setelements{
z2/.style={small,circ},
delta/.style={small,circ,all},
vertex/.style={tiny,circ,all},
}

\begin{document}

\title{The x+y Floquet code: A simple example for topological quantum computation in the path integral approach}
\author{Andreas Bauer}
\email{andib@mit.edu}
\affiliation{MIT Department of Mechanical Engineering, 77 Massachusetts Avenue, Cambridge, MA 02139, USA}

\begin{abstract}
The path-integral approach to topological quantum error correction provides a unified way to construct and analyze fault-tolerant circuits in spacetime.
In this work, we demonstrate its utility and versatility at hand of a simple example:
We construct a new fault-tolerant circuit for the toric-code phase by traversing its path integral on a $(x,y,z)$ cubic lattice in the $x+y$ direction.
The circuit acts on qubits on a square lattice, and alternates between horizontal nearest-neighbor $CX$ gates and vertical nearest-neighbor $ZZ$ and $XX$ measurements.
We show how to incorporate boundaries and corners into the fault-tolerant circuit and how to perform topologically protected logic gates.
As a specific example, we consider performing a fault-tolerant logical $ZZ$ measurement via lattice surgery of two spatial rectangular blocks of our fault-tolerant circuit.
\end{abstract}

\maketitle
\tableofcontents

\section{Introduction}
One of the leading candidates for fault-tolerantly storing and processing quantum information is the toric code or surface code \cite{Kitaev1997, Bravyi1998}.
The toric code is a topological code, which means that the logical information is stored in the ground space of a uniform model with geometrically local interactions in $2+1$ dimensions, representing a topological phase.
In order to turn the toric code into a concrete fault-tolerant protocol, we need to repeatedly measure its local generating stabilizers \cite{Dennis2001}.
Since it is hard to directly measure the weight-4 stabilizers in practice, we have to decompose them into sequences of 1 and 2-qubit operations.
The most straight-forward way is via one ancilla qubit per stabilizer, four controlled-$X$ ($CX$) gates, and one single-qubit measurement \cite{Dennis2001}.
However, there exist many more ways of compiling the stabilizer measurements, including rather exotic ones \cite{Gidney2022,Mceven2023}.
Strikingly, there also exist other protocols for topological quantum computation that behave in the same way as the toric code, but that cannot be understood directly as a way of compiling the toric-code stabilizers.
Let us give some examples for such protocols.

\textbf{Topological fault-tolerant protocols}.
The first example is measurement-based topological quantum computation \cite{Raussendorf2007}.
This is a $3+0$-dimensional protocol, where a $3$-dimensional resource state is prepared in constant depth, and then all qubits are measured and erased.
By preparing and measuring portions of the resource state on the fly, it can be turned into a $2+1$-dimensional protocol which behaves very similar to the toric code.
This also inspired the more recent idea of fusion-based quantum computation \cite{Bartolucci2021}, where small portions of the resource state are prepared and ``fused'' together via Bell-basis measurements.

Another example is the subsystem toric code \cite{Bravyi2012}, which gives rise a concrete protocol by alternatingly measuring its non-commuting gauge checks.
This protocol can be viewed as a dynamic alternative to the stabilizer toric code that involves only 3-qubit measurements instead of 4-qubit measurements.

The most recent and perhaps most striking example is the honeycomb Floquet code \cite{Hastings2021,Haah2021}, or its CSS version \cite{Kesselring2022,Davydova2022,Aasen2022}.
These are circuits of 2-body nearest-neighbor Pauli measurements acting on qubits on a hexagonal lattice, that protect logical information in a dynamical fashion.

\textbf{Different approaches to topological fault tolerance}.
The invention of Floquet codes has lead to renewed interest in dynamic codes and to a search for general methods to analyze and construct them.
There have been a few successful approaches in the literature.
The first approach is to look at the instantaneous stabilizer group (ISG) \cite{Hastings2021}.
The ISG is the stabilizer group describing the current code space after a sequence of measurements, which can be obtained by simulating the stabilizer circuit \cite{Aarenson2004}.
For example, for the honeycomb Floquet code, the ISG turns out to be equivalent to a toric-code stabilizer group.
However, the ISG does not always tell us the topological phase of a dynamic protocol:
As an example, consider a trivial protocol where we repeatedly prepare a toric code and immediately measure it out completely.
After the preparation step, the ISG is that of the toric code (together with its logical operators), even though the phase is trivial.
More intricate examples for Floquet codes with fluctuating ISG can be found in Refs.~\cite{Davydova2022,Zhang2022,Fuente2024,Ellison2022,Ellison2023,Dua2023}.
Most importantly, however, the ISG does not directly tell us how to come up with new dynamic protocols.

Another approach that has proven fruitful for constructing new codes is dynamic anyon condensation \cite{Kesselring2022,Davydova2023,Ellison2022}.
In a topological stabilizer code, we can condense a subgroup of anyons by measuring their short-string operators.
Dynamic protocols can then be obtained by successive condensations of different anyon subgroups of a larger parent topological phase.
For example, we can obtain the CSS honeycomb Floquet code by successive anyon condensations in a color code \cite{Kesselring2022}.
However, it is hard to define what a ``short string operator'' is, and if we choose the string operators too short or too long, this will not result in a fault-tolerant protocol.
Therefore, this method should be understood as a heuristic rather than a strict formalism.
In order to guarantee fault tolerance, we have to analyze how the short-string-operator measurements link up through spacetime.

Another interesting line of research has been focusing on the potential logical automorphism performed in every round of measurements of a dynamic code \cite{Aasen2022,Aasen2023}.
The prototypical example is the $e\leftrightarrow m$ automorphism of the Hastings-Haah honeycomb Floquet code \cite{Hastings2021}.
These automorphisms alone can be used to perform arbitrary logical Clifford operations inside a dynamic color code \cite{Davydova2023}.
However, the logical automorphism alone is neither necessary nor sufficient for fault tolerance, and disappears after doubling the time period.

In this work, we will use a fourth approach to topological fault-tolerant circuits, namely the path-integral approach proposed in Refs.~\cite{path_integral_qec, twisted_double_code}.
With less physical motivation and restricted to the context of Clifford circuits, an equivalent idea was proposed in Ref.~\cite{Bombin2023}, and successfully applied and generalized in Refs.~\cite{Teague2023,Fuente2024}.
Central to the path-integral approach is the realization that it is best to view fault-tolerant circuits holistically in spacetime
\footnote{
For the purpose of analyzing fault tolerance and decoding, this was already realized in Ref.~\cite{Dennis2001}.
}.
Post-selecting all measurement outcomes in the circuit to a trivial one (usually $+1$ out of $\pm 1$) yields a topological fixed-point path integral in spacetime.
Non-trivial ($-1$) measurement outcomes correspond to defects in the path integral, which in our case are abelian anyon worldlines.
The fault tolerance of the circuit then follows from the basic properties of the fixed-point path integral:
(1) An isolated ``error'' in spacetime surrounded by the path integral has no effect, (2) anyon worldlines must not terminate anywhere in spacetime, and (3) anyon worldlines can be arbitrarily deformed locally.
Here, by fault tolerance, we mean the existance of a fault-tolerant threshold for any kind of geometrically local noise in the circuit.
As shown in Refs.~\cite{path_integral_qec,Bombin2023}, the path-integral approach reveals the spacetime equivalence of a large group of fault-tolerant protocols, including the stabilizer toric code, subsystem toric code, measurement-based topological quantum computation, fusion-based topological quantum computation, and the (CSS) honeycomb Floquet code:
Their path integrals are all the same, apart from different choices of spacetime lattice and time direction.
Vice versa, we can use the approach to construct new fault-tolerant circuits:
We write a fixed-point path integral as a non-unitary circuit along some chosen time direction, and turn it into a full circuit of measurements by adding anyon worldlines.
As shown in Ref.~\cite{twisted_double_code}, this formalism extends beyond the scope of Pauli stabilizers or Clifford operations, and can be used to construct fault-tolerant circuits for all twisted abelian as well as selected non-abelian topological phases.

\textbf{Goals}.
The goals of this work are fourfold.
The first goal is to propose a new fault-tolerant circuit for topological quantum computation.
The overhead in terms of the number of qubits and 2-qubit operations per spacetime unit of code distance is similar to the stabilizer toric code or honeycomb Floquet code.
In a sense, the circuit interpolates between the toric-code stabilizer protocol involving $CX$ gates and single-qubit measurements, and the (CSS) honeycomb Floquet code involving two-qubit Pauli measurements:
It consists of a period-6 schedule of horizontal $CX$ gates and vertical $XX$ or $ZZ$ measurements.
We do not know whether there are experimental platforms where our new circuit has potential advantages over the stabilizer toric code or honeycomb Floquet code.
However, we expect our proposed circuit to have similar performance in terms of overhead and threshold.
To be maximally accommodating to experimentalists, it is important to have a large variety of theoretical architectures, and the proposed circuit is a natural alternative that seems to have been overlooked in the literature so far.

The second goal is to showcase the utility and versatility of the path-integral approach for constructing new fault-tolerant circuits.
While the proposed circuit arises straight-forwardly in the path-integral approach, it is unclear how one would derive it using other methods in the literature:
The instantaneous stabilizer group reveals that we have some sort of spacetime toric code, but does not help constructing it.
There is no automorphism performed by the circuit.
It is unclear how to interpret the circuit in terms of dynamic condensation since it involves $CX$ gates in addition to 2-qubit measurements.

The third goal is to demonstrate the ability of the path-integral approach to go beyond merely storing logical qubits on a torus, namely to construct protocols for other spatial geometries and logic gates.
We show that its geometric nature makes it well suited for constructing boundaries and corners for fault-tolerant circuits.
Specifically, we construct rough and smooth boundaries for our proposed fault-tolerant circuit, as well as corners separating these boundaries in spacetime.
This allows us to put the circuit on a rectangular spatial block of qubits, fault-tolerantly storing a single qubit analogous to the surface code.
In addition, we demonstrate how to construct protocols for fault-tolerant logic operations instead of just storage.
Specifically, we construct a protocol that performs lattice surgery \cite{Horsman2011} between two rectangular spatial blocks of our circuit, resulting in a logical $ZZ$ measurement.
Note that constructing boundaries \cite{Vuillot2021,Haah2021} and logic gates for Floquet codes or other non-stabilizer architectures \cite{Brown2018} is generally considered a not straight-forward task.

The fourth and final goal is to provide a more accessible introduction to the path-integral approach, compared to Refs.~\cite{path_integral_qec,twisted_double_code}.
We make an effort to keep the presentation simple by focusing on key practical aspects of the construction, and neglecting some deeper physical insights behind the path integral.

\section{The toric-code path integral}
\label{sec:path_integral}
In this section, we review the toric-code path integral and its anyon worldlines, as introduced in Ref.~\cite{path_integral_qec}.
To motivate this path integral, let us recall the toric code ground states \cite{Kitaev1997}.
The toric code is defined on qubits on the edges of a square lattice with periodic boundary conditions.
We will identify the qubit basis states $\ket0$ and $\ket1$ with elements of $\zz_2$, such that qubit configurations are elements of $\zz_2^{S_1}$.
Here, $S_i$ denotes the set of $i$-cells of the lattice.
The ground states (or code states) are the common $+1$ eigenstates of stabilizers located at all vertices and plaquettes.
A plaquette stabilizer is given by the product of Pauli-$Z$ operators acting on all edges of the plaquette.
It ensures that for all configurations with non-zero ground-state amplitude, the $\zz_2$ sum of the qubits on the edges of each plaquette is $0$.
In other words, all ground-state configurations must have an even number of $\ket1$ qubits around each plaquette.
The following depicts an example configuration,
\begin{equation}
\label{eq:closed_loop_config}
\begin{tikzpicture}[alignmid]
\draw[step=0.8,orange] (0,0) grid (4.8,3.2);
\begin{scope}[scale=0.8]
\draw[configcol,line width=0.1cm] (0,1)--++(90:1) (1,1)--++(90:1) (2,1)--++(90:1) (2,2)--++(0:1) (2,3)--++(0:1) (3,3)--++(90:1) (4,3)--++(90:1) (4,3)--++(0:1) (4,2)--++(0:1) (5,2)--++(-90:1) (6,2)--++(-90:1);
\draw[membranecol,line width=0.1cm] (0,1.5)--++(0:2.5)--++(90:2)--++(0:2)--++(-90:2)--++(0:1.5);
\end{scope}
\end{tikzpicture}
\;,
\end{equation}
where the square lattice is drawn in orange and edges with $\ket1$ qubits are marked by thick black lines.
As shown in blue, marking the according edges of the Poincar\'e dual lattice yields a closed-loop pattern.

A vertex stabilizer is given by the product of Pauli-$X$ operators acting on all edges incident to the vertex.
It ensures that a ground state $\psi$ assigns the same amplitude to all configurations related by $\zz_2$-adding $1$ to all edges around a vertex.
In other words, the ground state amplitude is invariant under deforming the closed-loop pattern by adding small local loops, such as
\begin{equation}
\label{eq:state_local_deformation}
\left\langle\psi\middle|
\begin{tikzpicture}[alignmid]
\draw[step=0.8,orange] (0,0) grid (1.6,1.6);
\begin{scope}[scale=0.8]
\draw[configcol,line width=0.1cm] (0,1)--++(90:1) (1,1)--++(90:1) (2,1)--++(90:1);
\draw[membranecol,line width=0.1cm] (0,1.5)--++(0:2);
\end{scope}
\end{tikzpicture}
\right\rangle
=
\left\langle\psi\middle|
\begin{tikzpicture}[alignmid]
\draw[step=0.8,orange] (0,0) grid (1.6,1.6);
\begin{scope}[scale=0.8]
\draw[configcol,line width=0.1cm] (0,1)--++(90:1) (0,1)--++(0:1) (1,1)--++(-90:1) (1,1)--++(0:1) (2,1)--++(90:1);
\draw[membranecol,line width=0.1cm] (0,1.5)--++(0:0.5)--++(-90:1)--++(0:1)--++(90:1)--++(0:0.5);
\end{scope}
\end{tikzpicture}
\right\rangle
\;.
\end{equation}
One specific ground state is the equal-weight superposition of all closed-loop configurations,
\begin{equation}
\label{eq:toric_code_ground_state}
\ket\psi=\sum_{\textcolor{black}{a}\in \zz_2^{S_1}} \prod_{f\in S_2} \delta_{\sum_{e\in S_1(f)}\textcolor{black}{a}(e)=0} \ket{\textcolor{black}{a}}\;,
\end{equation}
where $S_1(f)$ denotes the set of edges adjacent to the face $f$, and the $\delta$ symbol is $1$ if the condition in its subscript holds and $0$ otherwise.
More generally, there are four equivalence classes of closed-loop patterns under local deformations, which are determined by the $\mmod 2$ number of loops winding around the torus in horizontal and vertical direction each.
In the closed-loop picture, representatives of these \emph{homology classes}
\footnote{
They are the $\zz_2$-valued degree-1 homology classes of the torus, which form the group $\zz_2\times \zz_2$ as indicated by the subscripts.
}
are given by
\begin{equation}
\label{eq:toric_code_cohomology_classes}
\begin{gathered}
B_{00}=
\begin{tikzpicture}
\draw[orange] (0,0)edge[mark={bar,p=0.4}]++(1.5,0) (0,1)edge[mark={bar,p=0.4}]++(1.5,0) (0,0)edge[mark={dbar,p=0.4}]++(0,1) (1.5,0)edge[mark={dbar,p=0.4}]++(0,1);
\end{tikzpicture}
\;,
\quad
B_{10}=
\begin{tikzpicture}
\draw[orange] (0,0)edge[mark={bar,p=0.4}]++(1.5,0) (0,1)edge[mark={bar,p=0.4}]++(1.5,0) (0,0)edge[mark={dbar,p=0.4}]++(0,1) (1.5,0)edge[mark={dbar,p=0.4}]++(0,1);
\draw[membranecol,line width=0.1cm] (0,0.5)--++(1.5,0);
\end{tikzpicture}
\;,\\
B_{01}=
\begin{tikzpicture}
\draw[orange] (0,0)edge[mark={bar,p=0.4}]++(1.5,0) (0,1)edge[mark={bar,p=0.4}]++(1.5,0) (0,0)edge[mark={dbar,p=0.4}]++(0,1) (1.5,0)edge[mark={dbar,p=0.4}]++(0,1);
\draw[membranecol,line width=0.1cm] (0.75,0)--++(0,1);
\end{tikzpicture}
\;,
\quad
B_{11}=
\begin{tikzpicture}
\draw[orange] (0,0)edge[mark={bar,p=0.4}]++(1.5,0) (0,1)edge[mark={bar,p=0.4}]++(1.5,0) (0,0)edge[mark={dbar,p=0.4}]++(0,1) (1.5,0)edge[mark={dbar,p=0.4}]++(0,1);
\draw[membranecol,line width=0.1cm] (0.75,0)to[out=90,in=180](1.5,0.5) (0,0.5)to[out=0,in=-90](0.75,1);
\end{tikzpicture}
\;.
\end{gathered}
\end{equation}
For example, the configuration shown in Eq.~\eqref{eq:closed_loop_config} belongs to the $B_{10}$ homology class.
A general ground state only needs to assign equal amplitudes to all closed-loop patterns in the same homology class.
So the ground-state space is four-dimensional, with a canonical basis given by superpositions $\ket{B_{00}}$, $\ket{B_{10}}$, $\ket{B_{01}}$, $\ket{B_{11}}$ that only assign a non-zero weight to one of the four homology classes in Eq.~\eqref{eq:toric_code_cohomology_classes}.

With the toric-code ground states in mind, let us describe the toric code as a path integral.
We first need to briefly describe what we mean by a path integral in general:
A (discrete) path integral is a summation over discrete ``field'' configurations of discrete variables distributed over some spacetime lattice.
The summand is a product over weights that are distributed over the same lattice and depend on the configuration of the nearby variables.
The path integrals that we use have a physical interpretation as a discretized imaginary-time evolution.
Accordingly, the toric-code path integral lives on a 3-cellulation (for our purposes, a cubic lattice) representing a $2+1$-dimensional Euclidean spacetime.
There is one variable valued in $\zz_2=\{0,1\}$ at every edge, so the discrete field configurations are elements of $\zz_2^{S_1}$, where $S_1$ denotes the set of edges of the lattice.
There is one weight at every face, which depends on the variables on its four edges:
If their $\zz_2$ sum is $0$, the weight is one, otherwise the weight is zero.
In other words, the path integral is a sum over all edge configurations with an even number of $1$s around every face
\footnote{
Note that the number $Z$ we get on a closed lattice is not very relevant for our purpose, but rather the evaluation of the path integral on some lattice with boundary which we discuss below.
}
:
\begin{equation}
\label{eq:toric_code_path_integral}
Z=\sum_{\textcolor{black}{A}\in \zz_2^{S_1}} \prod_{f\in S_2} \delta_{\sum_{e\in S_1(f)}\textcolor{black}{A}(e)=0}\;.
\end{equation}
If we consider the faces Poincar\'e dual to the $1$ edges, then the non-zero weight configurations become closed-membrane patterns.
The following picture shows a path-integral variable configuration $A$ (thick black lines) on a patch of cubic lattice (in orange), together with the dual closed-membrane pattern (in blue):
\begin{equation}
\label{eq:toric_code_membranes}
\raisebox{-0.5\height}{\includegraphics[scale=0.25]{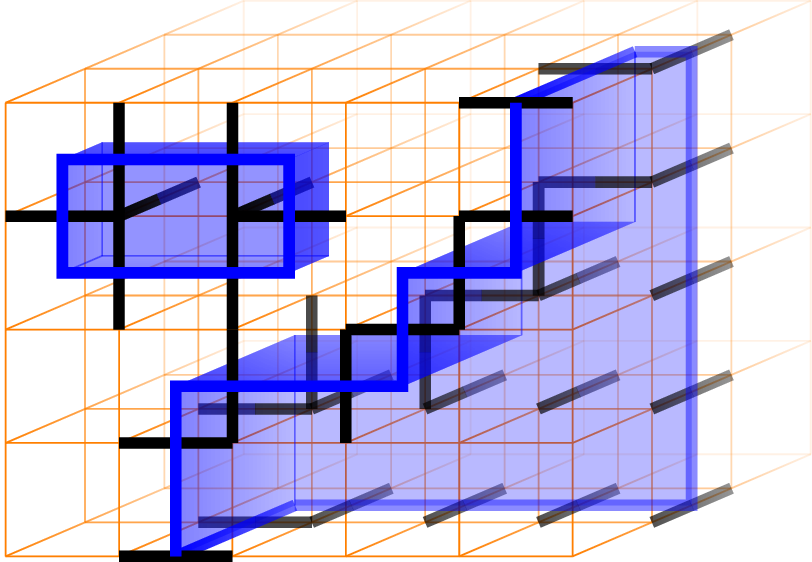}}\;.
\end{equation}
It is already rather evident that the path integral in Eq.~\eqref{eq:toric_code_path_integral} is the spacetime analogue of the ground state in Eq.~\eqref{eq:toric_code_ground_state}.
To see their precise relation we evaluate the path integral on a lattice with boundary, which we define as follows.
We fix a configuration of variables on the boundary edges, and sum only over variables at interior edges.
This results in one complex number for every fixed boundary configuration.
The collection of these complex numbers can be interpreted as the amplitudes of a state, supported on one qubit per boundary edge.
We will therefore refer to such boundaries as \emph{state boundaries}.
\footnote{
In Refs.~\cite{path_integral_qec, twisted_double_code}, these were called \emph{space boundaries}.
}.
It is not hard to see that if we evaluate the path integral in Eq.~\eqref{eq:toric_code_path_integral} on a lattice with boundary, we get the ground state in Eq.~\eqref{eq:toric_code_ground_state} up to normalization:
A closed-membrane configuration in the 3-dimensional spacetime restricts to a closed-loop configuration on the boundary, as can be seen in Eq.~\eqref{eq:toric_code_membranes}.
\footnote{
Technically, the boundary of the cubic block in Eq.~\eqref{eq:toric_code_membranes} is not a periodic-boundary square lattice, but the toric-code ground states (and toric-code path integral) can be analogously defined on arbitrary 2-cellulations (3-cellulations) of arbitrary 2-manifolds (3-manifolds).
Also note that which of the ground states we get at a state boundary depends on the topology of the spacetime 3-manifold.
}

For relating it to fault-tolerant circuits, it is useful to consider an alternative description of the path integral, namely as a tensor network
\footnote{Note that, unlike well-known tensor-network ansatzes like MPS or PEPS, a tensor-network path integral does not have open indices, except for where we terminate it at a 2-dimensional state boundary.}
.
The tensor network consists of two types of tensors, which we will refer to as the \emph{$\delta$-tensor},
\begin{equation}
\label{eq:delta_definition}
\begin{tikzpicture}
\atoms{delta}{0/}
\draw (0)edge[ind=$a$]++(0:0.5) (0)edge[ind=$b$]++(90:0.5) (0)edge[ind=$c$]++(180:0.5);
\node at (-90:0.5){$\ldots$};
\end{tikzpicture}
=
\delta_{a=b=c=\ldots}\;,
\end{equation}
and the \emph{$\zz_2$-tensor},
\begin{equation}
\begin{tikzpicture}
\atoms{z2}{0/}
\draw (0)edge[ind=$a$]++(0:0.5) (0)edge[ind=$b$]++(90:0.5) (0)edge[ind=$c$]++(180:0.5);
\node at (-90:0.5){$\ldots$};
\end{tikzpicture}
=
\delta_{a+b+c+\ldots=0}\;.
\end{equation}
These tensors are also known as phase-free $Z$ and $X$ spiders in the $ZX$ calculus \cite{Coecke2017,Wetering2020,Kissinger2022}.
For the path integral, we place one $\delta$-tensor at each edge, and one $\zz_2$-tensor at each face of the cubic lattice.
Then we add one bond (that is, one contracted index pair) connecting every pair of adjacent edge and face.
The following section shows a patch of the tensor network (in black and gray), supported on four cubes (in orange):
\begin{equation}
\begin{tikzpicture}
\atoms{void}{x/p={2,0}, y/p={0,2}, z/p={1.6,0.8}}
\foreach \x in {0,1,2}{
\foreach \y in {0,1,2}{
\foreach \z in {0,1}{
\atoms{void}{\x\y\z/p={$\x*(x)+\y*(y)+\z*(z)$}}
}}}
\draw[orange] (000)--(200) (010)--(210) (020)--(220) (000)--(020) (100)--(120) (200)--(220) (201)--(221) (021)--(221);
\draw[orange,back] (001)--(201) (011)--(211) (001)--(021) (101)--(121);
\draw[orange, back] (000)--(001) (100)--(101) (010)--(011) (110)--(111);
\draw[orange] (200)--(201) (210)--(211) (220)--(221) (020)--(021) (120)--(121);
\foreach \x/\y/\z in {1/0/0, 3/0/0, 1/2/0, 3/2/0, 1/4/0, 3/4/0, 0/1/0, 0/3/0, 2/1/0, 2/3/0, 4/1/0, 4/3/0, 4/0/1, 4/2/1, 0/4/1, 2/4/1, 4/4/1, 4/1/2, 4/3/2, 1/4/2, 3/4/2}{
\atoms{delta}{d\x\y\z/p={$0.5*\x*(x)+0.5*\y*(y)+0.5*\z*(z)$}}
}
\foreach \x/\y/\z in {1/0/2, 3/0/2, 1/2/2, 3/2/2, 0/1/2, 0/3/2, 2/1/2, 2/3/2, 2/2/1, 0/2/1, 0/0/1, 2/0/1}{
\atoms{delta,astyle=gray}{d\x\y\z/p={$0.5*\x*(x)+0.5*\y*(y)+0.5*\z*(z)$}}
}
\foreach \x/\y/\z in {1/1/0, 3/1/0, 1/3/0, 3/3/0, 1/4/1, 3/4/1, 4/1/1, 4/3/1}{
\atoms{z2}{z\x\y\z/p={$0.5*\x*(x)+0.5*\y*(y)+0.5*\z*(z)$}}
}
\foreach \x/\y/\z in {1/1/2, 3/1/2, 1/3/2, 3/3/2, 1/2/1, 3/2/1, 2/1/1, 2/3/1, 0/1/1, 0/3/1, 1/0/1, 3/0/1}{
\atoms{z2,astyle=gray}{z\x\y\z/p={$0.5*\x*(x)+0.5*\y*(y)+0.5*\z*(z)$}}
}
\draw (d010)--++($-0.2*(x)$) (d030)--++($-0.2*(x)$) (d410)--++($0.2*(x)$) (d430)--++($0.2*(x)$) (d340)--++($0.2*(y)$) (d100)--++($-0.2*(y)$) (d300)--++($-0.2*(y)$) (d401)--++($0.2*(x)$) (d401)--++($-0.2*(y)$) (d441)--++($0.2*(x)$) (d441)--++($0.2*(y)$) (d041)--++($-0.2*(x)$) (d041)--++($0.2*(y)$) (d142)--++($0.2*(y)$) (d342)--++($0.2*(y)$) (d142)--++($0.2*(z)$) (d342)--++($0.2*(z)$) (d030)--++($-0.2*(z)$) (d010)--++($-0.2*(z)$) (d100)--++($-0.2*(z)$) (d300)--++($-0.2*(z)$) (d410)--++($-0.2*(z)$) (d430)--++($-0.2*(z)$) (d340)--++($-0.2*(z)$) (d140)--++($-0.2*(z)$) (d140)--++($0.2*(y)$) (d421)--++($0.2*(x)$) (d210)--++($-0.2*(z)$) (d320)--++($-0.2*(z)$) (d230)--++($-0.2*(z)$) (d120)--++($-0.2*(z)$) (d412)--++($0.2*(z)$) (d412)--++($0.2*(x)$) (d432)--++($0.2*(x)$) (d432)--++($0.2*(z)$) (d241)--++($0.2*(y)$);
\draw[gray] (d201)--++($-0.2*(y)$) (d001)--++($-0.2*(x)$) (d001)--++($-0.2*(y)$) (d021)--++($-0.2*(x)$) (d032)--++($-0.2*(x)$) (d032)--++($0.2*(z)$) (d012)--++($0.2*(z)$) (d012)--++($-0.2*(x)$) (d102)--++($0.2*(z)$) (d102)--++($-0.2*(y)$) (d302)--++($0.2*(z)$) (d302)--++($-0.2*(y)$) (d212)--++($0.2*(z)$) (d232)--++($0.2*(z)$) (d122)--++($0.2*(z)$) (d322)--++($0.2*(z)$);
\foreach \x/\xx/\xxx in {0/1/2,2/3/4}{
\foreach \y/\yy/\yyy in {0/1/2,2/3/4}{
\foreach \z in {0}{
\draw (z\xx\yy\z)--(d\x\yy\z) (z\xx\yy\z)--(d\xxx\yy\z) (z\xx\yy\z)--(d\xx\y\z) (z\xx\yy\z)--(d\xx\yyy\z);
}
\foreach \z in {2}{
\draw[gray] (z\xx\yy\z)--(d\x\yy\z) (z\xx\yy\z)--(d\xxx\yy\z) (z\xx\yy\z)--(d\xx\y\z) (z\xx\yy\z)--(d\xx\yyy\z);
}}}
\foreach \x/\xx/\xxx in {0/1/2,2/3/4}{
\foreach \y in {0,2}{
\draw[gray] (z\xx\y1)--(d\x\y1) (z\xx\y1)--(d\xxx\y1) (z\xx\y1)--(d\xx\y0) (z\xx\y1)--(d\xx\y2);
}
\foreach \y in {4}{
\draw (z\xx\y1)--(d\x\y1) (z\xx\y1)--(d\xxx\y1) (z\xx\y1)--(d\xx\y0) (z\xx\y1)--(d\xx\y2);
}}
\foreach \y/\yy/\yyy in {0/1/2,2/3/4}{
\foreach \x in {0,2}{
\draw[gray] (z\x\yy1)--(d\x\y1) (z\x\yy1)--(d\x\yyy1) (z\x\yy1)--(d\x\yy0) (z\x\yy1)--(d\x\yy2);
}
\foreach \x in {4}{
\draw (z\x\yy1)--(d\x\y1) (z\x\yy1)--(d\x\yyy1) (z\x\yy1)--(d\x\yy0) (z\x\yy1)--(d\x\yy2);
}}
\end{tikzpicture}\;.
\end{equation}
When evaluating a tensor network, we sum over all configurations of each bond.
However, the $\delta$-tensors force all surrounding indices to take the same configuration, so the evaluation reduces to a sum over configurations of $\delta$-tensors.
The configuration of a $\delta$-tensor corresponds to the configuration of the path-integral variable at the same edge, and the $\zz_2$-tensor corresponds to the even-parity constraint at the same face.
This way we see that evaluating the tensor network indeed yields the path integral in Eq.~\eqref{eq:toric_code_path_integral}.
If we evaluate it on a lattice with boundary, we get a tensor supported on the open indices there, which again can be interpreted as a state supported on the boundary.
An advantage of the tensor-network formulation is that it makes the duality symmetry of the toric code evident:
The path integral is invariant under (1) swapping the primary lattice with its Poincar\'e dual and (2) exchanging $\delta$ and $\zz_2$-tensors.

In order to turn the toric-code path integral into a fault-tolerant circuit in Section~\ref{sec:circuit}, we will need to equip it with anyon worldlines
\footnote{For our use-case, it might be more fitting to call the anyon worldlines \emph{1-form symmetries} \cite{twisted_double_code}, since they will not occur at well-isolated regions but densely in spacetime.}
.
To motivate the form of these worldlines, let us first recall anyons in the toric-code ground states.
The toric code has two generating anyon types, namely $e$ anyons located at vertices, and $m$ anyons located at plaquettes.
To introduce an $m$ anyon, we modify the stabilizer code by flipping the sign of a plaquette stabilizer.
The ground state of the modified code is a sum over configurations where the $\zz_2$ sum of around the $m$ plaquettes is $1$ instead of $0$.
These configurations are not closed-loop patterns, but patterns of strings that terminate at the locations of $m$ anyons.
The following drawing shows one such configuration with two $m$ plaquettes marked in red,
\begin{equation}
\begin{tikzpicture}
\draw[step=0.8,orange] (0,0) grid (4.8,3.2);
\begin{scope}[scale=0.8]
\draw[configcol,line width=0.1cm] (0,1)--++(90:1) (1,1)--++(90:1) (2,1)--++(90:1) (2,2)--++(0:1) (2,3)--++(0:1) (3,3)--++(90:1) (4,3)--++(90:1) (5,2)--++(-90:1) (6,2)--++(-90:1);
\draw[line width=0.1cm,membranecol] (0,1.5)--++(0:2.5)--++(90:2)--++(0:2) (6,1.5)--++(180:1.5);
\end{scope}
\fill[red] (4.5*0.8,3.5*0.8)circle(0.15) (4.5*0.8,1.5*0.8)circle(0.15);
\end{tikzpicture}
\;.
\end{equation}
To introduce an $e$ anyon, we flip the sign of a vertex stabilizer.
The ground state of the modified code is not an equal-weight superposition anymore:
Instead, moving a loop past an $e$ anyon location yields a relative sign of $-1$.
The following picture illustrates this for an $e$ vertex marked in red,
\begin{equation}
%\label{eq:state_local_deformation}
\left\langle\psi\middle|
\begin{tikzpicture}[alignmid]
\draw[step=0.8,orange] (0,0) grid (1.6,1.6);
\begin{scope}[scale=0.8]
\draw[configcol,line width=0.1cm] (0,1)--++(90:1) (1,1)--++(90:1) (2,1)--++(90:1);
\draw[membranecol,line width=0.1cm] (0,1.5)--++(0:2);
\end{scope}
\fill[red] (1*0.8,1*0.8)circle(0.15);
\end{tikzpicture}
\right\rangle
=
(-1)\cdot
\left\langle\psi\middle|
\begin{tikzpicture}[alignmid]
\draw[step=0.8,orange] (0,0) grid (1.6,1.6);
\begin{scope}[scale=0.8]
\draw[configcol,line width=0.1cm] (0,1)--++(90:1) (0,1)--++(0:1) (1,1)--++(-90:1) (1,1)--++(0:1) (2,1)--++(90:1);
\draw[membranecol,line width=0.1cm] (0,1.5)--++(0:0.5)--++(-90:1)--++(0:1)--++(90:1)--++(0:0.5);
\end{scope}
\fill[red] (1*0.8,1*0.8)circle(0.15);
\end{tikzpicture}
\right\rangle
\;.
\end{equation}
All in all, one particular ground state is given by
\begin{equation}
\label{eq:toric_code_anyon_state}
\begin{multlined}
\ket\psi(w_e,w_m)\\
\propto\sum_{a\in \zz_2^{S_1}} \prod_{f\in S_2} \delta_{\sum_{e\in S_1(f)}a(e)=w_m(f)} (-1)^{\operatorname{wind}(a,w_e)} \ket a\;.
\end{multlined}
\end{equation}
Here, $w_m$ is the characteristic function $S_2\rightarrow\{0,1\}$ that assigns $1$ to all the plaquettes with $m$ anyons, and $w_e:S_0\rightarrow\{0,1\}$ is the analogous function for the $e$ anyons.
$\operatorname{wind}(a,w_e)$ is the (mod 2) number of times we need to cross an $e$ anyon in order to transform the string pattern into a reference pattern via local deformations as in Eq.~\eqref{eq:state_local_deformation}.
\footnote{
Different choices of the reference pattern only give rise to an additional global prefactor.
If there are no $m$ anyons, then a natural reference string pattern is the empty one, and $\operatorname{wind}(a,w_e)$ can be interpreted as the winding number of $a$ around $w_e$.
}

Having anyons in the ground state in mind, let us now show how to insert anyons into the path integral.
Since we are in spacetime, the anyons are no longer points, but become worldlines.
$m$ anyon worldlines are located on paths of faces whose Poincar\'e dual edges form a closed-loop configuration.
In the presence of such an $m$ anyon worldline, the path integral is modified as follows:
We sum over configurations with an odd instead of even parity of $1$ edges around every $m$-anyon face.
In the geometric picture, we do not sum over closed-membrane patterns, but patterns of surfaces that terminate at the $m$ anyon worldlines.

$e$ anyon worldlines are located on closed-loop configurations of edges of the primary lattice.
The modified path integral is not an equal-weight sum over all membrane patterns.
Instead, we get an additional weight of $(-1)$ for every edge that is both part of the $e$ anyon worldline and whose path-integral variable is in configuration $1$.
In the geometric picture, these edges correspond to points where the path-integral membranes intersect the $e$ anyon worldline.
The following picture illustrates a path-integral configuration with $m$ and $e$ worldlines, and the according $-1$ weights:
\begin{equation}
% this is in extra tex file and convertex to png since may viewers can't handle tikz shadings.
% convert -density 288 -quality 100 membrane_anyon_pic.pdf membran_anyon_pic.png
\raisebox{-0.5\height}{\includegraphics[scale=0.25]{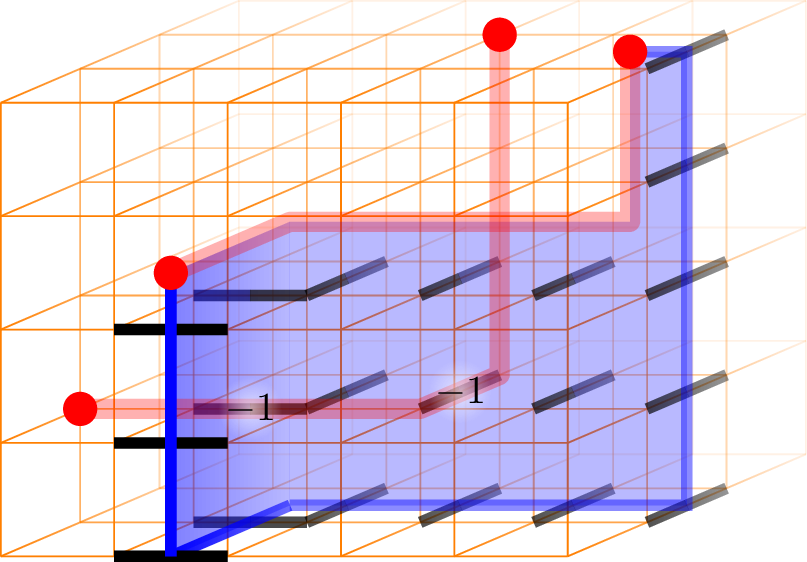}}\;.
\end{equation}
Overall, the path integral with anyons is given by
\begin{equation}
\label{eq:toric_code_anyon_integral}
\begin{multlined}
Z(W_e,W_m)\\
=\sum_{A\in \zz_2^{S_1}} \prod_{f\in S_2} \delta_{\sum_{e\in S_1(f)}A(e)=W_m(f)} \prod_{e\in S_1} (-1)^{A(e)\cdot W_e(e)}\;.
\end{multlined}
\end{equation}
Here, $W_m:S_2\rightarrow \{0,1\}$ and $W_e:S_1\rightarrow \{0,1\}$ are the characteristic functions encoding the locations of $m$ and $e$ worldlines.
Evaluating the path integral with anyon worldlines on a lattice with state boundary yields a toric-code ground state with anyons at the points where the worldlines terminate:
When we cross a string over an $e$ anyon in the ground-state configuration, this creates a spacetime intersection between the attached membrane and worldline.
Both the crossing and the intersection yield the same factor of $-1$.

Finally, $e$ and $m$ anyons can also be introduced in the tensor-network formulation of the path integral.
To this end, we use two new tensors
\footnote{
In the $ZX$ calculus, these would be spiders with a $\pi$ phase.
}
:
The \emph{charged $\delta$-tensor},
\begin{equation}
\label{eq:e_anyon_tensor}
\begin{tikzpicture}
\atoms{delta,bdastyle=red}{0/}
\draw (0)edge[ind=$a$]++(0:0.5) (0)edge[ind=$b$]++(90:0.5) (0)edge[ind=$c$]++(180:0.5);
\node at (-90:0.5){$\ldots$};
\end{tikzpicture}
=
(-1)^a \cdot \delta_{a=b=c=\ldots}\;,
\end{equation}
and the \emph{charged $\zz_2$-tensor},
\begin{equation}
\label{eq:m_anyon_tensor}
\begin{tikzpicture}
\atoms{z2,bdastyle=red}{0/}
\draw (0)edge[ind=$a$]++(0:0.5) (0)edge[ind=$b$]++(90:0.5) (0)edge[ind=$c$]++(180:0.5);
\node at (-90:0.5){$\ldots$};
\end{tikzpicture}
=
\delta_{a+b+c+\ldots=1}\;.
\end{equation}
Then we simply replace each $\delta$-tensor on an $e$-anyon worldline by a charged $\delta$-tensor, and each $\zz_2$-tensor on an $m$-anyon worldline by a charged $\zz_2$-tensor, for example:
\begin{equation}
\label{eq:anyon_configuration}
\begin{tikzpicture}
\atoms{void}{x/p={2,0}, y/p={0,2}, z/p={1.6,0.8}}
\foreach \x in {0,1,2}{
\foreach \y in {0,1,2}{
\foreach \z in {0,1}{
\atoms{void}{\x\y\z/p={$\x*(x)+\y*(y)+\z*(z)$}}
}}}
\foreach \x in {-1,1,3,5}{ % for m anyon worldlines
\foreach \y in {-1,1,3,5}{
\foreach \z in {-1,1,3}{
\atoms{void}{m\x\y\z/p={$0.5*\x*(x)+0.5*\y*(y)+0.5*\z*(z)$}}
}}}
\draw[orange] (000)--(200) (010)--(210) (020)--(220) (000)--(020) (100)--(120) (200)--(220) (201)--(221) (021)--(221);
\draw[orange,back] (001)--(201) (011)--(211) (001)--(021) (101)--(121);
\draw[orange, back] (000)--(001) (100)--(101) (010)--(011) (110)--(111);
\draw[orange] (200)--(201) (210)--(211) (220)--(221) (020)--(021) (120)--(121);
\draw[worldline] (m1-11)--(m111)--(m131)--(m331)--(m333);
\draw[worldline] ($(200)+(0:0.4)$)--(200)--(210)--(220)--(120)--(121)--++(0.4,0.2);
\foreach \x/\y/\z in {1/0/0, 3/0/0, 1/2/0, 3/2/0, 1/4/0, 3/4/0, 0/1/0, 0/3/0, 2/1/0, 2/3/0, 4/1/0, 4/3/0, 4/0/1, 4/2/1, 0/4/1, 2/4/1, 4/4/1, 4/1/2, 4/3/2, 1/4/2, 3/4/2}{
\atoms{delta}{d\x\y\z/p={$0.5*\x*(x)+0.5*\y*(y)+0.5*\z*(z)$}}
}
\foreach \x/\y/\z in {1/0/2, 3/0/2, 1/2/2, 3/2/2, 0/1/2, 0/3/2, 2/1/2, 2/3/2, 2/2/1, 0/2/1, 0/0/1, 2/0/1}{
\atoms{delta,astyle=gray}{d\x\y\z/p={$0.5*\x*(x)+0.5*\y*(y)+0.5*\z*(z)$}}
}
\foreach \x/\y/\z in {1/1/0, 3/1/0, 1/3/0, 3/3/0, 1/4/1, 3/4/1, 4/1/1, 4/3/1}{
\atoms{z2}{z\x\y\z/p={$0.5*\x*(x)+0.5*\y*(y)+0.5*\z*(z)$}}
}
\foreach \x/\y/\z in {1/1/2, 3/1/2, 1/3/2, 3/3/2, 1/2/1, 3/2/1, 2/1/1, 2/3/1, 0/1/1, 0/3/1, 1/0/1, 3/0/1}{
\atoms{z2,astyle=gray}{z\x\y\z/p={$0.5*\x*(x)+0.5*\y*(y)+0.5*\z*(z)$}}
}
\draw (d010)--++($-0.2*(x)$) (d030)--++($-0.2*(x)$) (d410)--++($0.2*(x)$) (d430)--++($0.2*(x)$) (d340)--++($0.2*(y)$) (d100)--++($-0.2*(y)$) (d300)--++($-0.2*(y)$) (d401)--++($0.2*(x)$) (d401)--++($-0.2*(y)$) (d441)--++($0.2*(x)$) (d441)--++($0.2*(y)$) (d041)--++($-0.2*(x)$) (d041)--++($0.2*(y)$) (d142)--++($0.2*(y)$) (d342)--++($0.2*(y)$) (d142)--++($0.2*(z)$) (d342)--++($0.2*(z)$) (d030)--++($-0.2*(z)$) (d010)--++($-0.2*(z)$) (d100)--++($-0.2*(z)$) (d300)--++($-0.2*(z)$) (d410)--++($-0.2*(z)$) (d430)--++($-0.2*(z)$) (d340)--++($-0.2*(z)$) (d140)--++($-0.2*(z)$) (d140)--++($0.2*(y)$) (d421)--++($0.2*(x)$) (d210)--++($-0.2*(z)$) (d320)--++($-0.2*(z)$) (d230)--++($-0.2*(z)$) (d120)--++($-0.2*(z)$) (d412)--++($0.2*(z)$) (d412)--++($0.2*(x)$) (d432)--++($0.2*(x)$) (d432)--++($0.2*(z)$) (d241)--++($0.2*(y)$);
\draw[gray] (d201)--++($-0.2*(y)$) (d001)--++($-0.2*(x)$) (d001)--++($-0.2*(y)$) (d021)--++($-0.2*(x)$) (d032)--++($-0.2*(x)$) (d032)--++($0.2*(z)$) (d012)--++($0.2*(z)$) (d012)--++($-0.2*(x)$) (d102)--++($0.2*(z)$) (d102)--++($-0.2*(y)$) (d302)--++($0.2*(z)$) (d302)--++($-0.2*(y)$) (d212)--++($0.2*(z)$) (d232)--++($0.2*(z)$) (d122)--++($0.2*(z)$) (d322)--++($0.2*(z)$);
\foreach \x/\xx/\xxx in {0/1/2,2/3/4}{
\foreach \y/\yy/\yyy in {0/1/2,2/3/4}{
\foreach \z in {0}{
\draw (z\xx\yy\z)--(d\x\yy\z) (z\xx\yy\z)--(d\xxx\yy\z) (z\xx\yy\z)--(d\xx\y\z) (z\xx\yy\z)--(d\xx\yyy\z);
}
\foreach \z in {2}{
\draw[gray] (z\xx\yy\z)--(d\x\yy\z) (z\xx\yy\z)--(d\xxx\yy\z) (z\xx\yy\z)--(d\xx\y\z) (z\xx\yy\z)--(d\xx\yyy\z);
}}}
\foreach \x/\xx/\xxx in {0/1/2,2/3/4}{
\foreach \y in {0,2}{
\draw[gray] (z\xx\y1)--(d\x\y1) (z\xx\y1)--(d\xxx\y1) (z\xx\y1)--(d\xx\y0) (z\xx\y1)--(d\xx\y2);
}
\foreach \y in {4}{
\draw (z\xx\y1)--(d\x\y1) (z\xx\y1)--(d\xxx\y1) (z\xx\y1)--(d\xx\y0) (z\xx\y1)--(d\xx\y2);
}}
\foreach \y/\yy/\yyy in {0/1/2,2/3/4}{
\foreach \x in {0,2}{
\draw[gray] (z\x\yy1)--(d\x\y1) (z\x\yy1)--(d\x\yyy1) (z\x\yy1)--(d\x\yy0) (z\x\yy1)--(d\x\yy2);
}
\foreach \x in {4}{
\draw (z\x\yy1)--(d\x\y1) (z\x\yy1)--(d\x\yyy1) (z\x\yy1)--(d\x\yy0) (z\x\yy1)--(d\x\yy2);
}}
\foreach \x/\y/\z in {1/0/1, 1/2/1, 2/3/1, 3/3/2}{
\atoms{z2,bdastyle=red}{d\x\y\z/p={$0.5*\x*(x)+0.5*\y*(y)+0.5*\z*(z)$}}
}
\foreach \x/\y/\z in {4/1/0, 4/3/0, 3/4/0, 2/4/1}{
\atoms{delta,bdastyle=red}{z\x\y\z/p={$0.5*\x*(x)+0.5*\y*(y)+0.5*\z*(z)$}}
}
\end{tikzpicture}\;.
\end{equation}
Finally, we note that the path integral and its anyon worldlines obey three properties that are crucial for fault-tolerance.
We will not explicitly prove these properties, but they follow either from topological invariance, from gauge invariance, or from the $ZX$ calculus rules, as shown in Refs.~\cite{path_integral_qec,twisted_double_code}.
First, any way of altering the path integral locally, in other words inserting any isolated ``defect'', only gives rise to a global prefactor when evaluating the path integral with state boundary
\footnote{
This holds as long as the defect is a constant minimum distance away from the state boundary.
Note that the prefactor can also be $0$.
}.
For example, if we replace the $\delta$-tensor at an edge by an arbitrary different tensor (marked with a cross below), the path integral evaluated on the four adjacent cubes will only change by a global prefactor,
\begin{equation}
\label{eq:fixed_point_property}
\begin{multlined}
\begin{tikzpicture}
\atoms{void}{x/p={1.5,0}, y/p={0,1.5}, z/p={1.2,0.6}}
%\atoms{void}{x/p={2,0}, y/p={0,2}, z/p={1.6,0.8}}
\foreach \x in {0,1,2}{
\foreach \y in {0,1,2}{
\foreach \z in {0,1}{
\atoms{void}{\x\y\z/p={$\x*(x)+\y*(y)+\z*(z)$}}
}}}
\draw[orange] (000)--(200) (010)--(210) (020)--(220) (000)--(020) (100)--(120) (200)--(220) (201)--(221) (021)--(221);
\draw[orange,back] (001)--(201) (011)--(211) (001)--(021) (101)--(121);
\draw[orange, back] (000)--(001) (100)--(101) (010)--(011) (110)--(111);
\draw[orange] (200)--(201) (210)--(211) (220)--(221) (020)--(021) (120)--(121);
\foreach \x/\y/\z in {1/0/0, 3/0/0, 1/2/0, 3/2/0, 1/4/0, 3/4/0, 0/1/0, 0/3/0, 2/1/0, 2/3/0, 4/1/0, 4/3/0, 4/0/1, 4/2/1, 0/4/1, 2/4/1, 4/4/1, 4/1/2, 4/3/2, 1/4/2, 3/4/2}{
\atoms{delta}{d\x\y\z/p={$0.5*\x*(x)+0.5*\y*(y)+0.5*\z*(z)$}}
}
\foreach \x/\y/\z in {1/0/2, 3/0/2, 1/2/2, 3/2/2, 0/1/2, 0/3/2, 2/1/2, 2/3/2, 2/2/1, 0/2/1, 0/0/1, 2/0/1}{
\atoms{delta,astyle=gray}{d\x\y\z/p={$0.5*\x*(x)+0.5*\y*(y)+0.5*\z*(z)$}}
}
\foreach \x/\y/\z in {1/1/0, 3/1/0, 1/3/0, 3/3/0, 1/4/1, 3/4/1, 4/1/1, 4/3/1}{
\atoms{z2}{z\x\y\z/p={$0.5*\x*(x)+0.5*\y*(y)+0.5*\z*(z)$}}
}
\foreach \x/\y/\z in {1/1/2, 3/1/2, 1/3/2, 3/3/2, 1/2/1, 3/2/1, 2/1/1, 2/3/1, 0/1/1, 0/3/1, 1/0/1, 3/0/1}{
\atoms{z2,astyle=gray}{z\x\y\z/p={$0.5*\x*(x)+0.5*\y*(y)+0.5*\z*(z)$}}
}
\draw (d010)--++($-0.2*(x)$) (d030)--++($-0.2*(x)$) (d410)--++($0.2*(x)$) (d430)--++($0.2*(x)$) (d340)--++($0.2*(y)$) (d100)--++($-0.2*(y)$) (d300)--++($-0.2*(y)$) (d401)--++($0.2*(x)$) (d401)--++($-0.2*(y)$) (d441)--++($0.2*(x)$) (d441)--++($0.2*(y)$) (d041)--++($-0.2*(x)$) (d041)--++($0.2*(y)$) (d142)--++($0.2*(y)$) (d342)--++($0.2*(y)$) (d142)--++($0.2*(z)$) (d342)--++($0.2*(z)$) (d030)--++($-0.2*(z)$) (d010)--++($-0.2*(z)$) (d100)--++($-0.2*(z)$) (d300)--++($-0.2*(z)$) (d410)--++($-0.2*(z)$) (d430)--++($-0.2*(z)$) (d340)--++($-0.2*(z)$) (d140)--++($-0.2*(z)$) (d140)--++($0.2*(y)$) (d421)--++($0.2*(x)$) (d210)--++($-0.2*(z)$) (d320)--++($-0.2*(z)$) (d230)--++($-0.2*(z)$) (d120)--++($-0.2*(z)$) (d412)--++($0.2*(z)$) (d412)--++($0.2*(x)$) (d432)--++($0.2*(x)$) (d432)--++($0.2*(z)$) (d241)--++($0.2*(y)$);
\draw[gray] (d201)--++($-0.2*(y)$) (d001)--++($-0.2*(x)$) (d001)--++($-0.2*(y)$) (d021)--++($-0.2*(x)$) (d032)--++($-0.2*(x)$) (d032)--++($0.2*(z)$) (d012)--++($0.2*(z)$) (d012)--++($-0.2*(x)$) (d102)--++($0.2*(z)$) (d102)--++($-0.2*(y)$) (d302)--++($0.2*(z)$) (d302)--++($-0.2*(y)$) (d212)--++($0.2*(z)$) (d232)--++($0.2*(z)$) (d122)--++($0.2*(z)$) (d322)--++($0.2*(z)$);
\foreach \x/\xx/\xxx in {0/1/2,2/3/4}{
\foreach \y/\yy/\yyy in {0/1/2,2/3/4}{
\foreach \z in {0}{
\draw (z\xx\yy\z)--(d\x\yy\z) (z\xx\yy\z)--(d\xxx\yy\z) (z\xx\yy\z)--(d\xx\y\z) (z\xx\yy\z)--(d\xx\yyy\z);
}
\foreach \z in {2}{
\draw[gray] (z\xx\yy\z)--(d\x\yy\z) (z\xx\yy\z)--(d\xxx\yy\z) (z\xx\yy\z)--(d\xx\y\z) (z\xx\yy\z)--(d\xx\yyy\z);
}}}
\foreach \x/\xx/\xxx in {0/1/2,2/3/4}{
\foreach \y in {0,2}{
\draw[gray] (z\xx\y1)--(d\x\y1) (z\xx\y1)--(d\xxx\y1) (z\xx\y1)--(d\xx\y0) (z\xx\y1)--(d\xx\y2);
}
\foreach \y in {4}{
\draw (z\xx\y1)--(d\x\y1) (z\xx\y1)--(d\xxx\y1) (z\xx\y1)--(d\xx\y0) (z\xx\y1)--(d\xx\y2);
}}
\foreach \y/\yy/\yyy in {0/1/2,2/3/4}{
\foreach \x in {0,2}{
\draw[gray] (z\x\yy1)--(d\x\y1) (z\x\yy1)--(d\x\yyy1) (z\x\yy1)--(d\x\yy0) (z\x\yy1)--(d\x\yy2);
}
\foreach \x in {4}{
\draw (z\x\yy1)--(d\x\y1) (z\x\yy1)--(d\x\yyy1) (z\x\yy1)--(d\x\yy0) (z\x\yy1)--(d\x\yy2);
}}
\atoms{circ,small,dcross,bdastyle=gray}{w/p={$0.5*2*(x)+0.5*2*(y)+0.5*1*(z)$}}
\end{tikzpicture}\\
\propto
\begin{tikzpicture}
\atoms{void}{x/p={1.5,0}, y/p={0,1.5}, z/p={1.2,0.6}}
%\atoms{void}{x/p={2,0}, y/p={0,2}, z/p={1.6,0.8}}
\foreach \x in {0,1,2}{
\foreach \y in {0,1,2}{
\foreach \z in {0,1}{
\atoms{void}{\x\y\z/p={$\x*(x)+\y*(y)+\z*(z)$}}
}}}
\draw[orange] (000)--(200) (010)--(210) (020)--(220) (000)--(020) (100)--(120) (200)--(220) (201)--(221) (021)--(221);
\draw[orange,back] (001)--(201) (011)--(211) (001)--(021) (101)--(121);
\draw[orange, back] (000)--(001) (100)--(101) (010)--(011) (110)--(111);
\draw[orange] (200)--(201) (210)--(211) (220)--(221) (020)--(021) (120)--(121);
\foreach \x/\y/\z in {1/0/0, 3/0/0, 1/2/0, 3/2/0, 1/4/0, 3/4/0, 0/1/0, 0/3/0, 2/1/0, 2/3/0, 4/1/0, 4/3/0, 4/0/1, 4/2/1, 0/4/1, 2/4/1, 4/4/1, 4/1/2, 4/3/2, 1/4/2, 3/4/2}{
\atoms{delta}{d\x\y\z/p={$0.5*\x*(x)+0.5*\y*(y)+0.5*\z*(z)$}}
}
\foreach \x/\y/\z in {1/0/2, 3/0/2, 1/2/2, 3/2/2, 0/1/2, 0/3/2, 2/1/2, 2/3/2, 2/2/1, 0/2/1, 0/0/1, 2/0/1}{
\atoms{delta,astyle=gray}{d\x\y\z/p={$0.5*\x*(x)+0.5*\y*(y)+0.5*\z*(z)$}}
}
\foreach \x/\y/\z in {1/1/0, 3/1/0, 1/3/0, 3/3/0, 1/4/1, 3/4/1, 4/1/1, 4/3/1}{
\atoms{z2}{z\x\y\z/p={$0.5*\x*(x)+0.5*\y*(y)+0.5*\z*(z)$}}
}
\foreach \x/\y/\z in {1/1/2, 3/1/2, 1/3/2, 3/3/2, 1/2/1, 3/2/1, 2/1/1, 2/3/1, 0/1/1, 0/3/1, 1/0/1, 3/0/1}{
\atoms{z2,astyle=gray}{z\x\y\z/p={$0.5*\x*(x)+0.5*\y*(y)+0.5*\z*(z)$}}
}
\draw (d010)--++($-0.2*(x)$) (d030)--++($-0.2*(x)$) (d410)--++($0.2*(x)$) (d430)--++($0.2*(x)$) (d340)--++($0.2*(y)$) (d100)--++($-0.2*(y)$) (d300)--++($-0.2*(y)$) (d401)--++($0.2*(x)$) (d401)--++($-0.2*(y)$) (d441)--++($0.2*(x)$) (d441)--++($0.2*(y)$) (d041)--++($-0.2*(x)$) (d041)--++($0.2*(y)$) (d142)--++($0.2*(y)$) (d342)--++($0.2*(y)$) (d142)--++($0.2*(z)$) (d342)--++($0.2*(z)$) (d030)--++($-0.2*(z)$) (d010)--++($-0.2*(z)$) (d100)--++($-0.2*(z)$) (d300)--++($-0.2*(z)$) (d410)--++($-0.2*(z)$) (d430)--++($-0.2*(z)$) (d340)--++($-0.2*(z)$) (d140)--++($-0.2*(z)$) (d140)--++($0.2*(y)$) (d421)--++($0.2*(x)$) (d210)--++($-0.2*(z)$) (d320)--++($-0.2*(z)$) (d230)--++($-0.2*(z)$) (d120)--++($-0.2*(z)$) (d412)--++($0.2*(z)$) (d412)--++($0.2*(x)$) (d432)--++($0.2*(x)$) (d432)--++($0.2*(z)$) (d241)--++($0.2*(y)$);
\draw[gray] (d201)--++($-0.2*(y)$) (d001)--++($-0.2*(x)$) (d001)--++($-0.2*(y)$) (d021)--++($-0.2*(x)$) (d032)--++($-0.2*(x)$) (d032)--++($0.2*(z)$) (d012)--++($0.2*(z)$) (d012)--++($-0.2*(x)$) (d102)--++($0.2*(z)$) (d102)--++($-0.2*(y)$) (d302)--++($0.2*(z)$) (d302)--++($-0.2*(y)$) (d212)--++($0.2*(z)$) (d232)--++($0.2*(z)$) (d122)--++($0.2*(z)$) (d322)--++($0.2*(z)$);
\foreach \x/\xx/\xxx in {0/1/2,2/3/4}{
\foreach \y/\yy/\yyy in {0/1/2,2/3/4}{
\foreach \z in {0}{
\draw (z\xx\yy\z)--(d\x\yy\z) (z\xx\yy\z)--(d\xxx\yy\z) (z\xx\yy\z)--(d\xx\y\z) (z\xx\yy\z)--(d\xx\yyy\z);
}
\foreach \z in {2}{
\draw[gray] (z\xx\yy\z)--(d\x\yy\z) (z\xx\yy\z)--(d\xxx\yy\z) (z\xx\yy\z)--(d\xx\y\z) (z\xx\yy\z)--(d\xx\yyy\z);
}}}
\foreach \x/\xx/\xxx in {0/1/2,2/3/4}{
\foreach \y in {0,2}{
\draw[gray] (z\xx\y1)--(d\x\y1) (z\xx\y1)--(d\xxx\y1) (z\xx\y1)--(d\xx\y0) (z\xx\y1)--(d\xx\y2);
}
\foreach \y in {4}{
\draw (z\xx\y1)--(d\x\y1) (z\xx\y1)--(d\xxx\y1) (z\xx\y1)--(d\xx\y0) (z\xx\y1)--(d\xx\y2);
}}
\foreach \y/\yy/\yyy in {0/1/2,2/3/4}{
\foreach \x in {0,2}{
\draw[gray] (z\x\yy1)--(d\x\y1) (z\x\yy1)--(d\x\yyy1) (z\x\yy1)--(d\x\yy0) (z\x\yy1)--(d\x\yy2);
}
\foreach \x in {4}{
\draw (z\x\yy1)--(d\x\y1) (z\x\yy1)--(d\x\yyy1) (z\x\yy1)--(d\x\yy0) (z\x\yy1)--(d\x\yy2);
}}
\end{tikzpicture}\;.
\end{multlined}
\end{equation}
Second, the path integral evaluates to $0$ if there is a spacetime point where an odd number of anyon worldlines meet, for example,
\begin{equation}
\label{eq:cocycle_constraint}
\begin{tikzpicture}
\atoms{void}{x/p={1.5,0}, y/p={0,1.5}, z/p={1.2,0.6}}
%\atoms{void}{x/p={2,0}, y/p={0,2}, z/p={1.6,0.8}}
\foreach \x in {0,1}{
\foreach \y in {0,1}{
\foreach \z in {0,1}{
\atoms{void}{\x\y\z/p={$\x*(x)+\y*(y)+\z*(z)$}}
}}}
\foreach \x in {-1,1,3}{
\foreach \y in {-1,1,3}{
\foreach \z in {-1,1,3}{
\atoms{void}{m\x\y\z/p={$0.5*\x*(x)+0.5*\y*(y)+0.5*\z*(z)$}}
}}}
\draw[orange] (000)--(100)--(110)--(010)--(000) (100)--(101)--(111)--(110) (010)--(011)--(111);
\draw[orange,back] (000)--(001)--(011) (001)--(101);
\draw[worldline] (m-111)--(m111) (m311)--(m111) (m111)--(m113);
\foreach \x/\y/\z in {1/0/0, 1/2/0, 0/1/0, 2/1/0, 2/0/1, 0/2/1, 2/2/1, 1/2/2, 2/1/2}{
\atoms{delta}{d\x\y\z/p={$0.5*\x*(x)+0.5*\y*(y)+0.5*\z*(z)$}}
}
\foreach \x/\y/\z in {1/0/2, 0/1/2, 0/0/1}{
\atoms{delta,astyle=gray}{d\x\y\z/p={$0.5*\x*(x)+0.5*\y*(y)+0.5*\z*(z)$}}
}
\foreach \x/\y/\z in {1/1/0, 1/2/1, 2/1/1}{
\atoms{z2}{z\x\y\z/p={$0.5*\x*(x)+0.5*\y*(y)+0.5*\z*(z)$}}
}
\foreach \x/\y/\z in {1/1/2, 1/0/1, 0/1/1}{
\atoms{z2,astyle=gray}{z\x\y\z/p={$0.5*\x*(x)+0.5*\y*(y)+0.5*\z*(z)$}}
}
\draw (z110)--(d100) (z110)--(d210) (z110)--(d120) (z110)--(d010) (z211)--(d201) (z211)--(d210) (z211)--(d212) (z211)--(d221) (z121)--(d221) (z121)--(d120) (z121)--(d021) (z121)--(d122);
\draw[gray] (z112)--(d102) (z112)--(d212) (z112)--(d122) (z112)--(d012) (z011)--(d001) (z011)--(d010) (z011)--(d012) (z011)--(d021) (z101)--(d201) (z101)--(d100) (z101)--(d001) (z101)--(d102);
\draw (d010)--++($-0.2*(x)$) (d010)--++($-0.2*(z)$) (d210)--++($0.2*(x)$) (d210)--++($-0.2*(z)$) (d212)--++($0.2*(x)$) (d212)--++($0.2*(z)$) (d100)--++($-0.2*(y)$) (d100)--++($-0.2*(x)$) (d201)--++($0.2*(x)$) (d201)--++($-0.2*(y)$) (d120)--++($-0.2*(z)$) (d120)--++($0.2*(y)$) (d221)--++($0.2*(x)$) (d221)--++($0.2*(y)$) (d122)--++($-0.2*(z)$) (d122)--++($0.2*(y)$) (d021)--++($-0.2*(x)$) (d021)--++($0.2*(y)$);
\draw[gray] (d001)--++($-0.2*(x)$) (d001)--++($-0.2*(y)$) (d102)--++($0.2*(z)$) (d102)--++($-0.2*(y)$) (d012)--++($-0.2*(x)$) (d012)--++($0.2*(z)$);
\foreach \x/\y/\z in {0/1/1, 2/1/1, 1/1/2}{
\atoms{z2,bdastyle=red}{d\x\y\z/p={$0.5*\x*(x)+0.5*\y*(y)+0.5*\z*(z)$}}
}
\end{tikzpicture}
=
0\;.
\end{equation}
Third, the path integral is invariant under local deformations of the anyon worldlines, such as,
\begin{equation}
\label{eq:anyon_worldline_invariance}
\begin{tikzpicture}
\atoms{void}{x/p={1.5,0}, y/p={0,1.5}, z/p={1.2,0.6}}
\foreach \x in {0,1}{
\foreach \y in {0,1}{
\foreach \z in {0,1}{
\atoms{void}{\x\y\z/p={$\x*(x)+\y*(y)+\z*(z)$}}
}}}
\foreach \x in {-1,1,3}{
\foreach \y in {-1,1,3}{
\foreach \z in {-1,1,3}{
\atoms{void}{m\x\y\z/p={$0.5*\x*(x)+0.5*\y*(y)+0.5*\z*(z)$}}
}}}
\draw[orange] (000)--(100)--(110)--(010)--(000) (100)--(101)--(111)--(110) (010)--(011)--(111);
\draw[orange,back] (000)--(001)--(011) (001)--(101);
\draw[worldline] ($(000)+(180:0.4)$)--(000)--(010)--(110)--(111)--++(0:0.4);
\foreach \x/\y/\z in {1/0/0, 1/2/0, 0/1/0, 2/1/0, 2/0/1, 0/2/1, 2/2/1, 1/2/2, 2/1/2}{
\atoms{delta}{d\x\y\z/p={$0.5*\x*(x)+0.5*\y*(y)+0.5*\z*(z)$}}
}
\foreach \x/\y/\z in {1/0/2, 0/1/2, 0/0/1}{
\atoms{delta,astyle=gray}{d\x\y\z/p={$0.5*\x*(x)+0.5*\y*(y)+0.5*\z*(z)$}}
}
\foreach \x/\y/\z in {1/1/0, 1/2/1, 2/1/1}{
\atoms{z2}{z\x\y\z/p={$0.5*\x*(x)+0.5*\y*(y)+0.5*\z*(z)$}}
}
\foreach \x/\y/\z in {1/1/2, 1/0/1, 0/1/1}{
\atoms{z2,astyle=gray}{z\x\y\z/p={$0.5*\x*(x)+0.5*\y*(y)+0.5*\z*(z)$}}
}
\draw (z110)--(d100) (z110)--(d210) (z110)--(d120) (z110)--(d010) (z211)--(d201) (z211)--(d210) (z211)--(d212) (z211)--(d221) (z121)--(d221) (z121)--(d120) (z121)--(d021) (z121)--(d122);
\draw[gray] (z112)--(d102) (z112)--(d212) (z112)--(d122) (z112)--(d012) (z011)--(d001) (z011)--(d010) (z011)--(d012) (z011)--(d021) (z101)--(d201) (z101)--(d100) (z101)--(d001) (z101)--(d102);
\draw (d010)--++($-0.2*(x)$) (d010)--++($-0.2*(z)$) (d210)--++($0.2*(x)$) (d210)--++($-0.2*(z)$) (d212)--++($0.2*(x)$) (d212)--++($0.2*(z)$) (d100)--++($-0.2*(y)$) (d100)--++($-0.2*(x)$) (d201)--++($0.2*(x)$) (d201)--++($-0.2*(y)$) (d120)--++($-0.2*(z)$) (d120)--++($0.2*(y)$) (d221)--++($0.2*(x)$) (d221)--++($0.2*(y)$) (d122)--++($-0.2*(z)$) (d122)--++($0.2*(y)$) (d021)--++($-0.2*(x)$) (d021)--++($0.2*(y)$);
\draw[gray] (d001)--++($-0.2*(x)$) (d001)--++($-0.2*(y)$) (d102)--++($0.2*(z)$) (d102)--++($-0.2*(y)$) (d012)--++($-0.2*(x)$) (d012)--++($0.2*(z)$);
\foreach \x/\y/\z in {0/1/0, 1/2/0, 2/2/1}{
\atoms{delta,bdastyle=red}{d\x\y\z/p={$0.5*\x*(x)+0.5*\y*(y)+0.5*\z*(z)$}}
}
\end{tikzpicture}
=
\begin{tikzpicture}
\atoms{void}{x/p={1.5,0}, y/p={0,1.5}, z/p={1.2,0.6}}
%\atoms{void}{x/p={2,0}, y/p={0,2}, z/p={1.6,0.8}}
\foreach \x in {0,1}{
\foreach \y in {0,1}{
\foreach \z in {0,1}{
\atoms{void}{\x\y\z/p={$\x*(x)+\y*(y)+\z*(z)$}}
}}}
\foreach \x in {-1,1,3}{
\foreach \y in {-1,1,3}{
\foreach \z in {-1,1,3}{
\atoms{void}{m\x\y\z/p={$0.5*\x*(x)+0.5*\y*(y)+0.5*\z*(z)$}}
}}}
\draw[orange] (000)--(100)--(110)--(010)--(000) (100)--(101)--(111)--(110) (010)--(011)--(111);
\draw[orange,back] (000)--(001)--(011) (001)--(101);
\draw[worldline] ($(000)+(180:0.4)$)--(000)--(010)--(011)--(111)--++(0:0.4);% (000)--(100)--(101)--(001)--cycle;
\foreach \x/\y/\z in {1/0/0, 1/2/0, 0/1/0, 2/1/0, 2/0/1, 0/2/1, 2/2/1, 1/2/2, 2/1/2}{
\atoms{delta}{d\x\y\z/p={$0.5*\x*(x)+0.5*\y*(y)+0.5*\z*(z)$}}
}
\foreach \x/\y/\z in {1/0/2, 0/1/2, 0/0/1}{
\atoms{delta,astyle=gray}{d\x\y\z/p={$0.5*\x*(x)+0.5*\y*(y)+0.5*\z*(z)$}}
}
\foreach \x/\y/\z in {1/1/0, 1/2/1, 2/1/1}{
\atoms{z2}{z\x\y\z/p={$0.5*\x*(x)+0.5*\y*(y)+0.5*\z*(z)$}}
}
\foreach \x/\y/\z in {1/1/2, 1/0/1, 0/1/1}{
\atoms{z2,astyle=gray}{z\x\y\z/p={$0.5*\x*(x)+0.5*\y*(y)+0.5*\z*(z)$}}
}
\draw (z110)--(d100) (z110)--(d210) (z110)--(d120) (z110)--(d010) (z211)--(d201) (z211)--(d210) (z211)--(d212) (z211)--(d221) (z121)--(d221) (z121)--(d120) (z121)--(d021) (z121)--(d122);
\draw[gray] (z112)--(d102) (z112)--(d212) (z112)--(d122) (z112)--(d012) (z011)--(d001) (z011)--(d010) (z011)--(d012) (z011)--(d021) (z101)--(d201) (z101)--(d100) (z101)--(d001) (z101)--(d102);
\draw (d010)--++($-0.2*(x)$) (d010)--++($-0.2*(z)$) (d210)--++($0.2*(x)$) (d210)--++($-0.2*(z)$) (d212)--++($0.2*(x)$) (d212)--++($0.2*(z)$) (d100)--++($-0.2*(y)$) (d100)--++($-0.2*(x)$) (d201)--++($0.2*(x)$) (d201)--++($-0.2*(y)$) (d120)--++($-0.2*(z)$) (d120)--++($0.2*(y)$) (d221)--++($0.2*(x)$) (d221)--++($0.2*(y)$) (d122)--++($-0.2*(z)$) (d122)--++($0.2*(y)$) (d021)--++($-0.2*(x)$) (d021)--++($0.2*(y)$);
\draw[gray] (d001)--++($-0.2*(x)$) (d001)--++($-0.2*(y)$) (d102)--++($0.2*(z)$) (d102)--++($-0.2*(y)$) (d012)--++($-0.2*(x)$) (d012)--++($0.2*(z)$);
\foreach \x/\y/\z in {0/1/0, 0/2/1, 1/2/2}{%, 1/0/0, 2/0/1, 1/0/2, 0/0/1}{
\atoms{delta,bdastyle=red}{d\x\y\z/p={$0.5*\x*(x)+0.5*\y*(y)+0.5*\z*(z)$}}
}
\end{tikzpicture}\;.
\end{equation}

\section{The \texorpdfstring{$x+y$}{x+y} fault-tolerant circuit}
\label{sec:circuit}
In this section, we construct a fault-tolerant circuit from the toric-code path integral and its anyon worldlines discussed in Section~\ref{sec:path_integral}.
To motivate the construction, we point out that every geometrically local quantum circuit is a discrete path integral.
As a simple example, consider a translation-invariant circuit given by applying a 2-qubit unitary $U$ in a brick-layer fashion to a ring of $2n$ qubits, for $T$ time periods.
If we let $U_i$ denote $U$ applied to the $i$th and $i+1$th qubit (understood $\mmod 2n$), the overall unitary operator $\mathbf U$ is given by
\begin{equation}
\label{eq:brick_ciruit_formula}
\mathbf U = \big(\prod_{0\leq i<n} U_{2i-1} \prod_{0\leq i<n} U_{2i} \big)^T\;.
\end{equation}
For $n=3$ and $T=3$, we get the following circuit diagram with left and right identified:
\begin{equation}
\label{eq:brickwork_circuit_diagram}
\begin{gathered}
\bra{c_{6,0}c_{6,1}c_{6,2}c_{6,3}c_{6,4}c_{6,5}}\mathbf U \ket{c_{0,0}c_{0,1}c_{0,2}c_{0,3}c_{0,4}c_{0,5}}\\
=
\begin{tikzpicture}
\foreach \y in {0,1,2}{
\foreach \x in {0,1,2}{
\atoms{small,circ,dot,lab={t=$U$,p=90:0.3}}{{a\x\y/p={\x*1.5,\y*1.2}}}
}};
\foreach \y in {0,1,2}{
\foreach \x in {0,1,2}{
\atoms{small,circ,dot,lab={t=$U$,p=90:0.3}}{{b\x\y/p={\x*1.5+0.75,\y*1.2+0.6}}}
}};
\draw (a00)--(b00) (a10)--(b00) (a10)--(b10) (a20)--(b10) (a20)--(b20);
\draw (a01)--(b00) (a11)--(b00) (a11)--(b10) (a21)--(b10) (a21)--(b20);
\draw (a01)--(b01) (a11)--(b01) (a11)--(b11) (a21)--(b11) (a21)--(b21);
\draw (a02)--(b01) (a12)--(b01) (a12)--(b11) (a22)--(b11) (a22)--(b21);
\draw (a02)--(b02) (a12)--(b02) (a12)--(b12) (a22)--(b12) (a22)--(b22);
\draw (a00)edge[ind=$c_{0,0}$]++(-135:0.4) (a00)edge[ind=$c_{0,1}$]++(-45:0.4) (a10)edge[ind=$c_{0,2}$]++(-135:0.4) (a10)edge[ind=$c_{0,3}$]++(-45:0.4) (a20)edge[ind=$c_{0,4}$]++(-135:0.4) (a20)edge[ind=$c_{0,5}$]++(-45:0.4);
\draw (b02)edge[ind=$c_{6,1}$]++(135:0.4) (b02)edge[ind=$c_{6,2}$]++(45:0.4) (b12)edge[ind=$c_{6,3}$]++(135:0.4) (b12)edge[ind=$c_{6,4}$]++(45:0.4) (b22)edge[ind=$c_{6,5}$]++(135:0.4) (b22)edge[ind=$c_{6,0}$]++(45:0.4);
\draw (a00)edge[]++(135:0.4) (a01)edge[]++(-135:0.4) (a01)edge[]++(135:0.4) (a02)edge[]++(-135:0.4) (a02)edge[]++(135:0.4);
\draw (b20)edge[]++(45:0.4) (b20)edge[]++(-45:0.4) (b21)edge[]++(-45:0.4) (b21)edge[]++(45:0.4) (b22)edge[]++(-45:0.4);
\end{tikzpicture}\;.
\end{gathered}
\end{equation}
Here, $c_0\coloneqq (c_{0,0},c_{0,1},\ldots)$ and $c_{2T}\coloneqq (c_{2T,0},c_{2T,1},\ldots)$ denote the qubit configurations at the input and output of $\mathbf U$, respectively.
Note that throughout this paper, time is assumed to go upwards, unless we show a cubic spacetime lattice, in which case time goes towards the top right in the $x+y$ direction.
By inserting resolutions of the identity between each layer of operators in Eq.~\eqref{eq:brick_ciruit_formula}, we can express the amplitude $\bra{c_{2T}}\mathbf U\ket{c_0}$ as a sum over all qubit configurations $\mathbf c = \{c_{t,i}\}_{0< t<2T,0\leq i<2n} \in \{0,1\}^{(2T-1)\times 2n}$ at all intermediate time steps,
\begin{equation}
\begin{gathered}
\bra{c_{2T}}\mathbf U\ket{c_0}\\
= \sum_{\mathbf c} \bra{c_{2T}}
\prod_{1\leq t<T} \Big(\prod_{0\leq i<n} U_{2i}\ket{c_{2t+1}}\bra{c_{2t+1}}\\
\cdot\prod_{0\leq i<n} U_{2i-1}\ket{c_{2t}}\bra{c_{2t}}\Big)\\
\cdot\prod_{0\leq i<n} U_{2i-1}\ket{c_1}\bra{c_1} \prod_{0\leq i<n} U_{2i-1} \ket{c_0}\\
=
\sum_{\mathbf c} \prod_{0\leq t<T,0\leq i<n} \bra{c_{2t+1,2i}c_{2t+1,2i+1}} U\ket{c_{2t,2i}c_{2t,2i+1}}\\
\bra{c_{2t+2,2i-1}c_{2t+2,2i}} U\ket{c_{2t+1,2i-1}c_{2t+1,2i}}
\;.
\end{gathered}
\end{equation}
This is a discrete path integral:
A sum over configurations of variables distributed over spacetime (the $c_{t,i}$), where each summand is a product of local weights (the $\bra{c_{\ldots} c_{\ldots}}U\ket{c_{\ldots} c_{\ldots}}$).
With respect to the circuit diagram in Eq.~\eqref{eq:brickwork_circuit_diagram}, the variables are distributed as follows:
\begin{equation}
\begin{tikzpicture}
\foreach \y in {0,1,2}{
\foreach \x in {0,1,2}{
\atoms{small,circ,dot}{{a\x\y/p={\x*1.5,\y*1.2}}}
}};
\foreach \y in {0,1,2}{
\foreach \x in {0,1,2}{
\atoms{small,circ,dot}{{b\x\y/p={\x*1.5+0.75,\y*1.2+0.6}}}
}};
\draw (a00)edge[mark={lab=$c_{1,1}$}](b00) (a10)edge[mark={lab=$c_{1,2}$}](b00) (a10)edge[mark={lab=$c_{1,3}$}](b10) (a20)edge[mark={lab=$c_{1,4}$}](b10) (a20)edge[mark={lab=$c_{1,5}$}](b20);
\draw (a01)edge[mark={lab=$c_{2,1}$}](b00) (a11)edge[mark={lab=$c_{2,2}$}](b00) (a11)edge[mark={lab=$c_{2,3}$}](b10) (a21)edge[mark={lab=$c_{2,4}$}](b10) (a21)edge[mark={lab=$c_{2,5}$}](b20);
\draw (a01)edge[mark={lab=$c_{3,1}$}](b01) (a11)edge[mark={lab=$c_{3,2}$}](b01) (a11)edge[mark={lab=$c_{3,3}$}](b11) (a21)edge[mark={lab=$c_{3,4}$}](b11) (a21)edge[mark={lab=$c_{3,5}$}](b21);
\draw (a02)edge[mark={lab=$c_{4,1}$}](b01) (a12)edge[mark={lab=$c_{4,2}$}](b01) (a12)edge[mark={lab=$c_{4,3}$}](b11) (a22)edge[mark={lab=$c_{4,4}$}](b11) (a22)edge[mark={lab=$c_{4,5}$}](b21);
\draw (a02)edge[mark={lab=$c_{5,1}$}](b02) (a12)edge[mark={lab=$c_{5,2}$}](b02) (a12)edge[mark={lab=$c_{5,3}$}](b12) (a22)edge[mark={lab=$c_{5,4}$}](b12) (a22)edge[mark={lab=$c_{5,5}$}](b22);
\draw (a00)edge[ind=$c_{0,0}$]++(-135:0.4) (a00)edge[ind=$c_{0,1}$]++(-45:0.4) (a10)edge[ind=$c_{0,2}$]++(-135:0.4) (a10)edge[ind=$c_{0,3}$]++(-45:0.4) (a20)edge[ind=$c_{0,4}$]++(-135:0.4) (a20)edge[ind=$c_{0,5}$]++(-45:0.4);
\draw (b02)edge[ind=$c_{6,1}$]++(135:0.4) (b02)edge[ind=$c_{6,2}$]++(45:0.4) (b12)edge[ind=$c_{6,3}$]++(135:0.4) (b12)edge[ind=$c_{6,4}$]++(45:0.4) (b22)edge[ind=$c_{6,5}$]++(135:0.4) (b22)edge[ind=$c_{6,0}$]++(45:0.4);
\draw (a00)edge[ind=$c_{1,0}$]++(135:0.4) (a01)edge[ind=$c_{2,0}$]++(-135:0.4) (a01)edge[ind=$c_{3,0}$]++(135:0.4) (a02)edge[ind=$c_{4,0}$]++(-135:0.4) (a02)edge[ind=$c_{5,0}$]++(135:0.4);
\draw (b20)edge[]++(45:0.4) (b20)edge[]++(-45:0.4) (b21)edge[]++(-45:0.4) (b21)edge[]++(45:0.4) (b22)edge[]++(-45:0.4);
\end{tikzpicture}\;.
\end{equation}
In the tensor-network formulation, the correspondence between circuits and path integrals is even easier to see:
The circuit diagram, such as in Eq.~\eqref{eq:brickwork_circuit_diagram}, can be used directly as the tensor-network diagram.

We will now go the reverse direction and turn our toric-code path integral into a fault-tolerant circuit.
While this is straight forward with the correspondence discussed above, the resulting circuit will not be unitary, in accordance with its physical interpretation as an imaginary and not a real-time evolution.
This is where the anyon worldlines come into play:
The non-unitary operators will be projectors associated with the $+1$ outcome of a measurement.
So the circuit post-selected to all $+1$ measurement outcomes equals the path integral.
The circuit post-selected to an arbitrary configuration of measurement outcomes equals the path integral together with a configuration of $e$ and $m$ anyon worldlines.

To turn the path integral into a circuit, we need to find
\begin{itemize}
\item a set of qubits distributed over a spatial lattice,
\item a way to reshape the path integral using rewrite rules, such that each tensor corresponds to one operator,
\item a 1-to-1 mapping between the bonds of the reshaped path integral and pairs of (1) a qubit and (2) a time step (i.e, an integer time coordinate),
\item such that the bonds of each tensor are mapped to only two consecutive time steps and to neighboring qubits.
\end{itemize}
There are many possible choices for the above, which yield different circuits.
We use the following intuitive procedure to find such a choice:
We start by choosing a time ($t$) direction in the 3D spacetime, and also a $t=0$ spatial plane.
Next, we identify according \emph{qubit worldlines}, which are sequences of bonds in the tensor network propagate (on average) in time direction, and \emph{qubit space planes}, which are sets of bonds that intersect with a (shifted) spatial plane.
To obtain a neat circuit, the qubit worldlines should incorporate most of the bonds, and each of them should be confined to a small spatial region.
The qubit worldlines become the set of bonds that we map to the same qubit at different time steps, and the qubit space planes become the bonds that we map to the different qubits at a fixed time step.
Then we project both the spacetime lattice and the qubit worldlines along the $t$ direction to obtain both a spatial lattice and a set of qubits.
Next, we group the tensors into operators (possibly after some splitting using the rewrite rules) such that every bond that is not part of a qubit worldline is internal to an operator.
Finally, we read off the sequence in which the operators are applied to the qubits when we traverse the tensor network in $t$ direction.

Let us start by choosing the $t$ direction.
We have described the $t=x$ and $t=x+y+z$ directions in Ref.~\cite{path_integral_qec}, where $x$, $y$, and $z$ are unit vectors of the cubic lattice:
These yield the stabilizer toric code and the CSS honeycomb Floquet code, respectively.
Here, we do the next obvious thing and traverse it in the $t=x+y$ direction, with the $t=0$ plane perpendicular spanned by the $z$ and $\overline x\coloneqq (y-x)/2$ directions.

Next, we choose the following two types of qubit worldlines:
``Green'' qubit worldlines consist of $x$ and $y$ bonds contained inside the $(x,y)$-faces.
``Purple'' qubit worldlines consist of $x$ and $y$ bonds contained inside the $(x,z)$ and $(y,z)$-faces.
The $z$ bonds are not part of any qubit worldline but will be internal to operators in the circuit.
The following picture shows one green and one purple worldline:
\begin{equation}
\begin{tikzpicture}
\draw (0,0)edge[mark={arr,e},ind=$x$]++(0:0.35) (0,0)edge[mark={arr,e},ind=$z$]++(25:0.25) (0,0)edge[mark={arr,e},ind=$y$]++(90:0.35) (0,0)edge[gray,mark={arr,e},ind=$t$]++(0.35,0.35) (0,0)edge[gray,mark={arr,e},ind=$\overline x$]++(-0.175,0.175);
\end{tikzpicture}
\begin{tikzpicture}
\atoms{void}{x/p={2,0}, y/p={0,2}, z/p={1.6,0.8}}
\foreach \x in {0,1,2}{
\foreach \y in {0,1,2}{
\foreach \z in {0,1}{
\atoms{void}{\x\y\z/p={$\x*(x)+\y*(y)+\z*(z)$}}
}}}
\draw[orange] (000)--(200) (010)--(210) (020)--(220) (000)--(020) (100)--(120) (200)--(220) (201)--(221) (021)--(221);
\draw[orange,back] (001)--(201) (011)--(211) (001)--(021) (101)--(121);
\draw[orange,back] (000)--(001) (100)--(101) (010)--(011) (110)--(111);
\draw[orange] (200)--(201) (210)--(211) (220)--(221) (020)--(021) (120)--(121);
\foreach \x/\y/\z in {1/0/0, 3/0/0, 1/2/0, 3/2/0, 1/4/0, 3/4/0, 0/1/0, 0/3/0, 2/1/0, 2/3/0, 4/1/0, 4/3/0, 4/0/1, 4/2/1, 0/4/1, 2/4/1, 4/4/1, 4/1/2, 4/3/2, 1/4/2, 3/4/2}{
\atoms{delta}{d\x\y\z/p={$0.5*\x*(x)+0.5*\y*(y)+0.5*\z*(z)$}}
}
\foreach \x/\y/\z in {1/0/2, 3/0/2, 1/2/2, 3/2/2, 0/1/2, 0/3/2, 2/1/2, 2/3/2, 2/2/1, 0/2/1, 0/0/1, 2/0/1}{
\atoms{delta,astyle=gray}{d\x\y\z/p={$0.5*\x*(x)+0.5*\y*(y)+0.5*\z*(z)$}}
}
\foreach \x/\y/\z in {1/1/0, 3/1/0, 1/3/0, 3/3/0, 1/4/1, 3/4/1, 4/1/1, 4/3/1}{
\atoms{z2}{z\x\y\z/p={$0.5*\x*(x)+0.5*\y*(y)+0.5*\z*(z)$}}
}
\foreach \x/\y/\z in {1/1/2, 3/1/2, 1/3/2, 3/3/2, 1/2/1, 3/2/1, 2/1/1, 2/3/1, 0/1/1, 0/3/1, 1/0/1, 3/0/1}{
\atoms{z2,astyle=gray}{z\x\y\z/p={$0.5*\x*(x)+0.5*\y*(y)+0.5*\z*(z)$}}
}
\draw (d010)--++($-0.2*(x)$) (d030)--++($-0.2*(x)$) (d410)--++($0.2*(x)$) (d430)--++($0.2*(x)$) (d340)--++($0.2*(y)$) (d100)--++($-0.2*(y)$) (d300)--++($-0.2*(y)$) (d401)--++($0.2*(x)$) (d401)--++($-0.2*(y)$) (d441)--++($0.2*(x)$) (d441)--++($0.2*(y)$) (d041)--++($-0.2*(x)$) (d041)--++($0.2*(y)$) (d142)--++($0.2*(y)$) (d342)--++($0.2*(y)$) (d142)--++($0.2*(z)$) (d342)--++($0.2*(z)$) (d030)--++($-0.2*(z)$) (d010)--++($-0.2*(z)$) (d100)--++($-0.2*(z)$) (d300)--++($-0.2*(z)$) (d410)--++($-0.2*(z)$) (d430)--++($-0.2*(z)$) (d340)--++($-0.2*(z)$) (d140)--++($-0.2*(z)$) (d140)--++($0.2*(y)$) (d421)--++($0.2*(x)$) (d210)--++($-0.2*(z)$) (d320)--++($-0.2*(z)$) (d230)--++($-0.2*(z)$) (d120)--++($-0.2*(z)$) (d412)--++($0.2*(z)$) (d412)--++($0.2*(x)$) (d432)--++($0.2*(x)$) (d432)--++($0.2*(z)$) (d241)--++($0.2*(y)$);
\draw[gray] (d201)--++($-0.2*(y)$) (d001)--++($-0.2*(x)$) (d001)--++($-0.2*(y)$) (d021)--++($-0.2*(x)$) (d032)--++($-0.2*(x)$) (d032)--++($0.2*(z)$) (d012)--++($0.2*(z)$) (d012)--++($-0.2*(x)$) (d102)--++($0.2*(z)$) (d102)--++($-0.2*(y)$) (d302)--++($0.2*(z)$) (d302)--++($-0.2*(y)$) (d212)--++($0.2*(z)$) (d232)--++($0.2*(z)$) (d122)--++($0.2*(z)$) (d322)--++($0.2*(z)$);
\foreach \x/\xx/\xxx in {0/1/2,2/3/4}{
\foreach \y/\yy/\yyy in {0/1/2,2/3/4}{
\foreach \z in {0}{
\draw (z\xx\yy\z)--(d\x\yy\z) (z\xx\yy\z)--(d\xxx\yy\z) (z\xx\yy\z)--(d\xx\y\z) (z\xx\yy\z)--(d\xx\yyy\z);
}
\foreach \z in {2}{
\draw[gray] (z\xx\yy\z)--(d\x\yy\z) (z\xx\yy\z)--(d\xxx\yy\z) (z\xx\yy\z)--(d\xx\y\z) (z\xx\yy\z)--(d\xx\yyy\z);
}}}
\foreach \x/\xx/\xxx in {0/1/2,2/3/4}{
\foreach \y in {0,2}{
\draw[gray] (z\xx\y1)--(d\x\y1) (z\xx\y1)--(d\xxx\y1) (z\xx\y1)--(d\xx\y0) (z\xx\y1)--(d\xx\y2);
}
\foreach \y in {4}{
\draw (z\xx\y1)--(d\x\y1) (z\xx\y1)--(d\xxx\y1) (z\xx\y1)--(d\xx\y0) (z\xx\y1)--(d\xx\y2);
}}
\foreach \y/\yy/\yyy in {0/1/2,2/3/4}{
\foreach \x in {0,2}{
\draw[gray] (z\x\yy1)--(d\x\y1) (z\x\yy1)--(d\x\yyy1) (z\x\yy1)--(d\x\yy0) (z\x\yy1)--(d\x\yy2);
}
\foreach \x in {4}{
\draw (z\x\yy1)--(d\x\y1) (z\x\yy1)--(d\x\yyy1) (z\x\yy1)--(d\x\yy0) (z\x\yy1)--(d\x\yy2);
}}
\draw[line width=0.3cm,\qubitb,opacity=0.5] (d100)--(z110)--(d210)--(z310)--(d320)--(z330)--(d430) (d430)--++($0.2*(x)$) (d100)--++($-0.2*(y)$);
\draw[line width=0.3cm,\qubita,opacity=0.5] (d001)--(z011)--(d021)--(z121)--(d221)--(z231)--(d241)--(z341)--(d441) (d441)--++($0.2*(y)$) (d001)--++($-0.2*(x)$);
\end{tikzpicture}\;.
\end{equation}
All other worldlines are obtained by shifting in the $x$, $y$, or $z$ direction.
An example of a qubit space plane is marked in the following picture:
\begin{equation}
\begin{tikzpicture}
\atoms{void}{x/p={2,0}, y/p={0,2}, z/p={1.6,0.8}}
\foreach \x in {0,1,2}{
\foreach \y in {0,1,2}{
\foreach \z in {0,1}{
\atoms{void}{\x\y\z/p={$\x*(x)+\y*(y)+\z*(z)$}}
}}}
\draw[orange] (000)--(200) (010)--(210) (020)--(220) (000)--(020) (100)--(120) (200)--(220) (201)--(221) (021)--(221);
\draw[orange,back] (001)--(201) (011)--(211) (001)--(021) (101)--(121);
\draw[orange,back] (000)--(001) (100)--(101) (010)--(011) (110)--(111);
\draw[orange] (200)--(201) (210)--(211) (220)--(221) (020)--(021) (120)--(121);
\foreach \x/\y/\z in {1/0/0, 3/0/0, 1/2/0, 3/2/0, 1/4/0, 3/4/0, 0/1/0, 0/3/0, 2/1/0, 2/3/0, 4/1/0, 4/3/0, 4/0/1, 4/2/1, 0/4/1, 2/4/1, 4/4/1, 4/1/2, 4/3/2, 1/4/2, 3/4/2}{
\atoms{delta}{d\x\y\z/p={$0.5*\x*(x)+0.5*\y*(y)+0.5*\z*(z)$}}
}
\foreach \x/\y/\z in {1/0/2, 3/0/2, 1/2/2, 3/2/2, 0/1/2, 0/3/2, 2/1/2, 2/3/2, 2/2/1, 0/2/1, 0/0/1, 2/0/1}{
\atoms{delta,astyle=gray}{d\x\y\z/p={$0.5*\x*(x)+0.5*\y*(y)+0.5*\z*(z)$}}
}
\foreach \x/\y/\z in {1/1/0, 3/1/0, 1/3/0, 3/3/0, 1/4/1, 3/4/1, 4/1/1, 4/3/1}{
\atoms{z2}{z\x\y\z/p={$0.5*\x*(x)+0.5*\y*(y)+0.5*\z*(z)$}}
}
\foreach \x/\y/\z in {1/1/2, 3/1/2, 1/3/2, 3/3/2, 1/2/1, 3/2/1, 2/1/1, 2/3/1, 0/1/1, 0/3/1, 1/0/1, 3/0/1}{
\atoms{z2,astyle=gray}{z\x\y\z/p={$0.5*\x*(x)+0.5*\y*(y)+0.5*\z*(z)$}}
}
\draw (d010)--++($-0.2*(x)$) (d030)--++($-0.2*(x)$) (d410)--++($0.2*(x)$) (d430)--++($0.2*(x)$) (d340)--++($0.2*(y)$) (d100)--++($-0.2*(y)$) (d300)--++($-0.2*(y)$) (d401)--++($0.2*(x)$) (d401)--++($-0.2*(y)$) (d441)--++($0.2*(x)$) (d441)--++($0.2*(y)$) (d041)--++($-0.2*(x)$) (d041)--++($0.2*(y)$) (d142)--++($0.2*(y)$) (d342)--++($0.2*(y)$) (d142)--++($0.2*(z)$) (d342)--++($0.2*(z)$) (d030)--++($-0.2*(z)$) (d010)--++($-0.2*(z)$) (d100)--++($-0.2*(z)$) (d300)--++($-0.2*(z)$) (d410)--++($-0.2*(z)$) (d430)--++($-0.2*(z)$) (d340)--++($-0.2*(z)$) (d140)--++($-0.2*(z)$) (d140)--++($0.2*(y)$) (d421)--++($0.2*(x)$) (d210)--++($-0.2*(z)$) (d320)--++($-0.2*(z)$) (d230)--++($-0.2*(z)$) (d120)--++($-0.2*(z)$) (d412)--++($0.2*(z)$) (d412)--++($0.2*(x)$) (d432)--++($0.2*(x)$) (d432)--++($0.2*(z)$) (d241)--++($0.2*(y)$);
\draw[gray] (d201)--++($-0.2*(y)$) (d001)--++($-0.2*(x)$) (d001)--++($-0.2*(y)$) (d021)--++($-0.2*(x)$) (d032)--++($-0.2*(x)$) (d032)--++($0.2*(z)$) (d012)--++($0.2*(z)$) (d012)--++($-0.2*(x)$) (d102)--++($0.2*(z)$) (d102)--++($-0.2*(y)$) (d302)--++($0.2*(z)$) (d302)--++($-0.2*(y)$) (d212)--++($0.2*(z)$) (d232)--++($0.2*(z)$) (d122)--++($0.2*(z)$) (d322)--++($0.2*(z)$);
\foreach \x/\xx/\xxx in {0/1/2,2/3/4}{
\foreach \y/\yy/\yyy in {0/1/2,2/3/4}{
\foreach \z in {0}{
\draw (z\xx\yy\z)--(d\x\yy\z) (z\xx\yy\z)--(d\xxx\yy\z) (z\xx\yy\z)--(d\xx\y\z) (z\xx\yy\z)--(d\xx\yyy\z);
}
\foreach \z in {2}{
\draw[gray] (z\xx\yy\z)--(d\x\yy\z) (z\xx\yy\z)--(d\xxx\yy\z) (z\xx\yy\z)--(d\xx\y\z) (z\xx\yy\z)--(d\xx\yyy\z);
}}}
\foreach \x/\xx/\xxx in {0/1/2,2/3/4}{
\foreach \y in {0,2}{
\draw[gray] (z\xx\y1)--(d\x\y1) (z\xx\y1)--(d\xxx\y1) (z\xx\y1)--(d\xx\y0) (z\xx\y1)--(d\xx\y2);
}
\foreach \y in {4}{
\draw (z\xx\y1)--(d\x\y1) (z\xx\y1)--(d\xxx\y1) (z\xx\y1)--(d\xx\y0) (z\xx\y1)--(d\xx\y2);
}}
\foreach \y/\yy/\yyy in {0/1/2,2/3/4}{
\foreach \x in {0,2}{
\draw[gray] (z\x\yy1)--(d\x\y1) (z\x\yy1)--(d\x\yyy1) (z\x\yy1)--(d\x\yy0) (z\x\yy1)--(d\x\yy2);
}
\foreach \x in {4}{
\draw (z\x\yy1)--(d\x\y1) (z\x\yy1)--(d\x\yyy1) (z\x\yy1)--(d\x\yy0) (z\x\yy1)--(d\x\yy2);
}}
\draw[line width=0.3cm,\qubitb,opacity=0.5] (d300)--(z310)--(d210) (d120)--(z130)--(d030) (d302)--(z312)--(d212) (d122)--(z132)--(d032);
\draw[line width=0.3cm,\qubita,opacity=0.5] (d041)--(z031) (z121)--(d221)--(z211) (z301)--(d401)--++($-0.2*(y)$) (d041)--++($-0.2*(x)$);
\end{tikzpicture}\;.
\end{equation}
Next, we project the cubic lattice onto the $t=0$ spatial plane along the $t=x+y$ direction, which yields a ``squeezed'' square lattice.
On this projected lattice, each purple qubit worldlines zigzags across a face between the centers of its two $z$ edges, so it is natural to put the associated qubit at the center of each face.
The green qubit worldlines zigzag along $\overline x$ edges between vertices, so it is natural to put the associated qubits to the centers of the $\overline x$ edges.
The following shows the qubits as green and purple dots on the projected lattice in orange:
\begin{equation}
\label{eq:spatial_lattice}
\begin{tikzpicture}
\draw (0,0)edge[mark={arr,e},ind=$z$]++(0:0.5) (0,0)edge[mark={arr,e},ind=$\overline x$]++(90:0.5*0.707107);
\end{tikzpicture}
\begin{tikzpicture}
\atoms{void}{x/p={1,0}, y/p={0,0.707107}}
\clip ($(x)+(y)+(-0.15,0.15)$) rectangle($5*(x)+5*(y)+(-0.15,-0.15)$);
\draw[orange] ($0*(y)$)--++($5*(x)$) ($1*(y)$)--++($5*(x)$) ($2*(y)$)--++($5*(x)$) ($3*(y)$)--++($5*(x)$) ($4*(y)$)--++($5*(x)$) ($5*(y)$)--++($5*(x)$);
\draw[orange] ($0*(x)$)--++($5*(y)$) ($1*(x)$)--++($5*(y)$) ($2*(x)$)--++($5*(y)$) ($3*(x)$)--++($5*(y)$) ($4*(x)$)--++($5*(y)$) ($5*(x)$)--++($5*(y)$);
\foreach \x in {0,1,2,3,4}{
\foreach \y in {0,1,2,3,4}{
\atoms{vertex,astyle={\qubitb}}{0/p={$\x*(x)+{\y+0.5}*(y)$}}
\atoms{vertex,astyle={\qubita}}{0/p={${\x+0.5}*(x)+{\y+0.5}*(y)$}}
}}
\end{tikzpicture}
\;.
\end{equation}
Next, we need to choose a way to group tensors together such that each group forms one circuit operator all of whose indices correspond to qubit worldlines.
As the time-perpendicular bonds in the tensor network are not part of any qubit worldlines, we need to group their two adjacent tensors into one operator.
Tensors that are not adjacent to any time-perpendicular bonds become operators on their own.
In total, there are four types of linear operators in the circuit, of which we mark one in blue in the following picture:
\begin{equation}
\label{eq:tensor_circuit_labeling}
\begin{tikzpicture}
\atoms{void}{x/p={2,0}, y/p={0,2}, z/p={1.6,0.8}}
\foreach \x in {0,1,2}{
\foreach \y in {0,1,2}{
\foreach \z in {0,1}{
\atoms{void}{\x\y\z/p={$\x*(x)+\y*(y)+\z*(z)$}}
}}}
\draw[orange] (000)--(200) (010)--(210) (020)--(220) (000)--(020) (100)--(120) (200)--(220) (201)--(221) (021)--(221);
\draw[orange,back] (001)--(201) (011)--(211) (001)--(021) (101)--(121);
\draw[orange,back] (000)--(001) (100)--(101) (010)--(011) (110)--(111);
\draw[orange] (200)--(201) (210)--(211) (220)--(221) (020)--(021) (120)--(121);
\foreach \x/\y/\z in {1/0/0, 3/0/0, 1/2/0, 3/2/0, 1/4/0, 3/4/0, 0/1/0, 0/3/0, 2/1/0, 2/3/0, 4/1/0, 4/3/0, 4/0/1, 4/2/1, 0/4/1, 2/4/1, 4/4/1, 4/1/2, 4/3/2, 1/4/2, 3/4/2}{
\atoms{delta}{d\x\y\z/p={$0.5*\x*(x)+0.5*\y*(y)+0.5*\z*(z)$}}
}
\foreach \x/\y/\z in {1/0/2, 3/0/2, 1/2/2, 3/2/2, 0/1/2, 0/3/2, 2/1/2, 2/3/2, 2/2/1, 0/2/1, 0/0/1, 2/0/1}{
\atoms{delta,astyle=gray}{d\x\y\z/p={$0.5*\x*(x)+0.5*\y*(y)+0.5*\z*(z)$}}
}
\foreach \x/\y/\z in {1/1/0, 3/1/0, 1/3/0, 3/3/0, 1/4/1, 3/4/1, 4/1/1, 4/3/1}{
\atoms{z2}{z\x\y\z/p={$0.5*\x*(x)+0.5*\y*(y)+0.5*\z*(z)$}}
}
\foreach \x/\y/\z in {1/1/2, 3/1/2, 1/3/2, 3/3/2, 1/2/1, 3/2/1, 2/1/1, 2/3/1, 0/1/1, 0/3/1, 1/0/1, 3/0/1}{
\atoms{z2,astyle=gray}{z\x\y\z/p={$0.5*\x*(x)+0.5*\y*(y)+0.5*\z*(z)$}}
}
\draw (d010)--++($-0.2*(x)$) (d030)--++($-0.2*(x)$) (d410)--++($0.2*(x)$) (d430)--++($0.2*(x)$) (d340)--++($0.2*(y)$) (d100)--++($-0.2*(y)$) (d300)--++($-0.2*(y)$) (d401)--++($0.2*(x)$) (d401)--++($-0.2*(y)$) (d441)--++($0.2*(x)$) (d441)--++($0.2*(y)$) (d041)--++($-0.2*(x)$) (d041)--++($0.2*(y)$) (d142)--++($0.2*(y)$) (d342)--++($0.2*(y)$) (d142)--++($0.2*(z)$) (d342)--++($0.2*(z)$) (d030)--++($-0.2*(z)$) (d010)--++($-0.2*(z)$) (d100)--++($-0.2*(z)$) (d300)--++($-0.2*(z)$) (d410)--++($-0.2*(z)$) (d430)--++($-0.2*(z)$) (d340)--++($-0.2*(z)$) (d140)--++($-0.2*(z)$) (d140)--++($0.2*(y)$) (d421)--++($0.2*(x)$) (d210)--++($-0.2*(z)$) (d320)--++($-0.2*(z)$) (d230)--++($-0.2*(z)$) (d120)--++($-0.2*(z)$) (d412)--++($0.2*(z)$) (d412)--++($0.2*(x)$) (d432)--++($0.2*(x)$) (d432)--++($0.2*(z)$) (d241)--++($0.2*(y)$);
\draw[gray] (d201)--++($-0.2*(y)$) (d001)--++($-0.2*(x)$) (d001)--++($-0.2*(y)$) (d021)--++($-0.2*(x)$) (d032)--++($-0.2*(x)$) (d032)--++($0.2*(z)$) (d012)--++($0.2*(z)$) (d012)--++($-0.2*(x)$) (d102)--++($0.2*(z)$) (d102)--++($-0.2*(y)$) (d302)--++($0.2*(z)$) (d302)--++($-0.2*(y)$) (d212)--++($0.2*(z)$) (d232)--++($0.2*(z)$) (d122)--++($0.2*(z)$) (d322)--++($0.2*(z)$);
\foreach \x/\xx/\xxx in {0/1/2,2/3/4}{
\foreach \y/\yy/\yyy in {0/1/2,2/3/4}{
\foreach \z in {0}{
\draw (z\xx\yy\z)--(d\x\yy\z) (z\xx\yy\z)--(d\xxx\yy\z) (z\xx\yy\z)--(d\xx\y\z) (z\xx\yy\z)--(d\xx\yyy\z);
}
\foreach \z in {2}{
\draw[gray] (z\xx\yy\z)--(d\x\yy\z) (z\xx\yy\z)--(d\xxx\yy\z) (z\xx\yy\z)--(d\xx\y\z) (z\xx\yy\z)--(d\xx\yyy\z);
}}}
\foreach \x/\xx/\xxx in {0/1/2,2/3/4}{
\foreach \y in {0,2}{
\draw[gray] (z\xx\y1)--(d\x\y1) (z\xx\y1)--(d\xxx\y1) (z\xx\y1)--(d\xx\y0) (z\xx\y1)--(d\xx\y2);
}
\foreach \y in {4}{
\draw (z\xx\y1)--(d\x\y1) (z\xx\y1)--(d\xxx\y1) (z\xx\y1)--(d\xx\y0) (z\xx\y1)--(d\xx\y2);
}}
\foreach \y/\yy/\yyy in {0/1/2,2/3/4}{
\foreach \x in {0,2}{
\draw[gray] (z\x\yy1)--(d\x\y1) (z\x\yy1)--(d\x\yyy1) (z\x\yy1)--(d\x\yy0) (z\x\yy1)--(d\x\yy2);
}
\foreach \x in {4}{
\draw (z\x\yy1)--(d\x\y1) (z\x\yy1)--(d\x\yyy1) (z\x\yy1)--(d\x\yy0) (z\x\yy1)--(d\x\yy2);
}}
\draw[line width=0.5cm,cyan,opacity=0.5,line cap=round] (d210-c)--(z211-c);
\draw[line width=0.5cm,cyan,opacity=0.5,line cap=round] (z231-c)--(d232-c);
\path[cyan,mark={slab=(c),sideoff=0.1}] (d210)--(z211);
\path[cyan,mark={slab=(d),sideoff=-0.1,r}] (z231)--(d232);
\fill[cyan,opacity=0.5] (z130)circle(0.3);
\fill[cyan,opacity=0.5] (d221)circle(0.3);
\path[cyan] (z130)++(45:0.5)node{(a)};
\path[cyan] (d221)++(-170:0.5)node{(b)};
\end{tikzpicture}\;.
\end{equation}
All other operators are obtained from the ones above by shifting by multiples of $x$, $y$, and $z$.
However, this does not quite yield a sensible decomposition yet, since each $\delta$-tensor on a $x$ or $y$-edge is part of both a (c) and a (d) operator.
The same holds for each $\zz_2$-tensor on a $(x,z)$ or $(y,z)$-face.
This can be resolved by splitting each of these 4-index $\delta$ and $\zz_2$-tensors into two 3-index tensors:
Along a straight line formed by $z$ bonds, alternating between 4-index $\zz_2$ and $\delta$-tensors, we choose the following splitting:
\begin{equation}
\label{eq:cd_resolution}
\begin{gathered}
\begin{tikzpicture}
\draw (0,0)edge[mark={arr,e},ind=$z$]++(0:0.35) (0,0)edge[mark={arr,e},ind=$t$]++(90:0.35);
\end{tikzpicture}
\begin{tikzpicture}
\atoms{z2}{z0/, z1/p={1.4,0}, z2/p={2.8,0}}
\atoms{delta}{d0/p={0.7,0}, d1/p={2.1,0}, d2/p={3.5,0}}
\draw (z0)--(d0)--(z1)--(d1)--(z2)--(d2) (z0)edge[mark={three dots,a}]++(180:0.3) (d2)edge[mark={three dots,a}]++(0:0.3);
\draw (z0)--++(-90:0.5) (z1)--++(-90:0.5) (z2)--++(-90:0.5) (d0)--++(-90:0.5) (d1)--++(-90:0.5) (d2)--++(-90:0.5);
\draw (z0)--++(90:0.5) (z1)--++(90:0.5) (z2)--++(90:0.5) (d0)--++(90:0.5) (d1)--++(90:0.5) (d2)--++(90:0.5);
\draw[line width=0.5cm,cyan,opacity=0.5,line cap=round] (d0-c)--(z1-c);
\draw[line width=0.5cm,cyan,opacity=0.5,line cap=round] (z1-c)--(d1-c);
\path[cyan,mark={slab=(c),sideoff=0.25}] (d0)--(z1);
\path[cyan,mark={slab=(d),sideoff=0.25}] (z1)--(d1);
\end{tikzpicture}
\\
=
\begin{tikzpicture}
\atoms{z2}{z0/p={0,0.6}, z1/p={1.4,0.6}, z2/p={2.8,0.6}}
\atoms{delta}{d0/p={0.7,0.6}, d1/p={2.1,0.6}, d2/p={3.5,0.6}}
\atoms{z2}{z0a/, z1a/p={1.4,0}, z2a/p={2.8,0}}
\atoms{delta}{d0a/p={0.7,0}, d1a/p={2.1,0}, d2a/p={3.5,0}}
\draw (z0)--(d0) (d0a)--(z1a) (z1)--(d1) (d1a)--(z2a) (z2)--(d2) (z0a)edge[mark={three dots,a}]++(180:0.3) (d2a)edge[mark={three dots,a}]++(0:0.3);
\draw (z0)--(z0a) (z1)--(z1a) (z2)--(z2a) (d0)--(d0a) (d1)--(d1a) (d2)--(d2a);
\draw (z0a)--++(-90:0.5) (z1a)--++(-90:0.5) (z2a)--++(-90:0.5) (d0a)--++(-90:0.5) (d1a)--++(-90:0.5) (d2a)--++(-90:0.5);
\draw (z0)--++(90:0.5) (z1)--++(90:0.5) (z2)--++(90:0.5) (d0)--++(90:0.5) (d1)--++(90:0.5) (d2)--++(90:0.5);
\draw[line width=0.5cm,cyan,opacity=0.5,line cap=round] (d0a-c)--(z1a-c) (z1-c)--(d1-c);
\path[cyan,mark={slab=(c),sideoff=-0.25,r}] (d0a)--(z1a);
\path[cyan,mark={slab=(d),sideoff=0.25}] (z1)--(d1);
\end{tikzpicture}
\;.
\end{gathered}
\end{equation}
After this, the $\zz_2$ and $\delta$-tensors come in pairs, and each pair forms a separate operator.
Each of the (a), (b), (c), or (d) tensors (or tensor pairs) now has two incoming and two outgoing qubit worldlines relative to the $t$ direction, and can accordingly be interpreted as a 2-qubit operator.

Let us now concretely write out these 2-qubit operators.
The (a) operator is proportional to the projector $\frac12(1+XX)$,
\begin{equation}
\begin{tikzpicture}
\draw (0,0)edge[mark={arr,e},ind=$\overline x$]++(0:0.35) (0,0)edge[mark={arr,e},ind=$t$]++(90:0.35);
\end{tikzpicture}
\begin{tikzpicture}
\atoms{z2}{0/}
\draw (0)edge[ind=$a$]++(-135:0.5) (0)edge[ind=$b$]++(-45:0.5) (0)edge[ind=$c$]++(135:0.5) (0)edge[ind=$d$]++(45:0.5);
\end{tikzpicture}
=
\delta_{a+b+c+d=0}
=
\bra{c,d} (1+XX) \ket{a,b}
\;.
\end{equation}
The (b) operator equals the projector $\frac12(1+ZZ)$,
\begin{equation}
\label{eq:delta_tensor_operator}
\begin{tikzpicture}
\draw (0,0)edge[mark={arr,e},ind=$\overline x$]++(0:0.35) (0,0)edge[mark={arr,e},ind=$t$]++(90:0.35);
\end{tikzpicture}
\begin{tikzpicture}
\atoms{delta}{0/}
\draw (0)edge[ind=$a$]++(-135:0.5) (0)edge[ind=$b$]++(-45:0.5) (0)edge[ind=$c$]++(135:0.5) (0)edge[ind=$d$]++(45:0.5);
\end{tikzpicture}
=
\delta_{a=b=c=d}
=
\bra{c,d} \frac12(1+ZZ) \ket{a,b}
\;.
\end{equation}
The (c) and (d) operators are equal to a $CX$ gate,
\begin{equation}
\begin{tikzpicture}
\draw (0,0)edge[mark={arr,e},ind=$z$]++(0:0.35) (0,0)edge[mark={arr,e},ind=$t$]++(90:0.35);
\end{tikzpicture}
\begin{tikzpicture}
\atoms{delta}{0/}
\atoms{z2}{1/p={0.8,0}}
\draw (0)--(1) (0)edge[ind=$a$]++(-90:0.5) (1)edge[ind=$b$]++(-90:0.5) (0)edge[ind=$c$]++(90:0.5) (1)edge[ind=$d$]++(90:0.5);
\end{tikzpicture}
=
\delta_{a=c=b+d}
=
\bra{c,d} CX \ket{a,b}
\;.
\end{equation}
While the $CX$ gates are already unitary, the projectors $\frac12(1+XX)$ and $\frac12(1+ZZ)$ are not.
Thus, we need to replace them by measurements, such that the projectors associated with the $+1$ outcome are precisely the ones above.
If we post-select the circuit to $+1$ measurement outcomes, then we have successfully executed the tensor-network path integral consisting of the $\frac12(1+XX)$ and $\frac12(1+ZZ)$ projectors.
If some measurement outcomes are not $+1$, then there will be ``defects'' in the executed path integral at the locations of these $\neq +1$ outcome.
We thus cannot choose the projectors for the other measurement outcomes arbitrarily, but they have to obey certain properties that allow the decoder to deal with them.
A choice that works is the obvious one, to replace the projectors by $XX$ and $ZZ$ measurements.
The other ($-1$) outcome then corresponds to the projectors $\frac12(1-XX)$ and $\frac12(1-ZZ)$.
These projectors can expressed in terms of the charged $\delta$ or $\zz_2$-tensors,
\begin{equation}
\begin{multlined}
\bra{c,d} \frac12(1-XX) \ket{a,b}
=
(-1)^{d+b}\cdot \delta_{a+b+c+d=0}\\
=
\begin{tikzpicture}
\atoms{z2}{0/}
\atoms{delta,bdastyle=red}{a/p={45:0.4}, b/p={-45:0.4}}
\draw (0)edge[ind=$a$]++(-135:0.5) (b)edge[ind=$b$]++(-45:0.3) (0)--(a) (0)--(b) (0)edge[ind=$c$]++(135:0.5) (a)edge[ind=$d$]++(45:0.3);
\end{tikzpicture}
\;,
\end{multlined}
\end{equation}
and
\begin{equation}
\label{eq:zz_minusone_projector}
\begin{multlined}
\bra{c,d} \frac12(1-ZZ) \ket{a,b}
=
\delta_{a=b=c+1=d+1}
=
\begin{tikzpicture}
\atoms{delta}{0/}
\atoms{z2,bdastyle=red}{a/p={45:0.4}, b/p={-45:0.4}}
\draw (0)edge[ind=$a$]++(-135:0.5) (b)edge[ind=$b$]++(-45:0.3) (0)--(a) (0)--(b) (0)edge[ind=$c$]++(135:0.5) (a)edge[ind=$d$]++(45:0.3);
\end{tikzpicture}
\;.
\end{multlined}
\end{equation}
In the tensor-network path integral, the additional charged tensors can be combined with the neighboring uncharged tensors, using identities
\begin{equation}
\label{eq:minusone_outcome_mapping}
\begin{gathered}
\begin{tikzpicture}
\atoms{void}{x/p={1.5,0}, y/p={0,1.5}, z/p={1.2,0.6}}
\foreach \x in {0,1}{
\foreach \y in {0,1}{
\foreach \z in {0,1}{
\atoms{void}{\x\y\z/p={$\x*(x)+\y*(y)+\z*(z)$}}
}}}
\draw[orange] (110)--(010) (011)--(111) (101)--(111)--(110)--(100) (000)--(100) (010)--(000) (100)--(101) (010)--(011);
\draw[orange,back] (001)--(101) (000)--(001) (001)--(011);
\atoms{z2}{d/p={$0.5*(x)+0.5*(y)$}}
\atoms{delta}{z0/p={$(x)+0.5*(y)$}, z2/p={$0.5*(x)$}}
\atoms{delta,bdastyle=red}{zm0/p={$0.75*(x)+0.5*(y)$}, zm1/p={$0.5*(x)+0.25*(y)$}}
\draw (d)--++($-0.4*(x)$) (d)--(zm1)--(z2) (d)--(zm0)--(z0) (d)--++($0.4*(y)$);
\draw (z0)--++($-0.3*(z)$) (z0)--++($0.3*(z)$) (z0)--++($0.3*(x)$) (z2)--++($-0.3*(y)$) (z2)--++($-0.3*(z)$) (z2)--++($0.3*(z)$);
\end{tikzpicture}
=
\begin{tikzpicture}
\atoms{void}{x/p={1.5,0}, y/p={0,1.5}, z/p={1.2,0.6}}
\foreach \x in {0,1}{
\foreach \y in {0,1}{
\foreach \z in {0,1}{
\atoms{void}{\x\y\z/p={$\x*(x)+\y*(y)+\z*(z)$}}
}}}
\draw[orange] (110)--(010) (011)--(111) (101)--(111)--(110)--(100) (000)--(100) (010)--(000) (100)--(101) (010)--(011);
\draw[orange,back] (001)--(101) (000)--(001) (001)--(011);
\draw[worldline] (0,0)--++(x)--++(y);
\atoms{z2}{d/p={$0.5*(x)+0.5*(y)$}}
\atoms{delta,bdastyle=red}{z0/p={$(x)+0.5*(y)$}, z2/p={$0.5*(x)$}}
\draw (d)--++($-0.4*(x)$) (d)--(z2) (d)--(z0) (d)--++($0.4*(y)$);
\draw (z0)--++($-0.3*(z)$) (z0)--++($0.3*(z)$) (z0)--++($0.3*(x)$) (z2)--++($-0.3*(y)$) (z2)--++($-0.3*(z)$) (z2)--++($0.3*(z)$);
\end{tikzpicture}
\;,\\
\begin{tikzpicture}
\atoms{void}{x/p={1.5,0}, y/p={0,1.5}, z/p={1.2,0.6}}
\foreach \x in {0,1}{
\foreach \y in {0,1}{
\foreach \z in {0,1}{
\atoms{void}{\x\y\z/p={$\x*(x)+\y*(y)+\z*(z)$}}
}}}
\draw[orange] (110)--(010) (011)--(111) (101)--(111)--(110)--(100) (000)--(100) (010)--(000) (100)--(101) (010)--(011);
\draw[orange,back] (001)--(101) (000)--(001) (001)--(011);
\atoms{delta}{d/p={$0.5*(z)+(y)$}}
\atoms{z2}{z0/p={$0.5*(x)+0.5*(z)+(y)$}, z2/p={$0.5*(y)+0.5*(z)$}}
\atoms{z2,bdastyle=red}{zm0/p={$0.25*(x)+0.5*(z)+(y)$}, zm1/p={$0.75*(y)+0.5*(z)$}}
\draw (d)--++($-0.4*(x)$) (d)--(zm1)--(z2) (d)--(zm0)--(z0) (d)--++($0.4*(y)$);
\draw (z0)--++($-0.3*(z)$) (z0)--++($0.3*(z)$) (z0)--++($0.3*(x)$) (z2)--++($-0.3*(y)$) (z2)--++($-0.3*(z)$) (z2)--++($0.3*(z)$);
\end{tikzpicture}
=
\begin{tikzpicture}
\atoms{void}{x/p={1.5,0}, y/p={0,1.5}, z/p={1.2,0.6}}
\foreach \x in {0,1}{
\foreach \y in {0,1}{
\foreach \z in {0,1}{
\atoms{void}{\x\y\z/p={$\x*(x)+\y*(y)+\z*(z)$}}
}}}
\draw[orange] (110)--(010) (011)--(111) (101)--(111)--(110)--(100) (000)--(100) (010)--(000) (100)--(101) (010)--(011);
\draw[orange,back] (001)--(101) (000)--(001) (001)--(011);
\draw[worldline] ($-0.5*(x)+0.5*(y)+0.5*(z)$)--++(x)--++(y);
\atoms{delta}{d/p={$0.5*(z)+(y)$}}
\atoms{z2,bdastyle=red}{z0/p={$0.5*(x)+0.5*(z)+(y)$}, z2/p={$0.5*(y)+0.5*(z)$}}
\draw (d)--++($-0.4*(x)$) (d)--(z2) (d)--(z0) (d)--++($0.4*(y)$);
\draw (z0)--++($-0.3*(z)$) (z0)--++($0.3*(z)$) (z0)--++($0.3*(x)$) (z2)--++($-0.3*(y)$) (z2)--++($-0.3*(z)$) (z2)--++($0.3*(z)$);
\end{tikzpicture}
\;.
\end{gathered}
\end{equation}
As shown, we find that each $-1$ outcome of a $XX$ measurement corresponds to two edges of the path integral carrying an $e$ anyon worldline.
Each $-1$ outcome of a $ZZ$ measurement corresponds to two faces (or Poincar\'e dual edges) of the path integral carrying an $m$ anyon worldline
\footnote{
We could have equivalently chosen the other two faces adjacent to the $\delta$-tensor to carry the $m$ anyon worldlines, and analogous for the $e$ anyon worldlines.
The above equations follow from the rewrite rules of the $ZX$-calculus, or from topological invariance as shown in Ref.~\cite{path_integral_qec}.
}.
The properties that allow the decoder deal with the $-1$ outcomes are precisely the properties of the anyon worldlines in Eqs.~\eqref{eq:cocycle_constraint} and \eqref{eq:anyon_worldline_invariance}.

After spelling out the different kinds of unitaries and measurements that make up the circuit, let us describe in which order they act on which qubits of the spatial lattice shown in Eq.~\eqref{eq:spatial_lattice}.
The $t=x+y$-coordinate of each tensor is either $k$, $k+\frac14$, $k+\frac12$, or $k+\frac34$, for some integer $k$.
The tensors at $k$ give rise to (a) and (b) operators.
All of these operators act on disjoint qubit pairs and can be applied in parallel within one layer:
Each (a) operator is an $XX$ measurement acting on a pair of green qubits with the same $z$ coordinate and $\overline x$ coordinates $2l+\frac12$ and $2l+\frac32$, for some integer $l$.
Each (b) operator is a $ZZ$ measurement acting on a pair of purple qubits with the same $z$ coordinate and $\overline x$ coordinates $2l-\frac12$ and $2l+\frac12$.
The tensors at $k+\frac12$ also give rise to (a) and (b) operators.
The action of these operators is the same as for those at $k$, apart from a shift by $\overline x$.

The tensors at $t$-coordinates $k+\frac14$ give rise to (c) and (d) operators.
As these operators result from resolving the $z$-lines of bonds as shown in Eq.~\eqref{eq:cd_resolution}, they yield two layers of operators:
We first act with all (c) operators, and then with all (d) operators.
Both operators are $CX$ gates acting on a green and a neighboring purple qubit, with the control on the green qubit.
For the (c) operators, the purple qubit has the larger $z$ coordinate, and for the (d) operators, the green qubit has the larger $z$ coordinate.
The tensors at $t$-coordinates $k+\frac34$ also give rise to two layers of (c) and (d) operators, which act in the same way as those $k+\frac14$.
So in total, one $t$ period of the circuit consists of 6 layers of operators:
\begin{equation}
\begin{tabular}{ccc}
\begin{tikzpicture}
\atoms{void}{x/p={1,0}, y/p={0,0.707107}}
\clip ($(x)+(y)+(-0.15,0.15)$) rectangle($4*(x)+4*(y)+(-0.15,-0.1)$);
\draw[orange] ($0*(y)$)--++($4*(x)$) ($1*(y)$)--++($4*(x)$) ($2*(y)$)--++($4*(x)$) ($3*(y)$)--++($4*(x)$) ($4*(y)$)--++($4*(x)$);
\draw[orange] ($1*(x)$)--++($4*(y)$) ($2*(x)$)--++($4*(y)$) ($3*(x)$)--++($4*(y)$) ($4*(x)$)--++($4*(y)$);
\foreach \x in {0,1,2,3,4}{
\foreach \y in {0,1,2,3,4}{
\atoms{vertex,astyle={\qubitb}}{g\x\y/p={$\x*(x)+{\y+0.5}*(y)$}}
\atoms{vertex,astyle={\qubita}}{r\x\y/p={${\x+0.5}*(x)+{\y+0.5}*(y)$}}
}}
\foreach \x in {1,2,3}{
\draw[line width=0.3cm,cyan,opacity=0.5,line cap=round] (g\x0-c)--(g\x1-c) (g\x2-c)--(g\x3-c) (r\x1-c)--(r\x2-c) (r\x3-c)--(r\x4-c);
\path (g\x2-c)--node[midway]{$\scriptstyle{M_{XX}}$} (g\x3-c);
\path (r\x1-c)--node[midway]{$\scriptstyle{M_{ZZ}}$} (r\x2-c);
}
\end{tikzpicture}
&
$\leftarrow$
&
\begin{tikzpicture}
\atoms{void}{x/p={1,0}, y/p={0,0.707107}}
\clip ($(x)+(y)+(-0.15,0.15)$) rectangle($4*(x)+4*(y)+(-0.15,-0.1)$);
\draw[orange] ($0*(y)$)--++($4*(x)$) ($1*(y)$)--++($4*(x)$) ($2*(y)$)--++($4*(x)$) ($3*(y)$)--++($4*(x)$) ($4*(y)$)--++($4*(x)$);
\draw[orange] ($1*(x)$)--++($4*(y)$) ($2*(x)$)--++($4*(y)$) ($3*(x)$)--++($4*(y)$) ($4*(x)$)--++($4*(y)$);
\foreach \x in {0,1,2,3,4}{
\foreach \y in {0,1,2,3,4}{
\atoms{vertex,astyle={\qubitb}}{g\x\y/p={$\x*(x)+{\y+0.5}*(y)$}}
\atoms{vertex,astyle={\qubita}}{r\x\y/p={${\x+0.5}*(x)+{\y+0.5}*(y)$}}
}}
\foreach \y in {1,2,3}{
\draw[line width=0.3cm,cyan,opacity=0.5,line cap=round] (r0\y-c)--(g1\y-c) (r1\y-c)--(g2\y-c) (r2\y-c)--(g3\y-c) (r3\y-c)--(g4\y-c);
\node at ($0.5*(r0\y-c)+0.5*(g1\y-c)+(90:0.15)$) {$\scriptstyle{CX}$};
\node at ($0.5*(r1\y-c)+0.5*(g2\y-c)+(90:0.15)$) {$\scriptstyle{CX}$};
\node at ($0.5*(r2\y-c)+0.5*(g3\y-c)+(90:0.15)$) {$\scriptstyle{CX}$};
\node at ($0.5*(r3\y-c)+0.5*(g4\y-c)+(90:0.15)$) {$\scriptstyle{CX}$};
}
\end{tikzpicture}
\\
$\downarrow$ & & $\uparrow$\\
\begin{tikzpicture}
\atoms{void}{x/p={1,0}, y/p={0,0.707107}}
\clip ($(x)+(y)+(-0.15,0.15)$) rectangle($4*(x)+4*(y)+(-0.15,-0.1)$);
\draw[orange] ($0*(y)$)--++($4*(x)$) ($1*(y)$)--++($4*(x)$) ($2*(y)$)--++($4*(x)$) ($3*(y)$)--++($4*(x)$) ($4*(y)$)--++($4*(x)$);
\draw[orange] ($1*(x)$)--++($4*(y)$) ($2*(x)$)--++($4*(y)$) ($3*(x)$)--++($4*(y)$) ($4*(x)$)--++($4*(y)$);
\foreach \x in {0,1,2,3,4}{
\foreach \y in {0,1,2,3,4}{
\atoms{vertex,astyle={\qubitb}}{g\x\y/p={$\x*(x)+{\y+0.5}*(y)$}}
\atoms{vertex,astyle={\qubita}}{r\x\y/p={${\x+0.5}*(x)+{\y+0.5}*(y)$}}
}}
\foreach \y in {1,2,3}{
\foreach \x in {1,2,3}{
\draw[line width=0.3cm,cyan,opacity=0.5,line cap=round] (g\x\y-c)--(r\x\y-c);
\node at ($0.5*(g\x\y-c)+0.5*(r\x\y-c)+(90:0.15)$) {$\scriptstyle{CX}$};
}}
\end{tikzpicture}
& &
\begin{tikzpicture}
\atoms{void}{x/p={1,0}, y/p={0,0.707107}}
\clip ($(x)+(y)+(-0.15,0.15)$) rectangle($4*(x)+4*(y)+(-0.15,-0.1)$);
\draw[orange] ($0*(y)$)--++($4*(x)$) ($1*(y)$)--++($4*(x)$) ($2*(y)$)--++($4*(x)$) ($3*(y)$)--++($4*(x)$) ($4*(y)$)--++($4*(x)$);
\draw[orange] ($1*(x)$)--++($4*(y)$) ($2*(x)$)--++($4*(y)$) ($3*(x)$)--++($4*(y)$) ($4*(x)$)--++($4*(y)$);
\foreach \x in {0,1,2,3,4}{
\foreach \y in {0,1,2,3,4}{
\atoms{vertex,astyle={\qubitb}}{g\x\y/p={$\x*(x)+{\y+0.5}*(y)$}}
\atoms{vertex,astyle={\qubita}}{r\x\y/p={${\x+0.5}*(x)+{\y+0.5}*(y)$}}
}}
\foreach \y in {1,2,3}{
\foreach \x in {1,2,3}{
\draw[line width=0.3cm,cyan,opacity=0.5,line cap=round] (g\x\y-c)--(r\x\y-c);
\node at ($0.5*(g\x\y-c)+0.5*(r\x\y-c)+(90:0.15)$) {$\scriptstyle{CX}$};
}}
\end{tikzpicture}
\\
$\downarrow$ && $\uparrow$\\
\begin{tikzpicture}
\atoms{void}{x/p={1,0}, y/p={0,0.707107}}
\clip ($(x)+(y)+(-0.15,0.15)$) rectangle($4*(x)+4*(y)+(-0.15,-0.1)$);
\draw[orange] ($0*(y)$)--++($4*(x)$) ($1*(y)$)--++($4*(x)$) ($2*(y)$)--++($4*(x)$) ($3*(y)$)--++($4*(x)$) ($4*(y)$)--++($4*(x)$);
\draw[orange] ($1*(x)$)--++($4*(y)$) ($2*(x)$)--++($4*(y)$) ($3*(x)$)--++($4*(y)$) ($4*(x)$)--++($4*(y)$);
\foreach \x in {0,1,2,3,4}{
\foreach \y in {0,1,2,3,4}{
\atoms{vertex,astyle={\qubitb}}{g\x\y/p={$\x*(x)+{\y+0.5}*(y)$}}
\atoms{vertex,astyle={\qubita}}{r\x\y/p={${\x+0.5}*(x)+{\y+0.5}*(y)$}}
}}
\foreach \y in {1,2,3}{
\draw[line width=0.3cm,cyan,opacity=0.5,line cap=round] (r0\y-c)--(g1\y-c) (r1\y-c)--(g2\y-c) (r2\y-c)--(g3\y-c) (r3\y-c)--(g4\y-c);
\node at ($0.5*(r0\y-c)+0.5*(g1\y-c)+(90:0.15)$) {$\scriptstyle{CX}$};
\node at ($0.5*(r1\y-c)+0.5*(g2\y-c)+(90:0.15)$) {$\scriptstyle{CX}$};
\node at ($0.5*(r2\y-c)+0.5*(g3\y-c)+(90:0.15)$) {$\scriptstyle{CX}$};
\node at ($0.5*(r3\y-c)+0.5*(g4\y-c)+(90:0.15)$) {$\scriptstyle{CX}$};
}
\end{tikzpicture}
&
$\rightarrow$
&
\begin{tikzpicture}
\atoms{void}{x/p={1,0}, y/p={0,0.707107}}
\clip ($(x)+(y)+(-0.15,0.15)$) rectangle($4*(x)+4*(y)+(-0.15,-0.1)$);
\draw[orange] ($0*(y)$)--++($4*(x)$) ($1*(y)$)--++($4*(x)$) ($2*(y)$)--++($4*(x)$) ($3*(y)$)--++($4*(x)$) ($4*(y)$)--++($4*(x)$);
\draw[orange] ($1*(x)$)--++($4*(y)$) ($2*(x)$)--++($4*(y)$) ($3*(x)$)--++($4*(y)$) ($4*(x)$)--++($4*(y)$);
\foreach \x in {0,1,2,3,4}{
\foreach \y in {0,1,2,3,4}{
\atoms{vertex,astyle={\qubitb}}{g\x\y/p={$\x*(x)+{\y+0.5}*(y)$}}
\atoms{vertex,astyle={\qubita}}{r\x\y/p={${\x+0.5}*(x)+{\y+0.5}*(y)$}}
}}
\foreach \x in {1,2,3}{
\draw[line width=0.3cm,cyan,opacity=0.5,line cap=round] (g\x1-c)--(g\x2-c) (g\x3-c)--(g\x4-c) (r\x0-c)--(r\x1-c) (r\x2-c)--(r\x3-c);
\path (g\x1-c)--node[midway]{$\scriptstyle{M_{XX}}$} (g\x2-c);
\path (r\x2-c)--node[midway]{$\scriptstyle{M_{ZZ}}$} (r\x3-c);
}
\end{tikzpicture}
\end{tabular}
\;.
\end{equation}

\section{Boundaries and lattice surgery}
In this section, we will define two different types of boundaries for our fault-tolerant circuit, as well as corners separating them.
We then use these boundaries and corners to perform a logical $ZZ$ measurement via lattice surgery of two spatial rectangular blocks of physical qubits, each storing a logical qubit.

\subsection{Boundaries and corners of the path integral}
Let us start describing boundaries and corners of the toric-code path integral, before turning them into boundaries and corners of the fault-tolerant circuit in Section~\ref{eq:dynamic_surface_code}.
Specifically, there are two topologically distinct boundaries, commonly referred to as \emph{rough} and \emph{smooth} boundaries, as well as one type of corner separating them.
Before discussing the path integral, let us first recall boundaries and corners on the level of the ground states \cite{Bravyi1998}.
Both rough and smooth boundaries can be defined on any square lattice with boundary.
For the smooth boundary, qubits are also associated to the boundary edges.
The stabilizers are defined on all plaquettes and vertices as in the bulk.
The only difference is that boundary vertices might have fewer adjacent edges than in the bulk and thus the according $X$ stabilizers act on fewer qubits.
For example, for a straight smooth boundary, the boundary vertex stabilizers act as $XXX$ on the two adjacent boundary edges and the one adjacent bulk edge.
The plaquette stabilizers in the bulk still impose that ground-state configurations are closed-loop patterns, but these closed loops may terminate at the boundary.
The following picture shows such an example configuration with smooth boundary at the bottom (drawn as a thick orange line):
\begin{equation}
\begin{tikzpicture}
\begin{scope}[scale=0.8]
\foreach \x in {0,1,2,3,4}{
\draw[orange] (\x,0)--++(0,3.3);
}
\foreach \y in {1,2,3}{
\draw[orange] (-0.3,\y)--++(4.6,0);
}
\draw[smoothbd] (-0.3,0)--(4.3,0);
\draw[configcol,line width=0.1cm] (0,0)--++(0:1) (0,1)--++(0:1) (0,2)--++(0:1) (1,2)--++(90:1) (2,2)--++(90:1) (2,2)--++(0:1) (3,2)--++(-90:1) (3,1)--++(0:1) (3,0)--++(0:1);
\draw[membranecol,line width=0.1cm] (0.5,0)--++(90:2.5)--++(0:2)--++(-90:1)--++(0:1)--++(-90:1.5);
\end{scope}
\end{tikzpicture}
\;.
\end{equation}
The boundary vertex stabilizers ensure that the amplitude is the same for all closed-loop patterns, no matter where they terminate on the boundary, for example:
\begin{equation}
\label{eq:state_local_boundary_deformation}
\left\langle\psi\middle|
\begin{tikzpicture}[scale=0.8]
\draw[orange] (0,0)--++(90:1.2) (1,0)--++(90:1.2) (2,0)--++(90:1.2) (-0.2,1)--++(0:2.4);
\draw[smoothbd] (-0.2,0)--++(0:2.4);
\draw[configcol,line width=0.1cm] (1,0)--++(0:1) (1,1)--++(0:1);
\draw[membranecol,line width=0.1cm] (1.5,0)--++(90:1.2);
\end{tikzpicture}
\right\rangle
=
\left\langle\psi\middle|
\begin{tikzpicture}[scale=0.8]
\draw[orange] (0,0)--++(90:1.2) (1,0)--++(90:1.2) (2,0)--++(90:1.2) (-0.2,1)--++(0:2.4);
\draw[smoothbd] (-0.2,0)--++(0:2.4);
\draw[configcol,line width=0.1cm] (0,0)--++(0:1) (1,0)--++(90:1) (1,1)--++(0:1);
\draw[membranecol,line width=0.1cm] (0.5,0)--++(90:0.5)--++(0:1)--++(90:0.7);
\end{tikzpicture}
\right\rangle
\;.
\end{equation}
In contrast, for a rough boundary, there are no qubits on the boundary edges, and no vertex stabilizers at the boundary vertices.
The remaining stabilizers are as in the bulk, except for the fact that stabilizers at a plaquettes that includes boundary edges acts on fewer qubits.
This is equivalent to taking a smooth boundary and forcing all boundary qubits to the $\ket0$ configuration.
As a consequence, ground-state closed-loop patterns must not terminate at a rough boundary, as shown in the following example with rough boundary at the bottom (drawn as a thick dotted orange line):
\begin{equation}
\begin{tikzpicture}
\begin{scope}[scale=0.8]
\foreach \x in {0,1,2,3,4}{
\draw[orange] (\x,0)--++(0,3.3);
}
\foreach \y in {1,2,3}{
\draw[orange] (-0.3,\y)--++(4.6,0);
}
\draw[roughbd] (-0.3,0)--(4.3,0);
\draw[configcol,line width=0.1cm] (0,1)--++(0:1) (0,2)--++(0:1) (1,2)--++(90:1) (2,2)--++(90:1) (2,2)--++(0:1) (3,2)--++(-90:1) (3,1)--++(0:1) (1,0)--++(90:1) (2,0)--++(90:1) (3,0)--++(90:1);
\draw[membranecol,line width=0.1cm] (0.5,0.5)--++(90:2)--++(0:2)--++(-90:1)--++(0:1)--++(-90:1)--cycle;
\end{scope}
\end{tikzpicture}
\;.
\end{equation}
Finally, corners are geometrically located at boundary vertices that separate a line of smooth boundary edges from a line of rough boundary edges.
They are defined by the fact that there is no $X$ stabilizer associated to the corner vertex itself, like for the rough boundary vertices.
Using corners, we can put the toric code on a rectangular patch with rough boundaries at the bottom and top and smooth boundaries on the left and right.
An example configuration is given by
\begin{equation}
\label{eq:surface_code_configuration}
\begin{tikzpicture}
\begin{scope}[scale=0.8]
\foreach \x in {1,2,3}{
\draw[orange] (\x,0)--++(0,3);
}
\foreach \y in {1,2}{
\draw[orange] (0,\y)--++(4,0);
}
\draw[roughbd] (0,0)--(4,0) (0,3)--++(4,0);
\draw[smoothbd] (0,0)--(0,3) (4,0)--++(0,3);
\draw[configcol,line width=0.1cm] (0,2)--++(0:1) (1,2)--++(90:1) (2,2)--++(90:1) (2,2)--++(0:1) (3,2)--++(-90:1) (0,1)--++(90:1) (4,1)--++(90:1);
\draw[membranecol,line width=0.1cm] (0,1.5)--++(0:0.5)--++(90:1)--++(0:2)--++(-90:1)--++(0:1.5);
\end{scope}
\end{tikzpicture}
\;.
\end{equation}
There are two equivalence classes of configurations under local deformations (that is, homology classes), determined by the $\mmod2$ number of strings crossing from the smooth boundary on the left to that on the right.
In the schematic closed-loop picture, two representatives look as follows:
\begin{equation}
\label{eq:surface_code_cohomology_classes}
B_0=
\begin{tikzpicture}
\draw[roughbd] (0,0)--++(1.5,0) (0,1)--++(1.5,0);
\draw[smoothbd] (0,0)--++(0,1) (1.5,0)--++(0,1);
\end{tikzpicture}
\;,
\quad
B_1=
\begin{tikzpicture}
\draw[roughbd] (0,0)--++(1.5,0) (0,1)--++(1.5,0);
\draw[smoothbd] (0,0)--++(0,1) (1.5,0)--++(0,1);
\draw[membranecol,line width=0.1cm] (0,0.5)--++(1.5,0);
\end{tikzpicture}
\;.
\end{equation}
For example, the configuration in Eq.~\eqref{eq:surface_code_configuration} belongs to the homology class $B_1$.
So we see that there is a 2-dimensional ground state space, and the homology classes provide a natural basis $\ket{B_0}$, $\ket{B_1}$.

With boundaries and corners for ground states in mind, let us now describe boundaries and corners for the path integral.
Note that in contrast to the state boundaries introduced in Section~\ref{sec:path_integral}, rough and smooth boundaries are \emph{physical boundaries}.
The most important difference is that when evaluating the path integral with physical boundaries, we do sum over all variables including the ones at the boundary.
\footnote{
So evaluating the path integral on a manifold with physical boundary yields a number $Z$.
We can also evaluate the path integral on a manifold with both physical and state boundaries, yielding a state supported on the state boundary.
Note that in this case the (2-dimensional) state boundary itself has a (1-dimensional) physical boundary.
}.
In general, physical boundaries may also come with special variables and weights at the boundary.
For the rough and smooth boundaries it will suffice to remove and truncate the bulk variables and weights.

In the spacetime path integral, rough and smooth boundaries are defined at a 2-dimensional boundary of the 3-dimensional cubic lattice.
For a smooth boundary, we include variables at the boundary edges, and impose the parity-even constraint at every boundary plaquette.
So the non-zero weight configurations on the Poincar\'e dual lattice are patterns of closed membranes that are allowed to terminate on the boundary.
Such membranes terminating at the boundary are depicted in Eq.~\eqref{eq:toric_code_membranes}.
However, in contrast to the state boundaries discussed there, evaluating the path integral with smooth physical boundaries means summing over all terminating closed-loop configurations at the boundary.
For a rough boundary, we do not include variables at the boundary edges.
As a consequence the parity-even constraints at plaquettes adjacent to boundary edges involve fewer variables.
The non-zero weight configurations on the Poincar\'e dual lattice are patterns of closed membranes that are not allowed to terminate on the boundary.
Furthermore, corners are defined at paths of edges separating areas of rough and smooth boundaries.
They are determined by the fact that there are no variables on the corner edges, just like for the rough boundary.
As a consequence, the closed-loop patterns on the smooth boundary must not terminate on the corner.

Let us next discuss smooth and rough boundaries in the tensor-network formulation.
For a smooth boundary, the tensor network is like in the bulk, including all $\delta$-tensors and $\zz_2$-tensors at the boundary edges and faces.
The only difference is that the boundary $\delta$-tensors may share fewer bonds with adjacent $\zz_2$-tensors.
For example, on a straight boundary segment of a cubic lattice, the boundary $\delta$-tensors have $3$ instead of $4$ indices.
For a rough boundary, we include neither the $\delta$-tensors nor the $\zz_2$-tensors at the boundary edges and faces.
This time, the $\zz_2$-tensors at faces adjacent to the boundary edges may have fewer bonds connected to neihboring $\delta$-tensors.
Finally, we do not include $\delta$-tensors at the corner edges.
The following shows an example of a spacetime patch with smooth boundary at the front (thick edges), and rough boundary on the right (dotted thick edges), separated by a vertical corner (very thick edges):
\begin{equation}
\label{eq:boundary_zx_toric_code}
\begin{tikzpicture}
\atoms{void}{x/p={2,0}, y/p={0,2}, z/p={1.6,0.8}}
\foreach \x in {0,1,2}{
\foreach \y in {0,1,2}{
\foreach \z in {0,1}{
\atoms{void}{\x\y\z/p={$\x*(x)+\y*(y)+\z*(z)$}}
}}}
\draw[smoothbd] (000)--(200) (010)--(210) (020)--(220) (000)--(020) (100)--(120);
\draw[corner] (200)--(220);
\draw[orange] (021)--(221) (020)--(021) (120)--(121);
\draw[orange,back] (001)--(201) (011)--(211) (001)--(021) (101)--(121);
\draw[orange, back] (000)--(001) (100)--(101) (010)--(011) (110)--(111);
\draw[roughbd] (200)--(201) (210)--(211) (220)--(221) (201)--(221);
\foreach \x/\y/\z in {1/0/0, 3/0/0, 1/2/0, 3/2/0, 1/4/0, 3/4/0, 0/1/0, 0/3/0, 2/1/0, 2/3/0, 0/4/1, 2/4/1, 1/4/2, 3/4/2}{
\atoms{delta}{d\x\y\z/p={$0.5*\x*(x)+0.5*\y*(y)+0.5*\z*(z)$}}
}
\foreach \x/\y/\z in {1/0/2, 3/0/2, 1/2/2, 3/2/2, 0/1/2, 0/3/2, 2/1/2, 2/3/2, 2/2/1, 0/2/1, 0/0/1, 2/0/1}{
\atoms{delta,astyle=gray}{d\x\y\z/p={$0.5*\x*(x)+0.5*\y*(y)+0.5*\z*(z)$}}
}
\foreach \x/\y/\z in {1/1/0, 3/1/0, 1/3/0, 3/3/0, 1/4/1, 3/4/1}{
\atoms{z2}{z\x\y\z/p={$0.5*\x*(x)+0.5*\y*(y)+0.5*\z*(z)$}}
}
\foreach \x/\y/\z in {1/1/2, 3/1/2, 1/3/2, 3/3/2, 1/2/1, 3/2/1, 2/1/1, 2/3/1, 0/1/1, 0/3/1, 1/0/1, 3/0/1}{
\atoms{z2,astyle=gray}{z\x\y\z/p={$0.5*\x*(x)+0.5*\y*(y)+0.5*\z*(z)$}}
}
\draw (d010)--++($-0.2*(x)$) (d030)--++($-0.2*(x)$) (d340)--++($0.2*(y)$) (d100)--++($-0.2*(y)$) (d300)--++($-0.2*(y)$) (d041)--++($-0.2*(x)$) (d041)--++($0.2*(y)$) (d142)--++($0.2*(y)$) (d342)--++($0.2*(y)$) (d142)--++($0.2*(z)$) (d342)--++($0.2*(z)$) (d140)--++($0.2*(y)$) (d241)--++($0.2*(y)$);
\draw[gray] (d201)--++($-0.2*(y)$) (d001)--++($-0.2*(x)$) (d001)--++($-0.2*(y)$) (d021)--++($-0.2*(x)$) (d032)--++($-0.2*(x)$) (d032)--++($0.2*(z)$) (d012)--++($0.2*(z)$) (d012)--++($-0.2*(x)$) (d102)--++($0.2*(z)$) (d102)--++($-0.2*(y)$) (d302)--++($0.2*(z)$) (d302)--++($-0.2*(y)$) (d212)--++($0.2*(z)$) (d232)--++($0.2*(z)$) (d122)--++($0.2*(z)$) (d322)--++($0.2*(z)$);
\foreach \x/\y/\col in {1/0/gray, 3/0/gray, 0/1/gray, 2/1/gray, 1/2/gray, 3/2/gray, 0/3/gray, 2/3/gray, 1/4/black, 3/4/black}{
\draw[\col] (d\x\y0)--(z\x\y1)--(d\x\y2);
}
\foreach \x/\z/\col in {1/0/black, 3/0/black, 0/1/gray, 2/1/gray, 1/2/gray, 3/2/gray}{
\draw[\col] (d\x0\z)--(z\x1\z)--(d\x2\z)--(z\x3\z)--(d\x4\z);
}
\foreach \y/\z/\col in {0/1/gray, 2/1/gray, 4/1/black, 1/0/black, 3/0/black, 1/2/gray, 3/2/gray}{
\draw[\col] (d0\y\z)--(z1\y\z)--(d2\y\z)--(z3\y\z);
}
\end{tikzpicture}\;.
\end{equation}
Note that we could apply tensor-network rewrite rules to obtain another tensor-network representation of a smooth boundary.
However, we cannot transform a smooth into a rough boundary by local rewrite rules, since they correspond to distinct topological boundary phases.
The boundaries are topologically protected in the sense that Eq.~\eqref{eq:fixed_point_property} also holds near the boundary, which implies that there are no constant-weight logical errors in the resulting circuits.
We also need to discuss how the anyons behave with respect to the boundaries.
The $m$ anyon worldlines may terminate on the smooth boundary, but not on the rough boundary.
The $e$ anyon worldlines may terminate on the rough boundary, but not on the smooth boundary.
The following picture shows one possible configuration for the tensor-network path integral:
\begin{equation}
%\label{eq:boundary_zx_toric_code}
\begin{tikzpicture}
\atoms{void}{x/p={2,0}, y/p={0,2}, z/p={1.6,0.8}}
\foreach \x in {0,1,2}{
\foreach \y in {0,1,2}{
\foreach \z in {0,1}{
\atoms{void}{\x\y\z/p={$\x*(x)+\y*(y)+\z*(z)$}}
}}}
\draw[smoothbd] (000)--(200) (010)--(210) (020)--(220) (000)--(020) (100)--(120);
\draw[corner] (200)--(220);
\draw[orange] (021)--(221) (020)--(021) (120)--(121);
\draw[orange,back] (001)--(201) (011)--(211) (001)--(021) (101)--(121);
\draw[orange, back] (000)--(001) (100)--(101) (010)--(011) (110)--(111);
\draw[roughbd] (200)--(201) (210)--(211) (220)--(221) (201)--(221);
\foreach \x/\y/\z in {1/0/0, 3/0/0, 1/2/0, 3/2/0, 1/4/0, 3/4/0, 0/1/0, 0/3/0, 2/1/0, 2/3/0, 0/4/1, 2/4/1, 1/4/2, 3/4/2}{
\atoms{delta}{d\x\y\z/p={$0.5*\x*(x)+0.5*\y*(y)+0.5*\z*(z)$}}
}
\foreach \x/\y/\z in {1/0/2, 3/0/2, 1/2/2, 3/2/2, 0/1/2, 0/3/2, 2/1/2, 2/3/2, 2/2/1, 0/2/1, 0/0/1, 2/0/1}{
\atoms{delta,astyle=gray}{d\x\y\z/p={$0.5*\x*(x)+0.5*\y*(y)+0.5*\z*(z)$}}
}
\foreach \x/\y/\z in {1/1/0, 3/1/0, 1/3/0, 3/3/0, 1/4/1, 3/4/1}{
\atoms{z2}{z\x\y\z/p={$0.5*\x*(x)+0.5*\y*(y)+0.5*\z*(z)$}}
}
\foreach \x/\y/\z in {1/1/2, 3/1/2, 1/3/2, 3/3/2, 1/2/1, 3/2/1, 2/1/1, 2/3/1, 0/1/1, 0/3/1, 1/0/1, 3/0/1}{
\atoms{z2,astyle=gray}{z\x\y\z/p={$0.5*\x*(x)+0.5*\y*(y)+0.5*\z*(z)$}}
}
\draw (d010)--++($-0.2*(x)$) (d030)--++($-0.2*(x)$) (d340)--++($0.2*(y)$) (d100)--++($-0.2*(y)$) (d300)--++($-0.2*(y)$) (d041)--++($-0.2*(x)$) (d041)--++($0.2*(y)$) (d142)--++($0.2*(y)$) (d342)--++($0.2*(y)$) (d142)--++($0.2*(z)$) (d342)--++($0.2*(z)$) (d140)--++($0.2*(y)$) (d241)--++($0.2*(y)$);
\draw[gray] (d201)--++($-0.2*(y)$) (d001)--++($-0.2*(x)$) (d001)--++($-0.2*(y)$) (d021)--++($-0.2*(x)$) (d032)--++($-0.2*(x)$) (d032)--++($0.2*(z)$) (d012)--++($0.2*(z)$) (d012)--++($-0.2*(x)$) (d102)--++($0.2*(z)$) (d102)--++($-0.2*(y)$) (d302)--++($0.2*(z)$) (d302)--++($-0.2*(y)$) (d212)--++($0.2*(z)$) (d232)--++($0.2*(z)$) (d122)--++($0.2*(z)$) (d322)--++($0.2*(z)$);
\foreach \x/\y/\col in {1/0/gray, 3/0/gray, 0/1/gray, 2/1/gray, 1/2/gray, 3/2/gray, 0/3/gray, 2/3/gray, 1/4/black, 3/4/black}{
\draw[\col] (d\x\y0)--(z\x\y1)--(d\x\y2);
}
\foreach \x/\z/\col in {1/0/black, 3/0/black, 0/1/gray, 2/1/gray, 1/2/gray, 3/2/gray}{
\draw[\col] (d\x0\z)--(z\x1\z)--(d\x2\z)--(z\x3\z)--(d\x4\z);
}
\foreach \y/\z/\col in {0/1/gray, 2/1/gray, 4/1/black, 1/0/black, 3/0/black, 1/2/gray, 3/2/gray}{
\draw[\col] (d0\y\z)--(z1\y\z)--(d2\y\z)--(z3\y\z);
}
\draw[worldline] ($0.5*(x)+0.5*(y)$)--++($0.5*(z)$)--++(x)--++($0.7*(z)$);
\draw[worldline] ($1.5*(x)+1*(y)+(z)$)--++($-1.7*(x)$);
\foreach \nod in {z110, z211, z312}{
\atoms{z2,bdastyle=red}{x/p={\nod}}
}
\foreach \nod in {d322, d122}{
\atoms{delta,bdastyle=red}{x/p={\nod}}
}
\end{tikzpicture}\;.
\end{equation}
If an $e$-anyon worldline terminates on the smooth boundary, or an $m$-anyon worldline on the rough boundary, the path integral becomes $0$ in analogy to Eq.~\eqref{eq:cocycle_constraint}, e.g.:
\begin{equation}
\label{eq:boundary_cocycle_constraint}
\begin{tikzpicture}
\atoms{void}{x/p={1.5,0}, y/p={0,1.5}, z/p={1.2,0.6}}
\foreach \x in {0,1}{
\foreach \y in {0,1}{
\foreach \z in {0,1}{
\atoms{void}{\x\y\z/p={$\x*(x)+\y*(y)+\z*(z)$}}
}}}
\draw[orange] (000)--(100) (110)--(010)--(000) (010)--(011)--(111);
\draw[roughbd] (100)--(101)--(111)--(110)--(100);
\draw[orange,back] (000)--(001)--(011) (001)--(101);
\draw[worldline] ($0.5*(x)+0.5*(y)+0.5*(z)$)--++($-0.7*(x)$);
\foreach \x/\y/\z in {1/0/0, 1/2/0, 0/1/0, 0/2/1, 1/2/2}{
\atoms{delta}{d\x\y\z/p={$0.5*\x*(x)+0.5*\y*(y)+0.5*\z*(z)$}}
}
\foreach \x/\y/\z in {1/0/2, 0/1/2, 0/0/1}{
\atoms{delta,astyle=gray}{d\x\y\z/p={$0.5*\x*(x)+0.5*\y*(y)+0.5*\z*(z)$}}
}
\foreach \x/\y/\z in {1/1/0, 1/2/1}{
\atoms{z2}{z\x\y\z/p={$0.5*\x*(x)+0.5*\y*(y)+0.5*\z*(z)$}}
}
\foreach \x/\y/\z in {1/1/2, 1/0/1, 0/1/1}{
\atoms{z2,astyle=gray}{z\x\y\z/p={$0.5*\x*(x)+0.5*\y*(y)+0.5*\z*(z)$}}
}
\draw (z110)--(d100) (z110)--(d120) (z110)--(d010) (z121)--(d120) (z121)--(d021) (z121)--(d122);
\draw[gray] (z112)--(d102) (z112)--(d122) (z112)--(d012) (z011)--(d001) (z011)--(d010) (z011)--(d012) (z011)--(d021) (z101)--(d100) (z101)--(d001) (z101)--(d102);
\draw (d010)--++($-0.2*(x)$) (d010)--++($-0.2*(z)$) (d100)--++($-0.2*(y)$) (d100)--++($-0.2*(x)$) (d120)--++($-0.2*(z)$) (d120)--++($0.2*(y)$) (d122)--++($-0.2*(z)$) (d122)--++($0.2*(y)$) (d021)--++($-0.2*(x)$) (d021)--++($0.2*(y)$);
\draw[gray] (d001)--++($-0.2*(x)$) (d001)--++($-0.2*(y)$) (d102)--++($0.2*(z)$) (d102)--++($-0.2*(y)$) (d012)--++($-0.2*(x)$) (d012)--++($0.2*(z)$);
\atoms{z2,bdastyle=red}{x/p={z011}}
\end{tikzpicture}
=
0\;.
\end{equation}
Finally, note that the tensor-network formulation reveals the duality symmetry of the construction under swapping between (1) primal and Poincar\'e dual lattice, (2) $\zz_2$ and $\delta$-tensors, (3) $e$ and $m$ anyons, and (4) rough and smooth boundaries.

\subsection{Circuit on a rectangular spatial block}
\label{eq:dynamic_surface_code}
Let us next discuss how to turn the boundaries and corners of the path integral into boundaries and corners of the fault-tolerant circuit.
Specifically, we will define spatial rough boundaries of the circuit along the $z$ direction, spatial smooth boundaries along the $\overline x$ direction, and corners separating them.
So we need a path integral with smooth boundaries along the $(\overline x,t)$-plane and rough boundaries along the $(z,t)$-plane.
So the smooth boundary simply consists of $(x,y)$-faces parallel to the $(\overline x,t)$-plane, as the one shown in Eq.~\eqref{eq:boundary_zx_toric_code}.
However, there are no faces parallel to the $(z,t)$-plane, so the rough boundary consists of $(x,z)$ and $(y,z)$-faces that extend in the $t$-direction in a zig-zag manner.
The corner separating the rough and smooth boundary also extends in a zigzag manner.
The following picture shows a section of this path integral near the corner:
\begin{equation}
\label{eq:boundary_path_integral}
\begin{tikzpicture}
\atoms{void}{x/p={2,0}, y/p={0,2}, z/p={1.6,0.8}}
\foreach \x in {0,1,2,3}{
\foreach \y in {0,1,2}{
\foreach \z in {0,1}{
\atoms{void}{\x\y\z/p={$\x*(x)+\y*(y)+\z*(z)$}}
}}}
\draw[orange] (021)--(321);
\draw[orange,back] (001)--(101) (011)--(211) (101)--(121) (001)--(021) (211)--(221);
\draw[orange, back] (000)--(001) (010)--(011) (110)--(111);
\draw[orange] (220)--(221) (020)--(021) (120)--(121) (310)--(320);
\draw[roughbd,back] (100)--(101) (210)--(211) (101)--(201)--(211)--(311);
\draw[smoothbd] (000)--(100) (010)--(210) (020)--(320) (000)--(020) (210)--(220) (100)--(120);
\draw[roughbd] (310)--(311) (200)--(201) (320)--(321) (311)--(321);
\draw[corner] (100)--(200)--(210)--(310)--(320);
\foreach \x/\y/\z in {1/0/0, 1/2/0, 3/2/0, 1/4/0, 3/4/0, 0/1/0, 0/3/0, 2/1/0, 2/3/0, 4/3/0, 0/4/1, 2/4/1, 4/4/1, 1/4/2, 3/4/2, 5/4/0, 5/4/2}{
\atoms{delta}{d\x\y\z/p={$0.5*\x*(x)+0.5*\y*(y)+0.5*\z*(z)$}}
}
\foreach \x/\y/\z in {1/0/2, 1/2/2, 3/2/2, 0/1/2, 0/3/2, 2/1/2, 2/3/2, 2/2/1, 0/2/1, 0/0/1, 4/3/2}{
\atoms{delta,astyle=gray}{d\x\y\z/p={$0.5*\x*(x)+0.5*\y*(y)+0.5*\z*(z)$}}
}
\foreach \x/\y/\z in {1/1/0, 3/1/0, 1/3/0, 3/3/0, 1/4/1, 3/4/1, 4/3/1, 5/3/0, 5/4/1}{
\atoms{z2}{z\x\y\z/p={$0.5*\x*(x)+0.5*\y*(y)+0.5*\z*(z)$}}
}
\foreach \x/\y/\z in {1/1/2, 3/1/2, 1/3/2, 3/3/2, 1/2/1, 3/2/1, 2/1/1, 2/3/1, 0/1/1, 0/3/1, 1/0/1, 5/3/2}{
\atoms{z2,astyle=gray}{z\x\y\z/p={$0.5*\x*(x)+0.5*\y*(y)+0.5*\z*(z)$}}
}
\draw (d010)--++($-0.2*(x)$) (d030)--++($-0.2*(x)$) (d340)--++($0.2*(y)$) (d100)--++($-0.2*(y)$) (d441)--++($0.2*(x)$) (d441)--++($0.2*(y)$) (d041)--++($-0.2*(x)$) (d041)--++($0.2*(y)$) (d142)--++($0.2*(y)$) (d342)--++($0.2*(y)$) (d142)--++($0.2*(z)$) (d342)--++($0.2*(z)$) (d140)--++($0.2*(y)$) (d241)--++($0.2*(y)$) (d540)--++($0.2*(y)$) (d542)--++($0.2*(y)$) (d542)--++($0.2*(z)$);
\draw[gray] (d001)--++($-0.2*(x)$) (d001)--++($-0.2*(y)$) (d021)--++($-0.2*(x)$) (d032)--++($-0.2*(x)$) (d032)--++($0.2*(z)$) (d012)--++($0.2*(z)$) (d012)--++($-0.2*(x)$) (d102)--++($0.2*(z)$) (d102)--++($-0.2*(y)$) (d212)--++($0.2*(z)$) (d232)--++($0.2*(z)$) (d122)--++($0.2*(z)$) (d432)--++($0.2*(z)$) (d322)--++($0.2*(z)$);
\foreach \x/\y/\col in {1/0/gray, 0/1/gray, 2/1/gray, 1/2/gray, 0/3/gray, 2/3/gray, 3/2/gray, 4/3/gray, 1/4/black, 3/4/black, 5/4/black}{
\draw[\col] (d\x\y0)--(z\x\y1)--(d\x\y2);
}
\draw (d041)--(z141)--(d241)--(z341)--(d441)--(z541) (d030)--(z130)--(d230)--(z330)--(d430)--(z530) (d010)--(z110)--(d210)--(z310);
\draw[gray] (d001)--(z101) (d021)--(z121)--(d221)--(z321) (d012)--(z112)--(d212)--(z312) (d032)--(z132)--(d232)--(z332)--(d432)--(z532);
\draw (d100)--(z110)--(d120)--(z130)--(d140) (z310)--(d320)--(z330)--(d340) (z530)--(d540);
\draw[gray] (d001)--(z011)--(d021)--(z031)--(d041) (z211)--(d221)--(z231)--(d241) (z431)--(d441) (d102)--(z112)--(d122)--(z132)--(d142) (z312)--(d322)--(z332)--(d342) (z532)--(d542);
\draw[line width=0.3cm,\qubitb,opacity=0.5] (d100)--(z110)--(d210)--(z310)--(d320)--(z330)--(d430)--(z530)--(d540) (d100)--++($-0.2*(y)$) (d540)--++($0.2*(y)$);
\draw[line width=0.3cm,\qubita,opacity=0.5] (d001)--(z101) (z211)--(d221)--(z321) (z431)--(d441)--(z541);
\fill[cyan,opacity=0.5] (z310)circle(0.3);
\draw[line width=0.5cm,cyan,opacity=0.5,line cap=round,line join=round] (z321-c)-|(z431-c);
\path[cyan] (z310)++(0:0.5)node{(e)};
\path[cyan] (z321)++(0:1.45)node{(f)};
\end{tikzpicture}\;.
\end{equation}
The smooth boundary consisting of $(x,y)$-faces is in front, and the rough boundary consisting of $(x,z)$ and $(y,z)$-faces is at the bottom right.

As a next step, we need to define qubit worldlines near the boundaries.
The qubit worldlines near the (smooth) $(\overline x,t)$-boundary look like these in the bulk, as they run parallel to the $(\overline x,t)$-plane.
The qubit worldlines near the (rough) $(z,t)$-boundary also look like these in the bulk, but have missing tensors or indices as shown in Eq.~\eqref{eq:boundary_path_integral}.
Next, we consider the rectangular lattice with boundary obtained from projecting the spacetime lattice along $t$.
On this rectangular lattice, the qubit worldlines give rise to the following qubit locations
\footnote{
To be precise, the lattice shown below is not exactly the projection of Eq.~\eqref{eq:boundary_path_integral}, since we have stripped an extra row of ``empty'' rectangles at the bottom.
}:
\begin{equation}
\label{eq:spatial_boundary_lattice}
\begin{tikzpicture}
\draw (0,0)edge[mark={arr,e},ind=$z$]++(0:0.5) (0,0)edge[mark={arr,e},ind=$\overline x$]++(90:0.5*0.707107);
\end{tikzpicture}
\begin{tikzpicture}
\atoms{void}{x/p={1,0}, y/p={0,0.707107}}
\clip (-0.15,-0.15) rectangle($4*(x)+4*(y)+(-0.15,-0.15)$);
\draw[roughbd] ($0*(y)$)--++($5*(x)$);
\draw[orange] ($1*(y)$)--++($5*(x)$) ($2*(y)$)--++($5*(x)$) ($3*(y)$)--++($5*(x)$) ($4*(y)$)--++($5*(x)$) ($5*(y)$)--++($5*(x)$);
\draw[smoothbd] ($0*(x)$)--++($5*(y)$);
\draw[orange] ($1*(x)$)--++($5*(y)$) ($2*(x)$)--++($5*(y)$) ($3*(x)$)--++($5*(y)$) ($4*(x)$)--++($5*(y)$) ($5*(x)$)--++($5*(y)$);
\foreach \x in {0,1,2,3}{
\foreach \y in {0,1,2,3}{
\atoms{vertex,astyle={\qubitb}}{0/p={$\x*(x)+{\y+0.5}*(y)$}}
\atoms{vertex,astyle={\qubita}}{0/p={${\x+0.5}*(x)+{\y+0.5}*(y)$}}
}}
\end{tikzpicture}
\;.
\end{equation}
Let us next discuss how tensors near the boundary become operators in the circuit.
The (a) tensors of Eq.~\eqref{eq:tensor_circuit_labeling} on the smooth $(\overline x,t)$ boundary become $XX$ measurements as before.
The only way in which the smooth $(\overline x,t)$ boundary differs from the bulk is that the $\delta$-tensors only contribute to a single (c) operator or $CX$ gate.
On the rough boundary, the $\zz_2$-tensors labeled (e) in Eq.~\eqref{eq:tensor_circuit_labeling} are 2-index tensors that become identity operators:
\begin{equation}
\label{eq:single_qubit_identity}
\begin{tikzpicture}
\atoms{z2}{0/}
\draw (0)edge[ind=$a$]++(-90:0.5) (0)edge[ind=$c$]++(90:0.5);
\fill[cyan,opacity=0.5] (0)circle(0.3);
\path[cyan] (0)++(0:0.5)node{(e)};
\end{tikzpicture}
=
\begin{tikzpicture}
\draw (0,0)edge[ind=$a$,startind=$b$]++(-90:0.5);
\end{tikzpicture}
=
\bra{b} \idop \ket{a}
\;.
\end{equation}
Pairs of $\zz_2$ tensors on the rough $(z,t)$ boundary are combined into single tensors labeled (f) in Eq.~\eqref{eq:tensor_circuit_labeling}.
These tensors are also part of the (c) and (d) operators from Eq.~\eqref{eq:tensor_circuit_labeling}, and they are split up accordingly,
\begin{equation}
\label{eq:single_qubit_mz}
\begin{tikzpicture}
\atoms{z2}{0/}
\atoms{delta}{a0/p={-0.8,0}, b0/p={0.8,0}}
\atoms{z2}{1/p={0,1}}
\atoms{delta}{a1/p={-0.8,1}, b1/p={0.8,1}}
\draw (0)--(a0) (0)--(b0) (0)edge[ind=$a$]++(-90:0.5);
\draw (1)--(a1) (1)--(b1) (1)edge[ind=$b$]++(90:0.5);
\draw[line width=0.5cm,cyan,opacity=0.5,line cap=round] (0-c)--(1-c);
\path[cyan] (0,0.5)++(0:0.5)node{(f)};
\draw[line width=0.5cm,cyan,opacity=0.5,line cap=round] (0-c)--(a0-c);
\draw[line width=0.5cm,cyan,opacity=0.5,line cap=round] (0-c)--(b0-c);
\draw[line width=0.5cm,cyan,opacity=0.5,line cap=round] (1-c)--(a1-c);
\draw[line width=0.5cm,cyan,opacity=0.5,line cap=round] (1-c)--(b1-c);
\path[cyan] ($0.5*(0)+0.5*(a0)+(-90:0.5)$) node{(c)};
\path[cyan] ($0.5*(0)+0.5*(b0)+(-90:0.5)$) node{(d)};
\path[cyan] ($0.5*(1)+0.5*(a1)+(90:0.5)$) node{(c)};
\path[cyan] ($0.5*(1)+0.5*(b1)+(90:0.5)$) node{(d)};
\end{tikzpicture}
=
\begin{tikzpicture}
\atoms{z2}{0/, 2/p={0,-0.6}, 1/p={0,-1.2}}
\atoms{delta}{b/p={0.8,-0.6}, a/p={-0.8,-1.2}}
\draw (1)--(a) (2)--(b) (0)--(2) (2)--(1) (1)edge[ind=$a$]++(-90:0.5);
\atoms{z2}{0x/p={0,0.6}, 2x/p={0,1.2}, 1x/p={0,1.8}}
\atoms{delta}{bx/p={-0.8,1.2}, ax/p={0.8,1.8}}
\draw (1x)--(ax) (2x)--(bx) (0x)--(2x) (2x)--(1x) (1x)edge[ind=$b$]++(90:0.5);
\draw[line width=0.5cm,cyan,opacity=0.5,line cap=round] (0-c)--(0x-c);
\path[cyan] (0,0.3)++(0:0.5)node{(f)};
\draw[line width=0.5cm,cyan,opacity=0.5,line cap=round] (1-c)--(a-c);
\draw[line width=0.5cm,cyan,opacity=0.5,line cap=round] (2-c)--(b-c);
\draw[line width=0.5cm,cyan,opacity=0.5,line cap=round] (1x-c)--(ax-c);
\draw[line width=0.5cm,cyan,opacity=0.5,line cap=round] (2x-c)--(bx-c);
\path[cyan] ($0.5*(1)+0.5*(a)+(-90:0.5)$) node{(c)};
\path[cyan] ($0.5*(2)+0.5*(b)+(-90:0.5)$) node{(d)};
\path[cyan] ($0.5*(1x)+0.5*(ax)+(90:0.5)$) node{(d)};
\path[cyan] ($0.5*(2x)+0.5*(bx)+(90:0.5)$) node{(c)};
\end{tikzpicture}
\;.
\end{equation}
The resulting (f) operator is the projector $\ket0\bra0$.
\begin{equation}
\begin{tikzpicture}
\atoms{z2}{0/, 0x/p={0,0.6}}
\draw (0)edge[ind=$a$]++(-90:0.5) (0x)edge[ind=$b$]++(90:0.5);
\end{tikzpicture}
=
\delta_{a=0} \delta_{b=0}
=
\braket{b|0}\braket{0|a}
\;.
\end{equation}
Since a projector is not a valid quantum operation, we view it as a $+1$-post-selected single-qubit $Z$ measurement $M_Z$.
The $\ket1\bra1$ projector corresponding to the $-1$ measurement outcome can be expressed using charged $\zz_2$-tensors as
\begin{equation}
\label{eq:zmeas_minusone}
\braket{b|1}\braket{1|a}
=
\delta_{a=1}\delta_{b=1}
=
\begin{tikzpicture}
\atoms{z2,bdastyle=red}{0/, 0x/p={0,0.6}}
\draw (0)edge[ind=$a$]++(-90:0.5) (0x)edge[ind=$b$]++(90:0.5);
\end{tikzpicture}
\;.
\end{equation}
In order to examine the effect of the $-1$ measurement outcome on the path integral, we plug in Eq.~\eqref{eq:zmeas_minusone} instead of Eq.~\eqref{eq:single_qubit_mz}, and fuse the resulting charged tensors with the adjacent uncharged tensors.
We find that the $-1$ measurement outcome corresponds to two $m$ anyon segments being present at two faces adjacent to the rough $(z,t)$-boundary:
\begin{equation}
\begin{tikzpicture}
\atoms{void}{x/p={1.5,0}, y/p={0,1.5}, z/p={1.2,0.6}}
\foreach \x in {0,1}{
\foreach \y in {0,1}{
\foreach \z in {0,1}{
\atoms{void}{\x\y\z/p={$\x*(x)+\y*(y)+\z*(z)$}}
}}}
\draw[orange] (110)--(010) (011)--(111) (101)--(111)--(110)--(100) (000)--(100) (010)--(000) (010)--(011);
\draw[roughbd] (100)--(101);
\draw[orange,back] (001)--(101) (000)--(001) (001)--(011);
\atoms{z2}{z0/p={$0.3*(x)+0.5*(z)$}, z1/p={$0.5*(x)+0.5*(z)$}, z2/p={$(x)+0.5*(y)+0.5*(z)$}, z3/p={$(x)+0.7*(y)+0.5*(z)$}}
\atoms{z2,bdastyle=red}{zm0/p={$0.75*(x)+0.5*(z)$}, zm1/p={$(x)+0.25*(y)+0.5*(z)$}}
\draw (z0)--(z1)--(zm0) (zm1)--(z2)--(z3);
\draw (z0)--++($-0.3*(x)$) (z3)--++($0.3*(y)$) (z0)--++($-0.3*(z)$) (z1)--++($0.3*(z)$) (z2)--++($-0.3*(z)$) (z3)--++($0.3*(z)$);
\end{tikzpicture}
=
\begin{tikzpicture}
\atoms{void}{x/p={1.5,0}, y/p={0,1.5}, z/p={1.2,0.6}}
\foreach \x in {0,1}{
\foreach \y in {0,1}{
\foreach \z in {0,1}{
\atoms{void}{\x\y\z/p={$\x*(x)+\y*(y)+\z*(z)$}}
}}}
\draw[orange] (110)--(010) (011)--(111) (101)--(111)--(110)--(100) (000)--(100) (010)--(000) (010)--(011);
\draw[roughbd] (100)--(101);
\draw[orange,back] (001)--(101) (000)--(001) (001)--(011);
\draw[worldline] ($0.5*(x)+0.5*(z)-0.5*(y)$)--++(y)--++(x);
\atoms{z2,bdastyle=red}{z1/p={$0.5*(x)+0.5*(z)$}, z2/p={$(x)+0.5*(y)+0.5*(z)$}}
\draw (z1)--++($-0.3*(x)$) (z2)--++($0.3*(y)$) (z1)--++($-0.3*(z)$) (z1)--++($0.3*(z)$) (z2)--++($-0.3*(z)$) (z2)--++($0.3*(z)$);
\end{tikzpicture}
\;.
\end{equation}
Let us now discuss in which order the above operators act on the qubits on the spatial lattice shown in Eq.~\eqref{eq:spatial_boundary_lattice}, by looking at Eq.~\eqref{eq:boundary_path_integral}.
The (e) single-qubit identity operations in Eq.~\eqref{eq:single_qubit_identity} act at the same time as a layer of $XX$ and $ZZ$ measurements, namely the one where the green qubits nearest to the boundary are not acted on by a bulk $XX$ measurement.
The (f) single-qubit $Z$ measurements in Eq.~\eqref{eq:single_qubit_mz} act at the same time as the other layer of $XX$ and $ZZ$ measurements, where the purple qubits nearest to the boundary are not acted on by $ZZ$ measurements.
All in all, the circuit is given by:
\begin{equation}
\label{eq:boundary_circuit}
\begin{tabular}{ccc}
\begin{tikzpicture}
\atoms{void}{x/p={1,0}, y/p={0,0.707107}}
\clip (-0.35,-0.05) rectangle($3*(x)+3*(y)+(-0.15,-0.15)$);
\draw[roughbd] ($0*(y)$)--++($5*(x)$);
\draw[orange] ($1*(y)$)--++($5*(x)$) ($2*(y)$)--++($5*(x)$) ($3*(y)$)--++($5*(x)$) ($4*(y)$)--++($5*(x)$) ($5*(y)$)--++($5*(x)$);
\draw[smoothbd] ($0*(x)$)--++($5*(y)$);
\draw[orange] ($1*(x)$)--++($5*(y)$) ($2*(x)$)--++($5*(y)$) ($3*(x)$)--++($5*(y)$) ($4*(x)$)--++($5*(y)$) ($5*(x)$)--++($5*(y)$);
\foreach \x in {0,1,2,3}{
\foreach \y in {0,1,2,3}{
\atoms{vertex,astyle={\qubitb}}{g\x\y/p={$\x*(x)+{\y+0.5}*(y)$}}
\atoms{vertex,astyle={\qubita}}{r\x\y/p={${\x+0.5}*(x)+{\y+0.5}*(y)$}}
}}
\foreach \x in {0,1,2}{
\draw[line width=0.3cm,cyan,opacity=0.5,line cap=round] (g\x0-c)--(g\x1-c) (g\x2-c)--(g\x3-c) (r\x1-c)--(r\x2-c);
\path (g\x0-c)--node[midway]{$\scriptstyle{M_{XX}}$} (g\x1-c);
\path (r\x1-c)--node[midway]{$\scriptstyle{M_{ZZ}}$} (r\x2-c);
\fill[cyan,opacity=0.5] (r\x0)circle(0.2);
\path (r\x0)++(-90:0.2)node{$\scriptstyle{M_Z}$};
}
\end{tikzpicture}
&
$\leftarrow$
&
\begin{tikzpicture}
\atoms{void}{x/p={1,0}, y/p={0,0.707107}}
\clip (-0.35,-0.05) rectangle($3*(x)+3*(y)+(-0.15,-0.1)$);
\draw[roughbd] ($0*(y)$)--++($5*(x)$);
\draw[orange] ($1*(y)$)--++($5*(x)$) ($2*(y)$)--++($5*(x)$) ($3*(y)$)--++($5*(x)$) ($4*(y)$)--++($5*(x)$) ($5*(y)$)--++($5*(x)$);
\draw[smoothbd] ($0*(x)$)--++($5*(y)$);
\draw[orange] ($1*(x)$)--++($5*(y)$) ($2*(x)$)--++($5*(y)$) ($3*(x)$)--++($5*(y)$) ($4*(x)$)--++($5*(y)$) ($5*(x)$)--++($5*(y)$);
\foreach \x in {0,1,2,3}{
\foreach \y in {0,1,2,3}{
\atoms{vertex,astyle={\qubitb}}{g\x\y/p={$\x*(x)+{\y+0.5}*(y)$}}
\atoms{vertex,astyle={\qubita}}{r\x\y/p={${\x+0.5}*(x)+{\y+0.5}*(y)$}}
}}
\foreach \y in {0,1,2}{
\draw[line width=0.3cm,cyan,opacity=0.5,line cap=round] (r0\y-c)--(g1\y-c) (r1\y-c)--(g2\y-c) (r2\y-c)--(g3\y-c);
\node at ($0.5*(r0\y-c)+0.5*(g1\y-c)+(90:0.15)$) {$\scriptstyle{CX}$};
\node at ($0.5*(r1\y-c)+0.5*(g2\y-c)+(90:0.15)$) {$\scriptstyle{CX}$};
\node at ($0.5*(r2\y-c)+0.5*(g3\y-c)+(90:0.15)$) {$\scriptstyle{CX}$};
}
\end{tikzpicture}
\\
$\downarrow$ & & $\uparrow$\\
\begin{tikzpicture}
\atoms{void}{x/p={1,0}, y/p={0,0.707107}}
\clip (-0.35,-0.05) rectangle($3*(x)+3*(y)+(-0.15,-0.1)$);
\draw[roughbd] ($0*(y)$)--++($5*(x)$);
\draw[orange] ($1*(y)$)--++($5*(x)$) ($2*(y)$)--++($5*(x)$) ($3*(y)$)--++($5*(x)$) ($4*(y)$)--++($5*(x)$) ($5*(y)$)--++($5*(x)$);
\draw[smoothbd] ($0*(x)$)--++($5*(y)$);
\draw[orange] ($1*(x)$)--++($5*(y)$) ($2*(x)$)--++($5*(y)$) ($3*(x)$)--++($5*(y)$) ($4*(x)$)--++($5*(y)$) ($5*(x)$)--++($5*(y)$);
\foreach \x in {0,1,2,3}{
\foreach \y in {0,1,2,3}{
\atoms{vertex,astyle={\qubitb}}{g\x\y/p={$\x*(x)+{\y+0.5}*(y)$}}
\atoms{vertex,astyle={\qubita}}{r\x\y/p={${\x+0.5}*(x)+{\y+0.5}*(y)$}}
}}
\foreach \y in {0,1,2}{
\draw[line width=0.3cm,cyan,opacity=0.5,line cap=round] (g0\y-c)--(r0\y-c) (g1\y-c)--(r1\y-c) (g2\y-c)--(r2\y-c);
\node at ($0.5*(g0\y-c)+0.5*(r0\y-c)+(90:0.15)$) {$\scriptstyle{CX}$};
\node at ($0.5*(g1\y-c)+0.5*(r1\y-c)+(90:0.15)$) {$\scriptstyle{CX}$};
\node at ($0.5*(g2\y-c)+0.5*(r2\y-c)+(90:0.15)$) {$\scriptstyle{CX}$};
}
\end{tikzpicture}
& &
\begin{tikzpicture}
\atoms{void}{x/p={1,0}, y/p={0,0.707107}}
\clip (-0.35,-0.05) rectangle($3*(x)+3*(y)+(-0.15,-0.1)$);
\draw[roughbd] ($0*(y)$)--++($5*(x)$);
\draw[orange] ($1*(y)$)--++($5*(x)$) ($2*(y)$)--++($5*(x)$) ($3*(y)$)--++($5*(x)$) ($4*(y)$)--++($5*(x)$) ($5*(y)$)--++($5*(x)$);
\draw[smoothbd] ($0*(x)$)--++($5*(y)$);
\draw[orange] ($1*(x)$)--++($5*(y)$) ($2*(x)$)--++($5*(y)$) ($3*(x)$)--++($5*(y)$) ($4*(x)$)--++($5*(y)$) ($5*(x)$)--++($5*(y)$);
\foreach \x in {0,1,2,3}{
\foreach \y in {0,1,2,3}{
\atoms{vertex,astyle={\qubitb}}{g\x\y/p={$\x*(x)+{\y+0.5}*(y)$}}
\atoms{vertex,astyle={\qubita}}{r\x\y/p={${\x+0.5}*(x)+{\y+0.5}*(y)$}}
}}
\foreach \y in {0,1,2}{
\draw[line width=0.3cm,cyan,opacity=0.5,line cap=round] (g0\y-c)--(r0\y-c) (g1\y-c)--(r1\y-c) (g2\y-c)--(r2\y-c);
\node at ($0.5*(g0\y-c)+0.5*(r0\y-c)+(90:0.15)$) {$\scriptstyle{CX}$};
\node at ($0.5*(g1\y-c)+0.5*(r1\y-c)+(90:0.15)$) {$\scriptstyle{CX}$};
\node at ($0.5*(g2\y-c)+0.5*(r2\y-c)+(90:0.15)$) {$\scriptstyle{CX}$};
}
\end{tikzpicture}
\\
$\downarrow$ && $\uparrow$\\
\begin{tikzpicture}
\atoms{void}{x/p={1,0}, y/p={0,0.707107}}
\clip (-0.35,-0.05) rectangle($3*(x)+3*(y)+(-0.15,-0.1)$);
\draw[roughbd] ($0*(y)$)--++($5*(x)$);
\draw[orange] ($1*(y)$)--++($5*(x)$) ($2*(y)$)--++($5*(x)$) ($3*(y)$)--++($5*(x)$) ($4*(y)$)--++($5*(x)$) ($5*(y)$)--++($5*(x)$);
\draw[smoothbd] ($0*(x)$)--++($5*(y)$);
\draw[orange] ($1*(x)$)--++($5*(y)$) ($2*(x)$)--++($5*(y)$) ($3*(x)$)--++($5*(y)$) ($4*(x)$)--++($5*(y)$) ($5*(x)$)--++($5*(y)$);
\foreach \x in {0,1,2,3}{
\foreach \y in {0,1,2,3}{
\atoms{vertex,astyle={\qubitb}}{g\x\y/p={$\x*(x)+{\y+0.5}*(y)$}}
\atoms{vertex,astyle={\qubita}}{r\x\y/p={${\x+0.5}*(x)+{\y+0.5}*(y)$}}
}}
\foreach \y in {0,1,2}{
\draw[line width=0.3cm,cyan,opacity=0.5,line cap=round] (r0\y-c)--(g1\y-c) (r1\y-c)--(g2\y-c) (r2\y-c)--(g3\y-c);
\node at ($0.5*(r0\y-c)+0.5*(g1\y-c)+(90:0.15)$) {$\scriptstyle{CX}$};
\node at ($0.5*(r1\y-c)+0.5*(g2\y-c)+(90:0.15)$) {$\scriptstyle{CX}$};
\node at ($0.5*(r2\y-c)+0.5*(g3\y-c)+(90:0.15)$) {$\scriptstyle{CX}$};
}
\end{tikzpicture}
&
$\rightarrow$
&
\begin{tikzpicture}
\atoms{void}{x/p={1,0}, y/p={0,0.707107}}
\clip (-0.35,-0.05) rectangle($3*(x)+3*(y)+(-0.15,-0.15)$);
\draw[roughbd] ($0*(y)$)--++($5*(x)$);
\draw[orange] ($1*(y)$)--++($5*(x)$) ($2*(y)$)--++($5*(x)$) ($3*(y)$)--++($5*(x)$) ($4*(y)$)--++($5*(x)$) ($5*(y)$)--++($5*(x)$);
\draw[smoothbd] ($0*(x)$)--++($5*(y)$);
\draw[orange] ($1*(x)$)--++($5*(y)$) ($2*(x)$)--++($5*(y)$) ($3*(x)$)--++($5*(y)$) ($4*(x)$)--++($5*(y)$) ($5*(x)$)--++($5*(y)$);
\foreach \x in {0,1,2,3}{
\foreach \y in {0,1,2,3}{
\atoms{vertex,astyle={\qubitb}}{g\x\y/p={$\x*(x)+{\y+0.5}*(y)$}}
\atoms{vertex,astyle={\qubita}}{r\x\y/p={${\x+0.5}*(x)+{\y+0.5}*(y)$}}
}}
\foreach \x in {0,1,2}{
\draw[line width=0.3cm,cyan,opacity=0.5,line cap=round] (g\x1-c)--(g\x2-c) (r\x0-c)--(r\x1-c) (r\x2-c)--(r\x3-c);
\path (g\x1-c)--node[midway]{$\scriptstyle{M_{XX}}$} (g\x2-c);
\path (r\x0-c)--node[midway]{$\scriptstyle{M_{ZZ}}$} (r\x1-c);
}
\end{tikzpicture}
\end{tabular}
\;.
\end{equation}
By mirroring the boundaries shown above, we can define the circuit on a spatial $L_1\times L_2$ block of rectangular lattice, with smooth boundaries at the top and bottom and rough boundaries on the left and right.

\subsection{Logical \texorpdfstring{$ZZ$}{ZZ} measurement}
\label{sec:lattice_surgery}
Finally, let us construct a fault-tolerant circuit to implement a logical $ZZ$ measurement by performing lattice surgery of two spatial rectangular circuit blocks along their smooth boundaries.
Just like for the ordinary surface code, lattice surgery is a process where we put two spatial $L\times L$ blocks of qubits next to each other, merge them at their smooth boundaries into a single rectangular block, wait for some time $\propto L$, and then split it into two blocks again.
The spacetime volume corresponding to this process is given by two nearby cubic spacetime blocks connected via a ``bridge'' for some time $\propto L$:
\begin{equation}
\label{eq:lattice_surgary_topology}
\begin{tikzpicture}
\draw (-0.7,0)edge[mark={arr,e},ind=$z$]++(0:0.35) (-0.7,0)edge[mark={arr,e},ind=$\overline x$]++(25:0.25) (-0.7,0)edge[mark={arr,e},ind=$t$]++(90:0.35);
\atoms{void}{x/p={2,0}, y/p={0,2}, z/p={1.6,0.8}}
\fill[roughbdfillback] (z)--++(x)--++($0.5*(y)$)--++($0.4*(x)$)--++($-0.5*(y)$)--++(x)--++($2*(y)$)--++($-1*(x)$)--++($-0.5*(y)$)--++($-0.4*(x)$)--++($0.5*(y)$)--++($-1*(x)$)--cycle;
\draw[very thick] ($(x)+(z)$)--++($0.5*(y)$)--++($0.4*(x)$)--++($-0.5*(y)$);
\draw[very thick] ($2*(y)+(x)+(z)$)--++($-0.5*(y)$)--++($0.4*(x)$)--++($0.5*(y)$);
\draw[very thick] (z)--++($2*(y)$) ($2.4*(x)+(z)$)--++($2*(y)$);
\fill[red,opacity=0.6] (0,0)--++(x)--++(z)--++($-1*(x)$)--cycle;
\fill[red,opacity=0.6] ($1.4*(x)$)--++(x)--++(z)--++($-1*(x)$)--cycle;
\fill[black,opacity=0.6] (0,0)--++(z)--++($2*(y)$)--++($-1*(z)$)--cycle;
\fill[black,opacity=0.6] ($2.4*(x)$)--++(z)--++($2*(y)$)--++($-1*(z)$)--cycle;
\fill[black,opacity=0.6] (x)--++(z)--++($0.5*(y)$)--++($-1*(z)$)--cycle;
\fill[black,opacity=0.6] ($1.4*(x)$)--++(z)--++($0.5*(y)$)--++($-1*(z)$)--cycle;
\fill[black,opacity=0.6] ($(x)+0.5*(y)$)--++(z)--++($0.4*(x)$)--++($-1*(z)$)--cycle;
\fill[black,opacity=0.6] ($2*(y)+(x)$)--++(z)--++($-0.5*(y)$)--++($-1*(z)$)--cycle;
\fill[black,opacity=0.6] ($2*(y)+1.4*(x)$)--++(z)--++($-0.5*(y)$)--++($-1*(z)$)--cycle;
\fill[black,opacity=0.6] ($(x)+1.5*(y)$)--++(z)--++($0.4*(x)$)--++($-1*(z)$)--cycle;
\fill[red,opacity=0.6] ($2*(y)$)--++(x)--++(z)--++($-1*(x)$)--cycle;
\fill[red,opacity=0.6] ($2*(y)+1.4*(x)$)--++(x)--++(z)--++($-1*(x)$)--cycle;
\path[roughbdfill] (0,0)--++(x)--++($0.5*(y)$)--++($0.4*(x)$)--++($-0.5*(y)$)--++(x)--++($2*(y)$)--++($-1*(x)$)--++($-0.5*(y)$)--++($-0.4*(x)$)--++($0.5*(y)$)--++($-1*(x)$)--cycle;
\draw[very thick] (x)--++($0.5*(y)$)--++($0.4*(x)$)--++($-0.5*(y)$);
\draw[very thick] ($2*(y)+(x)$)--++($-0.5*(y)$)--++($0.4*(x)$)--++($0.5*(y)$);
\draw[very thick] (0,0)--++($2*(y)$) ($2.4*(x)$)--++($2*(y)$);
\draw[purple] (0,-0.2)edge[<->,mark={slab=$L$,r}]++(x) ($1.4*(x)+(0,-0.2)$)edge[<->,mark={slab=$L$,r}]++(x) ($(x)+(0,-0.2)$)edge[<->,mark={slab=$1$,r}]++($0.4*(x)$) ($2.4*(x)+(0,-0.2)$)edge[<->,mark={slab=$L$,r}]++(z);
\draw[dashed,purple] ($(x)+0.5*(y)$)edge[ind=$T_0$]++($-1.2*(x)$) ($(x)+1.5*(y)$)edge[ind=$T_1$]++($-1.2*(x)$);
\draw[purple] ($-0.2*(x)+0.5*(y)$)edge[<->,mark={slab=$L$}]++(y);
\fill[cyan,opacity=0.5,rc] ($0.9*(x)+0.3*(y)$)rectangle++($0.6*(x)+0.4*(y)$);
\end{tikzpicture}
\;.
\end{equation}
Here, the smooth boundaries on the left, right, and middle are shaded in gray, the rough boundaries at the front and back are dotted, and corners between rough and smooth boundaries are solid black lines.
The two $L\times L$ red boundaries at each the bottom and the top are not physical boundaries, but correspond to the initial and final blocks of qubits that constitute the inputs and outputs of the circuit.
For the corresponding path integral, these are state boundaries.
In this context, the tensor resulting from evaluating the path integral is interpreted as an operator from the initial to the final qubits, rather than a state supported on both the initial and final qubits.

To implement the above protocol, we need to choose a microscopic spacetime lattice representing the above spacetime manifold.
We start with the spacetime block for the rectangular circuit of Section~\ref{eq:dynamic_surface_code}, a patch of which is depicted in Eq.~\eqref{eq:boundary_path_integral}.
We use $L$ unit cells in the $\overline x$ direction and $2L+1$ unit cells in $z$ direction.
Then we remove all cubes with $z$ coordinates between $L$ and $L+1$ and $t$ coordinates outside of $[T_0, T_1]$.
This effectively splits the single spacetime block into two before time $T_0$ and after time $T_1$.
The boundary that is created from removing the cubes is smooth.
The following picture shows a section of this lattice and the according the tensor-network path integral at $T_0$, near the area marked in blue in Eq.~\eqref{eq:lattice_surgary_topology}:
\begin{equation}
\label{eq:surgery_path_integral}
\begin{tikzpicture}
\atoms{void}{x/p={2,0}, y/p={0,2}, z/p={1.6,0.8}}
\foreach \x in {0,1,2,3}{
\foreach \y in {-1,0,1}{
\foreach \z in {0,1}{
\atoms{void}{\x\y\z/p={$\x*(x)+\y*(y)+\z*(z)$}}
}}}
\draw[corner] (1-10)--(2-10)--(200) (2-11)--(201);
\draw[corner,back] (1-11)--(2-11) (200)--(201);
\draw[smoothbd] (000)--(100)--(110)--(010)--(000) (100)--(1-10) (200)--(100) (010)--(011) (110)--(111) (011)--(111) (000)--(0-10)--(1-10);
\draw[smoothbd,back] (100)--(101)--(001)--(011) (111)--(101) (201)--(101)--(1-11) (001)--(0-11)--(1-11);
\draw[roughbd, back] (201)--(301);
\draw[roughbd] (200)--(300)--(301)--(311)--(310)--(300);
\draw[orange] (200)--(210) (110)--(210) (210)--(211) (210)--(310) (111)--(311);
\draw[orange,back] (201)--(211);
\foreach \x/\y/\z in {1/0/0, 3/0/0, 1/2/0, 3/2/0, 0/1/0, 2/1/0, 4/1/0, 4/2/1, 2/-1/0, 0/-1/0, 1/-2/0, 5/2/0, 5/2/2, 1/2/2, 3/2/2, 2/2/1, 0/2/1}{
\atoms{delta}{d\x\y\z/p={$0.5*\x*(x)+0.5*\y*(y)+0.5*\z*(z)$}}
}
\foreach \x/\y/\z in {1/0/2, 3/0/2, 0/1/2, 2/1/2, 2/0/1, 2/-1/2, 0/-1/2, 1/-2/2, 4/1/2}{
\atoms{delta,astyle=gray}{d\x\y\z/p={$0.5*\x*(x)+0.5*\y*(y)+0.5*\z*(z)$}}
}
\foreach \x/\y/\z in {1/1/0, 3/1/0, 3/-1/0, 1/-1/0, 5/1/0, 5/2/1, 3/2/1, 1/2/1}{
\atoms{z2}{z\x\y\z/p={$0.5*\x*(x)+0.5*\y*(y)+0.5*\z*(z)$}}
}
\foreach \x/\y/\z in {1/1/2, 3/1/2, 2/1/1, 3/0/1, 3/-1/2, 1/-1/2, 5/1/2, 4/1/1}{
\atoms{z2,astyle=gray}{z\x\y\z/p={$0.5*\x*(x)+0.5*\y*(y)+0.5*\z*(z)$}}
}
\draw (d010)--++($-0.2*(x)$) (d320)--++($0.2*(y)$) (d421)--++($0.2*(y)$) (d021)--++($0.2*(y)$) (d122)--++($0.2*(y)$) (d322)--++($0.2*(y)$) (d010)--++($-0.2*(z)$) (d100)--++($-0.2*(z)$) (d300)--++($-0.2*(z)$) (d410)--++($-0.2*(z)$) (d120)--++($0.2*(y)$) (d210)--++($-0.2*(z)$) (d320)--++($-0.2*(z)$) (d120)--++($-0.2*(z)$) (d0-10)--++($-0.2*(z)$) (d1-20)--++($-0.2*(z)$) (d2-10)--++($-0.2*(z)$) (d520)--++($-0.2*(z)$) (d122)--++($0.2*(z)$) (d322)--++($0.2*(z)$) (d522)--++($0.2*(z)$) (d221)--++($0.2*(y)$) (d520)--++($0.2*(y)$) (d522)--++($0.2*(y)$) (d0-10)--++($-0.2*(x)$) (d1-20)--++($-0.2*(y)$);
\draw[gray] (d012)--++($0.2*(z)$) (d012)--++($-0.2*(x)$) (d102)--++($0.2*(z)$) (d302)--++($0.2*(z)$) (d212)--++($0.2*(z)$) (d412)--++($0.2*(z)$) (d0-12)--++($0.2*(z)$) (d1-22)--++($0.2*(z)$) (d2-12)--++($0.2*(z)$) (d0-12)--++($-0.2*(x)$) (d1-22)--++($-0.2*(y)$);
\draw (d1-20)--(z1-10)--(d100)--(z110)--(d120) (z3-10)--(d300)--(z310)--(d320) (z510)--(d520) (d0-10)--(z1-10)--(d2-10)--(z3-10) (d010)--(z110)--(d210)--(z310)--(d410)--(z510);
\draw[gray] (d1-22)--(z1-12)--(d102)--(z112)--(d122) (z3-12)--(d302)--(z312)--(d322) (z512)--(d522) (d0-12)--(z1-12)--(d2-12)--(z3-12) (d012)--(z112)--(d212)--(z312)--(d412)--(z512);
\draw (d120)--(z121)--(d122) (d320)--(z321)--(d322) (d520)--(z521)--(d522) (d021)--(z121)--(d221)--(z321)--(d421)--(z521);
\draw[gray] (d210)--(z211)--(d212) (d300)--(z301)--(d302) (d410)--(z411)--(d412) (d201)--(z211)--(d221) (d201)--(z301) (z411)--(d421);
\draw[line width=0.3cm,\qubitb,opacity=0.5] (d0-10)--(z1-10)--(d100)--(z110)--(d210)--(z310)--(d320) (d0-10)--++($-0.2*(x)$) (d320)--++($0.2*(y)$);
\draw[line width=0.3cm,\qubita,opacity=0.5] (d201)--(z301) (z411)--(d421)--(z521);
\fill[cyan,opacity=0.5] (d201)circle(0.3);
\path[cyan] (d201)++(180:0.5)node{(g)};
\draw (6.5,-1.5)edge[mark={arr,e},ind=$x$]++(0:0.35) (6.5,-1.5)edge[mark={arr,e},ind=$z$]++(25:0.25) (6.5,-1.5)edge[mark={arr,e},ind=$y$]++(90:0.35) (6.5,-1.5)edge[gray,mark={arr,e},ind=$t$]++(0.35,0.35) (6.5,-1.5)edge[gray,mark={arr,e},ind=$\overline x$]++(-0.175,0.175);
\end{tikzpicture}\;.
\end{equation}
Be aware that the coordinate system of the microscopic lattice shown above equals that in Eq.~\eqref{eq:boundary_path_integral} and differs from that in Eq.~\eqref{eq:lattice_surgary_topology}.
The three cubes at the bottom left are cubes that have been removed.
Boundary edges are drawn in thick, and corner edges or extra thick.
The zig-zag sequence of extra-thick corner edges towards the bottom right corresponds to the U-shaped corner in front marked blue in Eq.~\eqref{eq:lattice_surgary_topology}.
The tensor-network path integral at $T_1$ or at the back in Eq.~\eqref{eq:lattice_surgary_topology} is the same as shown above, apart from a reflection along the $t$ or $\overline x$ direction.

Let us now turn the path integral on the above lattice into a circuit.
Before $T_0$ and after $T_1$, the circuit is the one of two separate spatial $L\times L$ blocks described in Section~\ref{eq:dynamic_surface_code}.
Between $T_0$ and $T_1$, the circuit is the one for a single spatial $L\times (2L+1)$ block.
All that remains to be constructed is the sequence of operations that merges two blocks to a single block at $T_0$, and the sequence that splits the single block into two blocks at $T_1$.
We start by discussing how the qubit worldlines behave at these merging and splitting transitions.
The qubit worldlines of one of the individual blocks transfer to the merged block, such as the green worldline shown in Eq.~\eqref{eq:surgery_path_integral}.
However, the purple qubit worldlines of the merged block with $z$-coordinate $L+\frac12$ appear at time $T_0$ and disappear at time $T_1$.
Thus, the set of all qubits used in the protocol is that of the merged $L\times (2L+1)$ block, but the $z=L+\frac12$-qubits are unused at times outside of $[T_0,T_1]$.

Let us start by constructing the merging transition at $T_0$.
The purple $z=L+\frac12$ qubit worldlines originate pairwise at $\delta$-tensors at $z$-edges of the smooth boundary, one of which we have marked as (g) in blue in Eq.~\eqref{eq:surgery_path_integral}.
These (g) tensors can be interpreted as linear operators $\cc^1\rightarrow \cc^{2\times 2}$
\footnote{
Strictly speaking, the qubits already exist before time $T_0$, so we should define operators $\cc^{2\times 2}\rightarrow \cc^{2\times 2}$.
However, since they are unused before $T_0$, we can pretend that they are created just at time $T_0$.
}:
\begin{equation}
\begin{tikzpicture}
\atoms{delta}{0/}
\draw (0)edge[ind=$a$]++(135:0.5) (0)edge[ind=$b$]++(45:0.5);
\fill[cyan,opacity=0.5] (0)circle(0.3);
\path[cyan] (0)++(0:0.5)node{(g)};
\end{tikzpicture}
=
\bra{ab}(\ket{00}+\ket{11})
\;.
\end{equation}
We see that this operator is already an isometry, namely the preparation of two qubits in a Bell state.
However, it turns out more elegant to split the 2-index $\delta$-tensor into a 4-index $\delta$-tensor and two 1-index $\delta$-tensors,
\begin{equation}
\label{eq:t0_transition_operator}
\begin{tikzpicture}
\atoms{delta}{0/}
\draw (0)edge[ind=$a$]++(135:0.5) (0)edge[ind=$b$]++(45:0.5);
\fill[cyan,opacity=0.5] (0)circle(0.3);
\path[cyan] (0)++(0:0.5)node{(g)};
\end{tikzpicture}
=
\begin{tikzpicture}
\atoms{delta}{0/, 1/p={-135:0.5}, 2/p={-45:0.5}}
\draw (0)edge[ind=$a$]++(135:0.5) (0)edge[ind=$b$]++(45:0.5);
\draw (0)--(1) (0)--(2);
\fill[cyan,opacity=0.5] (0)circle(0.3);
\path[cyan] (0)++(0:0.5)node{(b)};
\end{tikzpicture}
\;.
\end{equation}
The 4-index $\delta$-tensor is interpreted as a $\frac12(1+ZZ)$ projector just like the bulk (b) tensors in Eq.~\eqref{eq:tensor_circuit_labeling}.
The 1-index $\delta$-tensors are preparations of a single qubit in a $\ket+$ state:
\begin{equation}
\begin{tikzpicture}
\atoms{delta}{0/}
\draw (0)edge[ind=$a$]++(90:0.5);
\end{tikzpicture}
=
\braket{a|+}
\;.
\end{equation}
While the latter operator is already an isometry, the $\frac12(1+ZZ)$ projector can be viewed as a $+1$ post-selected $ZZ$ measurement.
To see the effect of a $-1$ outcome on the path integral, we plug in the tensor-network representation of $\frac12(1-ZZ)$ shown in Eq.~\eqref{eq:zz_minusone_projector}, and fuse the charged $\zz_2$-tensors with the nearby uncharged $\zz_2$-tensors.
In contrast to the (b) operators in the bulk in Eq.~\eqref{eq:minusone_outcome_mapping}, a $-1$ outcome corresponds to an $m$ anyon being present only at a single smooth boundary face:
\begin{equation}
\begin{tikzpicture}
\atoms{void}{x/p={1.5,0}, y/p={0,1.5}, z/p={1.2,0.6}}
\foreach \x in {0,1}{
\foreach \y in {0,1}{
\foreach \z in {0,1}{
\atoms{void}{\x\y\z/p={$\x*(x)+\y*(y)+\z*(z)$}}
}}}
\draw[orange] (110)--(010) (011)--(111) (101)--(111)--(110)--(100);
\draw[smoothbd] (000)--(100) (010)--(000) (100)--(101) (010)--(011);
\draw[smoothbd,back] (000)--(001) (001)--(011) (001)--(101);
\atoms{delta}{d/p={$0.5*(z)$}}
\atoms{z2}{z0/p={$0.5*(x)+0.5*(z)$}, z1/p={$0.5*(y)+0.5*(z)$}}
\atoms{z2,bdastyle=red}{zm0/p={$0.25*(x)+0.5*(z)$}, zm1/p={$-0.25*(y)+0.5*(z)$}}
\atoms{delta}{d0/p={$-0.5*(x)+0.5*(z)$}, d1/p={$-0.5*(y)+0.5*(z)$}}
\draw (d)--(d0) (d)--(zm1)--(d1) (d)--(zm0)--(z0) (d)--(z1);
\draw (z0)--++($-0.3*(z)$) (z0)--++($0.3*(z)$) (z0)--++($0.3*(x)$) (z1)--++($0.3*(y)$) (z1)--++($-0.3*(z)$) (z1)--++($0.3*(z)$);
\end{tikzpicture}
=
\begin{tikzpicture}
\atoms{void}{x/p={1.5,0}, y/p={0,1.5}, z/p={1.2,0.6}}
\foreach \x in {0,1}{
\foreach \y in {0,1}{
\foreach \z in {0,1}{
\atoms{void}{\x\y\z/p={$\x*(x)+\y*(y)+\z*(z)$}}
}}}
\draw[orange] (110)--(010) (011)--(111) (101)--(111)--(110)--(100);
\draw[smoothbd] (000)--(100) (010)--(000) (100)--(101) (010)--(011);
\draw[smoothbd,back] (000)--(001) (001)--(011) (001)--(101);
\draw[worldline] ($0.5*(x)+0.5*(z)$)--++($0.5*(y)$);
\atoms{delta}{d/p={$0.5*(z)$}}
\atoms{z2}{{z0/p={$0.5*(x)+0.5*(z)$},bdastyle=red}, z1/p={$0.5*(y)+0.5*(z)$}}
\draw (d)--(z0) (d)--(z1);
\draw (z0)--++($-0.3*(z)$) (z0)--++($0.3*(z)$) (z0)--++($0.3*(x)$) (z1)--++($0.3*(y)$) (z1)--++($-0.3*(z)$) (z1)--++($0.3*(z)$);
\end{tikzpicture}
\;.
\end{equation}
Next, we determine in which order the above operators act on which qubits.
The (g) tensors in Eq.~\eqref{eq:surgery_path_integral} have $t$-coordinate $k+0$ or $k+\frac12$ for some integer $k$.
Thus, the (b) $ZZ$ measurement in Eq.~\eqref{eq:t0_transition_operator} acts together with one of the two layers of bulk (a) and (b) $XX$ or $ZZ$ measurements.
We could make the transition at either of these two layers, but for the lattice shown in Eq.~\eqref{eq:surgery_path_integral}, this is the layer that contains the (c) identity operations on the green qubits near the boundary.
Accordingly, the $\ket+$ state preparations in Eq.~\eqref{eq:t0_transition_operator} are performed together with the preceeding layer of $CX$ gates.
So all in all, the transition consists in preparing the unused qubits in a $\ket+$ state, and switching from two blocks to a single block of qubits at the right time.
The following picture shows the sequence of operations performed around this merging transition:
\begin{equation}
\begin{tabular}{ccc}
\begin{tikzpicture}
\atoms{void}{x/p={1,0}, y/p={0,0.707107}}
\clip ($-2.5*(x)+(0.15,-0.15)$) rectangle($1.5*(x)+4*(y)+(-0.15,-0.15)$);
\draw[roughbd] ($0*(y)$)--++($4*(x)$) ($-1*(x)$)--++($-4*(x)$);
\draw[orange] ($1*(y)$)--++($4*(x)$) ($2*(y)$)--++($4*(x)$) ($3*(y)$)--++($4*(x)$) ($4*(y)$)--++($4*(x)$) ($5*(y)$)--++($4*(x)$);
\draw[orange] ($1*(y)-(x)$)--++($-4*(x)$) ($2*(y)-(x)$)--++($-4*(x)$) ($3*(y)-(x)$)--++($-4*(x)$) ($4*(y)-(x)$)--++($-4*(x)$) ($5*(y)-(x)$)--++($-4*(x)$);
\draw[smoothbd] ($0*(x)$)--++($5*(y)$) ($-1*(x)$)--++($5*(y)$);
\draw[orange] ($1*(x)$)--++($5*(y)$) ($-2*(x)$)--++($5*(y)$);
\foreach \x in {-3,-2,-1,0,1,2}{
\foreach \y in {0,1,2,3,4}{
\atoms{vertex,astyle={\qubitb}}{g\x\y/p={$\x*(x)+{\y+0.5}*(y)$}}
\atoms{vertex,astyle={\qubita}}{r\x\y/p={${\x+0.5}*(x)+{\y+0.5}*(y)$}}
}}
\foreach \x in {-2,-1,0,1}{
\draw[line width=0.3cm,cyan,opacity=0.5,line cap=round] (g\x0-c)--(g\x1-c) (g\x2-c)--(g\x3-c);
\path (g\x0-c)--node[midway]{$\scriptstyle{M_{XX}}$} (g\x1-c);
\path (g\x2-c)--node[midway]{$\scriptstyle{M_{XX}}$} (g\x3-c);
}
\foreach \x in {-2,0}{
\draw[line width=0.3cm,cyan,opacity=0.5,line cap=round] (r\x1-c)--(r\x2-c) (r\x3-c)--(r\x4-c);
\path (r\x1-c)--node[midway]{$\scriptstyle{M_{ZZ}}$} (r\x2-c);
\fill[cyan,opacity=0.5] (r\x0)circle(0.2);
\path (r\x0)++(-90:0.2)node{$\scriptstyle{M_Z}$};
}
\end{tikzpicture}
&$\rightarrow$&
\begin{tikzpicture}
\atoms{void}{x/p={1,0}, y/p={0,0.707107}}
\clip ($-2.5*(x)+(0.15,-0.15)$) rectangle($1.5*(x)+4*(y)+(-0.15,-0.15)$);
\draw[roughbd] ($0*(y)$)--++($4*(x)$) ($-1*(x)$)--++($-4*(x)$);
\draw[orange] ($1*(y)$)--++($4*(x)$) ($2*(y)$)--++($4*(x)$) ($3*(y)$)--++($4*(x)$) ($4*(y)$)--++($4*(x)$) ($5*(y)$)--++($4*(x)$);
\draw[orange] ($1*(y)-(x)$)--++($-4*(x)$) ($2*(y)-(x)$)--++($-4*(x)$) ($3*(y)-(x)$)--++($-4*(x)$) ($4*(y)-(x)$)--++($-4*(x)$) ($5*(y)-(x)$)--++($-4*(x)$);
\draw[smoothbd] ($0*(x)$)--++($5*(y)$) ($-1*(x)$)--++($5*(y)$);
\draw[orange] ($1*(x)$)--++($5*(y)$) ($-2*(x)$)--++($5*(y)$);
\foreach \x in {-3,-2,-1,0,1,2}{
\foreach \y in {0,1,2,3,4}{
\atoms{vertex,astyle={\qubitb}}{g\x\y/p={$\x*(x)+{\y+0.5}*(y)$}}
\atoms{vertex,astyle={\qubita}}{r\x\y/p={${\x+0.5}*(x)+{\y+0.5}*(y)$}}
}}
\foreach \y in {0,1,2,3}{
\draw[line width=0.3cm,cyan,opacity=0.5,line cap=round] (g-2\y-c)--(r-2\y-c) (g0\y-c)--(r0\y-c) (g1\y-c)--(r1\y-c);
\node at ($0.5*(g-2\y-c)+0.5*(r-2\y-c)+(90:0.15)$) {$\scriptstyle{CX}$};
\node at ($0.5*(g0\y-c)+0.5*(r0\y-c)+(90:0.15)$) {$\scriptstyle{CX}$};
\node at ($0.5*(g1\y-c)+0.5*(r1\y-c)+(90:0.15)$) {$\scriptstyle{CX}$};
}
\end{tikzpicture}
\\
&&$\downarrow$\\
\begin{tikzpicture}
\atoms{void}{x/p={1,0}, y/p={0,0.707107}}
\clip ($-2.5*(x)+(0.15,-0.15)$) rectangle($1.5*(x)+4*(y)+(-0.15,-0.15)$);
\draw[roughbd] ($-4*(x)$)--++($8*(x)$);
\draw[orange] ($1*(y)-4*(x)$)--++($8*(x)$) ($2*(y)-4*(x)$)--++($8*(x)$) ($3*(y)-4*(x)$)--++($8*(x)$);
\draw[orange] ($1*(x)$)--++($5*(y)$) ($-2*(x)$)--++($5*(y)$) ($-1*(x)$)--++($5*(y)$) ($0*(x)$)--++($5*(y)$);
% \draw[roughbd] ($0*(y)$)--++($4*(x)$) ($-1*(x)$)--++($-4*(x)$);
% \draw[orange] ($1*(y)$)--++($4*(x)$) ($2*(y)$)--++($4*(x)$) ($3*(y)$)--++($4*(x)$) ($4*(y)$)--++($4*(x)$) ($5*(y)$)--++($4*(x)$);
% \draw[orange] ($1*(y)-(x)$)--++($-4*(x)$) ($2*(y)-(x)$)--++($-4*(x)$) ($3*(y)-(x)$)--++($-4*(x)$) ($4*(y)-(x)$)--++($-4*(x)$) ($5*(y)-(x)$)--++($-4*(x)$);
% \draw[smoothbd] ($0*(x)$)--++($5*(y)$) ($-1*(x)$)--++($5*(y)$);
% \draw[orange] ($1*(x)$)--++($5*(y)$) ($-2*(x)$)--++($5*(y)$);
\foreach \x in {-3,-2,-1,0,1,2}{
\foreach \y in {0,1,2,3,4}{
\atoms{vertex,astyle={\qubitb}}{g\x\y/p={$\x*(x)+{\y+0.5}*(y)$}}
\atoms{vertex,astyle={\qubita}}{r\x\y/p={${\x+0.5}*(x)+{\y+0.5}*(y)$}}
}}
\foreach \x in {-2,-1,0,1}{
\draw[line width=0.3cm,cyan,opacity=0.5,line cap=round] (g\x1-c)--(g\x2-c) (g\x3-c)--(g\x4-c);
\path (g\x1-c)--node[midway]{$\scriptstyle{M_{XX}}$} (g\x2-c);
}
\foreach \x in {-2,-1,0}{
\draw[line width=0.3cm,cyan,opacity=0.5,line cap=round] (r\x0-c)--(r\x1-c) (r\x2-c)--(r\x3-c);
\path (r\x0-c)--node[midway]{$\scriptstyle{M_{ZZ}}$} (r\x1-c);
\path (r\x2-c)--node[midway]{$\scriptstyle{M_{ZZ}}$} (r\x3-c);
}
\end{tikzpicture}
&$\leftarrow$&
\begin{tikzpicture}
\atoms{void}{x/p={1,0}, y/p={0,0.707107}}
\clip ($-2.5*(x)+(0.15,-0.15)$) rectangle($1.5*(x)+4*(y)+(-0.15,-0.15)$);
\draw[roughbd] ($0*(y)$)--++($4*(x)$) ($-1*(x)$)--++($-4*(x)$);
\draw[orange] ($1*(y)$)--++($4*(x)$) ($2*(y)$)--++($4*(x)$) ($3*(y)$)--++($4*(x)$) ($4*(y)$)--++($4*(x)$) ($5*(y)$)--++($4*(x)$);
\draw[orange] ($1*(y)-(x)$)--++($-4*(x)$) ($2*(y)-(x)$)--++($-4*(x)$) ($3*(y)-(x)$)--++($-4*(x)$) ($4*(y)-(x)$)--++($-4*(x)$) ($5*(y)-(x)$)--++($-4*(x)$);
\draw[smoothbd] ($0*(x)$)--++($5*(y)$) ($-1*(x)$)--++($5*(y)$);
\draw[orange] ($1*(x)$)--++($5*(y)$) ($-2*(x)$)--++($5*(y)$);
\foreach \x in {-3,-2,-1,0,1,2}{
\foreach \y in {0,1,2,3,4}{
\atoms{vertex,astyle={\qubitb}}{g\x\y/p={$\x*(x)+{\y+0.5}*(y)$}}
\atoms{vertex,astyle={\qubita}}{r\x\y/p={${\x+0.5}*(x)+{\y+0.5}*(y)$}}
}}
\foreach \y in {0,1,2,3}{
\draw[line width=0.3cm,cyan,opacity=0.5,line cap=round] (r-3\y-c)--(g-2\y-c) (r-2\y-c)--(g-1\y-c) (r0\y-c)--(g1\y-c);
\node at ($0.5*(r-3\y-c)+0.5*(g-2\y-c)+(90:0.15)$) {$\scriptstyle{CX}$};
\node at ($0.5*(r-2\y-c)+0.5*(g-1\y-c)+(90:0.15)$) {$\scriptstyle{CX}$};
\node at ($0.5*(r0\y-c)+0.5*(g1\y-c)+(90:0.15)$) {$\scriptstyle{CX}$};
\fill[cyan,opacity=0.5] (r-1\y)circle(0.2);
\path (r-1\y)++(-90:0.2)node{$\scriptstyle{\ket+}$};
}
\end{tikzpicture}
\end{tabular}
\;.
\end{equation}
Next, we construct the sequence of operations that transition from a single merged block back to two separate blocks at $T_1$.
Reflecting Eq.~\eqref{eq:surgery_path_integral} along the $t$ direction, we find that the purple $z=L+\frac12$ qubit worldlines disappear at a 2-index $\delta$-tensor.
The according operator is a projection onto the Bell state:
\begin{equation}
\begin{tikzpicture}
\atoms{delta}{0/}
\draw (0)edge[ind=$a$]++(-135:0.5) (0)edge[ind=$b$]++(-45:0.5);
\fill[cyan,opacity=0.5] (0)circle(0.3);
\path[cyan] (0)++(0:0.5)node{(g)};
\end{tikzpicture}
=
(\bra{00}+\bra{11})\ket{ab}
\;.
\end{equation}
We could view it as the $(+1,+1)$-postselected projector of a Bell measurement.
However, we will go a slightly different route and split it up into one 4-index $\delta$-tensor and two 1-index $\delta$-tensors,
\begin{equation}
\label{eq:t1_transition_measurement}
\begin{tikzpicture}
\atoms{delta}{0/}
\draw (0)edge[ind=$a$]++(-135:0.5) (0)edge[ind=$b$]++(-45:0.5);
\fill[cyan,opacity=0.5] (0)circle(0.3);
\path[cyan] (0)++(0:0.5)node{(g)};
\end{tikzpicture}
=
\begin{tikzpicture}
\atoms{delta}{0/, 1/p={135:0.5}, 2/p={45:0.5}}
\draw (0)edge[ind=$a$]++(-135:0.5) (0)edge[ind=$b$]++(-45:0.5);
\draw (0)--(1) (0)--(2);
\fill[cyan,opacity=0.5] (0)circle(0.3);
\path[cyan] (0)++(0:0.5)node{(b)};
\end{tikzpicture}
\;.
\end{equation}
We view the 4-index $\delta$-tensor as a $\frac12(1+ZZ)$ projector like the (b) bulk operators shown in Eq.~\eqref{eq:delta_tensor_operator}.
The 1-index $\delta$-tensor can be viewed as an operator $\cc^2\rightarrow\cc^1$ given by the inner product of the qubit with $\ket+$
\footnote{
Again, the $z=L+\frac12$ qubits do not actually disappear at time $T_1$, but we can pretend that they do since they are unused after $T_1$.
}:
\begin{equation}
\begin{tikzpicture}
\atoms{delta}{0/}
\draw (0)edge[ind=$a$]++(-90:0.5);
\end{tikzpicture}
=
\braket{+|a}
\;.
\end{equation}
Like in the bulk, the $\frac12(1+ZZ)$ projectors are turned into $ZZ$ measurements.
To study the effect of a $-1$ outcome of the $ZZ$-measurement on the path integral, we plug in Eq.~\eqref{eq:zz_minusone_projector}, and fuse the charged $\zz_2$-tensors with the surrounding uncharged ones.
In contrast to the (b) tensors in the bulk in Eq.~\eqref{eq:minusone_outcome_mapping}, a $-1$ outcome only corresponds to an $m$ anyon being present at a single face of the smooth boundary:
\begin{equation}
\begin{tikzpicture}
\atoms{void}{x/p={1.5,0}, y/p={0,1.5}, z/p={1.2,0.6}}
\foreach \x in {0,1}{
\foreach \y in {0,1}{
\foreach \z in {0,1}{
\atoms{void}{\x\y\z/p={$\x*(x)+\y*(y)+\z*(z)$}}
}}}
\draw[orange] (000)--(100) (010)--(000);
\draw[smoothbd] (110)--(010) (011)--(111) (101)--(111)--(110)--(100) (100)--(101) (010)--(011);
\draw[orange,back] (001)--(101) (000)--(001) (001)--(011);
\atoms{delta}{d/p={$0.5*(z)+(x)+(y)$}}
\atoms{z2}{z0/p={$0.5*(x)+0.5*(z)+(y)$}, z1/p={$0.5*(y)+0.5*(z)+(x)$}}
\atoms{z2,bdastyle=red}{zm0/p={$(x)+0.5*(z)+0.75*(y)$}, zm1/p={$(y)+0.5*(z)+1.25*(x)$}}
\atoms{delta}{d0/p={$1.5*(y)+0.5*(z)+(x)$}, d1/p={$1.5*(x)+0.5*(z)+(y)$}}
\draw (d)--(d0) (d)--(zm1)--(d1) (d)--(zm0)--(z1) (d)--(z0);
\draw (z0)--++($-0.3*(z)$) (z0)--++($0.3*(z)$) (z0)--++($-0.3*(x)$) (z1)--++($-0.3*(y)$) (z1)--++($-0.3*(z)$) (z1)--++($0.3*(z)$);
\end{tikzpicture}
=
\begin{tikzpicture}
\atoms{void}{x/p={1.5,0}, y/p={0,1.5}, z/p={1.2,0.6}}
\foreach \x in {0,1}{
\foreach \y in {0,1}{
\foreach \z in {0,1}{
\atoms{void}{\x\y\z/p={$\x*(x)+\y*(y)+\z*(z)$}}
}}}
\draw[orange] (000)--(100) (010)--(000);
\draw[smoothbd] (110)--(010) (011)--(111) (101)--(111)--(110)--(100) (100)--(101) (010)--(011);
\draw[orange,back] (001)--(101) (000)--(001) (001)--(011);
\draw[worldline] ($(x)+0.5*(z)+0.5*(y)$)--++($-0.5*(x)$);
\atoms{delta}{d/p={$0.5*(z)+(x)+(y)$}}
\atoms{z2}{z0/p={$0.5*(x)+0.5*(z)+(y)$}, {z1/p={$0.5*(y)+0.5*(z)+(x)$},bdastyle=red}}
\draw (d)--(z1) (d)--(z0);
\draw (z0)--++($-0.3*(z)$) (z0)--++($0.3*(z)$) (z0)--++($-0.3*(x)$) (z1)--++($-0.3*(y)$) (z1)--++($-0.3*(z)$) (z1)--++($0.3*(z)$);
\end{tikzpicture}
\;.
\end{equation}
The $\bra+$ inner products are turned into destructive single-qubit $X$ measurements.
Whereas Eq.~\eqref{eq:t1_transition_measurement} represents the measurement post-selected onto the $+1$ outcome, the $-1$ outcome can be represented diagrammatically as:
\begin{equation}
\label{eq:single_qubit_destructive_x_minus}
\braket{-|a}
=
\begin{tikzpicture}
\atoms{delta,bdastyle=red}{0/}
\draw (0)edge[ind=$a$]++(-90:0.5);
\end{tikzpicture}
\;.
\end{equation}
To study the effect of a $-1$ outcome on the path integral, we plug in Eq.~\eqref{eq:single_qubit_destructive_x_minus} instead of Eq.~\eqref{eq:t1_transition_measurement}, and fuse the charged $\delta$-tensor with the adjacent uncharged one:
\begin{equation}
\begin{tikzpicture}
\atoms{void}{x/p={1.5,0}, y/p={0,1.5}, z/p={1.2,0.6}}
\foreach \x in {0,1}{
\foreach \y in {0,1}{
\foreach \z in {0,1}{
\atoms{void}{\x\y\z/p={$\x*(x)+\y*(y)+\z*(z)$}}
}}}
\draw[orange] (000)--(100) (010)--(000);
\draw[smoothbd] (110)--(010) (011)--(111) (101)--(111)--(110)--(100) (100)--(101) (010)--(011);
\draw[orange,back] (001)--(101) (000)--(001) (001)--(011);
\atoms{delta}{d/p={$0.5*(z)+(x)+(y)$}}
\atoms{z2}{z0/p={$0.5*(x)+0.5*(z)+(y)$}, z1/p={$0.5*(y)+0.5*(z)+(x)$}}
\atoms{delta}{d0/p={$1.5*(y)+0.5*(z)+(x)$}, {d1/p={$1.5*(x)+0.5*(z)+(y)$},bdastyle=red}}
\draw (d)--(d0) (d)--(d1) (d)--(z1) (d)--(z0);
\draw (z0)--++($-0.3*(z)$) (z0)--++($0.3*(z)$) (z0)--++($-0.3*(x)$) (z1)--++($-0.3*(y)$) (z1)--++($-0.3*(z)$) (z1)--++($0.3*(z)$);
\end{tikzpicture}
=
\begin{tikzpicture}
\atoms{void}{x/p={1.5,0}, y/p={0,1.5}, z/p={1.2,0.6}}
\foreach \x in {0,1}{
\foreach \y in {0,1}{
\foreach \z in {0,1}{
\atoms{void}{\x\y\z/p={$\x*(x)+\y*(y)+\z*(z)$}}
}}}
\draw[orange] (000)--(100) (010)--(000);
\draw[smoothbd] (110)--(010) (011)--(111) (101)--(111)--(110)--(100) (100)--(101) (010)--(011);
\draw[orange,back] (001)--(101) (000)--(001) (001)--(011);
\draw[worldline] ($(x)+(y)$)--++(z);
\atoms{delta,bdastyle=red}{d/p={$0.5*(z)+(x)+(y)$}}
\atoms{z2}{z0/p={$0.5*(x)+0.5*(z)+(y)$}, z1/p={$0.5*(y)+0.5*(z)+(x)$}}
\draw (d)--(z1) (d)--(z0);
\draw (z0)--++($-0.3*(z)$) (z0)--++($0.3*(z)$) (z0)--++($-0.3*(x)$) (z1)--++($-0.3*(y)$) (z1)--++($-0.3*(z)$) (z1)--++($0.3*(z)$);
\end{tikzpicture}
\;.
\end{equation}
We find that a $-1$ measurement on either of the two qubits corresponds to an $e$ anyon worldline present at the according edge of the smooth boundary.

Next, let us determine on which qubits and in which order the operators above act.
The (b) operator (or $ZZ$ measurement) in Eq.~\eqref{eq:t1_transition_measurement} acts together with all bulk (b) and (a) operators, which constitutes the last layer of the circuit for the merged block.
More precisely, there are two such layers in each round, and both could be chosen as the last layer of the merged block.
For the path integral shown in Eq.~\eqref{eq:surgery_path_integral}, mirrored along $t$, it is the layer that includes identity gates on the green qubits near the rough boundary.
The single-qubit $X$ measurements happen in following layer of $CX$ gates.
The following picture shows the operators applied before, at, and after the splitting transition at $T_1$:
\begin{equation}
\begin{tabular}{ccc}
\begin{tikzpicture}
\atoms{void}{x/p={1,0}, y/p={0,0.707107}}
\clip ($-2.5*(x)+(0.15,-0.15)$) rectangle($1.5*(x)+4*(y)+(-0.15,-0.15)$);
\draw[roughbd] ($-4*(x)$)--++($8*(x)$);
\draw[orange] ($1*(y)-4*(x)$)--++($8*(x)$) ($2*(y)-4*(x)$)--++($8*(x)$) ($3*(y)-4*(x)$)--++($8*(x)$);
\draw[orange] ($1*(x)$)--++($5*(y)$) ($-2*(x)$)--++($5*(y)$) ($-1*(x)$)--++($5*(y)$) ($0*(x)$)--++($5*(y)$);
\foreach \x in {-3,-2,-1,0,1,2}{
\foreach \y in {0,1,2,3,4}{
\atoms{vertex,astyle={\qubitb}}{g\x\y/p={$\x*(x)+{\y+0.5}*(y)$}}
\atoms{vertex,astyle={\qubita}}{r\x\y/p={${\x+0.5}*(x)+{\y+0.5}*(y)$}}
}}
\foreach \x in {-2,-1,0,1}{
\draw[line width=0.3cm,cyan,opacity=0.5,line cap=round] (g\x1-c)--(g\x2-c) (g\x3-c)--(g\x4-c);
\path (g\x1-c)--node[midway]{$\scriptstyle{M_{XX}}$} (g\x2-c);
}
\foreach \x in {-2,-1,0}{
\draw[line width=0.3cm,cyan,opacity=0.5,line cap=round] (r\x0-c)--(r\x1-c) (r\x2-c)--(r\x3-c);
\path (r\x0-c)--node[midway]{$\scriptstyle{M_{ZZ}}$} (r\x1-c);
\path (r\x2-c)--node[midway]{$\scriptstyle{M_{ZZ}}$} (r\x3-c);
}
\end{tikzpicture}
&$\rightarrow$&
\begin{tikzpicture}
\atoms{void}{x/p={1,0}, y/p={0,0.707107}}
\clip ($-2.5*(x)+(0.15,-0.15)$) rectangle($1.5*(x)+4*(y)+(-0.15,-0.15)$);
\draw[roughbd] ($-4*(x)$)--++($8*(x)$);
\draw[orange] ($1*(y)-4*(x)$)--++($8*(x)$) ($2*(y)-4*(x)$)--++($8*(x)$) ($3*(y)-4*(x)$)--++($8*(x)$);
\draw[orange] ($1*(x)$)--++($5*(y)$) ($-2*(x)$)--++($5*(y)$) ($-1*(x)$)--++($5*(y)$) ($0*(x)$)--++($5*(y)$);
\foreach \x in {-3,-2,-1,0,1,2}{
\foreach \y in {0,1,2,3,4}{
\atoms{vertex,astyle={\qubitb}}{g\x\y/p={$\x*(x)+{\y+0.5}*(y)$}}
\atoms{vertex,astyle={\qubita}}{r\x\y/p={${\x+0.5}*(x)+{\y+0.5}*(y)$}}
}}
\foreach \y in {0,1,2,3}{
\draw[line width=0.3cm,cyan,opacity=0.5,line cap=round] (g-2\y-c)--(r-2\y-c) (g0\y-c)--(r0\y-c) (g1\y-c)--(r1\y-c);
\node at ($0.5*(g-2\y-c)+0.5*(r-2\y-c)+(90:0.15)$) {$\scriptstyle{CX}$};
\node at ($0.5*(g0\y-c)+0.5*(r0\y-c)+(90:0.15)$) {$\scriptstyle{CX}$};
\node at ($0.5*(g1\y-c)+0.5*(r1\y-c)+(90:0.15)$) {$\scriptstyle{CX}$};
\fill[cyan,opacity=0.5] (r-1\y)circle(0.2);
\path (r-1\y)++(-90:0.2)node{$\scriptstyle{M_X}$};
}
\end{tikzpicture}
\\
&&$\downarrow$\\
\begin{tikzpicture}
\atoms{void}{x/p={1,0}, y/p={0,0.707107}}
\clip ($-2.5*(x)+(0.15,-0.15)$) rectangle($1.5*(x)+4*(y)+(-0.15,-0.15)$);
\draw[roughbd] ($0*(y)$)--++($4*(x)$) ($-1*(x)$)--++($-4*(x)$);
\draw[orange] ($1*(y)$)--++($4*(x)$) ($2*(y)$)--++($4*(x)$) ($3*(y)$)--++($4*(x)$) ($4*(y)$)--++($4*(x)$) ($5*(y)$)--++($4*(x)$);
\draw[orange] ($1*(y)-(x)$)--++($-4*(x)$) ($2*(y)-(x)$)--++($-4*(x)$) ($3*(y)-(x)$)--++($-4*(x)$) ($4*(y)-(x)$)--++($-4*(x)$) ($5*(y)-(x)$)--++($-4*(x)$);
\draw[smoothbd] ($0*(x)$)--++($5*(y)$) ($-1*(x)$)--++($5*(y)$);
\draw[orange] ($1*(x)$)--++($5*(y)$) ($-2*(x)$)--++($5*(y)$);
\foreach \x in {-3,-2,-1,0,1,2}{
\foreach \y in {0,1,2,3,4}{
\atoms{vertex,astyle={\qubitb}}{g\x\y/p={$\x*(x)+{\y+0.5}*(y)$}}
\atoms{vertex,astyle={\qubita}}{r\x\y/p={${\x+0.5}*(x)+{\y+0.5}*(y)$}}
}}
\foreach \x in {-2,-1,0,1}{
\draw[line width=0.3cm,cyan,opacity=0.5,line cap=round] (g\x0-c)--(g\x1-c) (g\x2-c)--(g\x3-c);
\path (g\x0-c)--node[midway]{$\scriptstyle{M_{XX}}$} (g\x1-c);
\path (g\x2-c)--node[midway]{$\scriptstyle{M_{XX}}$} (g\x3-c);
}
\foreach \x in {-2,0}{
\draw[line width=0.3cm,cyan,opacity=0.5,line cap=round] (r\x1-c)--(r\x2-c) (r\x3-c)--(r\x4-c);
\path (r\x1-c)--node[midway]{$\scriptstyle{M_{ZZ}}$} (r\x2-c);
\fill[cyan,opacity=0.5] (r\x0)circle(0.2);
\path (r\x0)++(-90:0.2)node{$\scriptstyle{M_Z}$};
}
\end{tikzpicture}
&$\leftarrow$&
\begin{tikzpicture}
\atoms{void}{x/p={1,0}, y/p={0,0.707107}}
\clip ($-2.5*(x)+(0.15,-0.15)$) rectangle($1.5*(x)+4*(y)+(-0.15,-0.15)$);
\draw[roughbd] ($0*(y)$)--++($4*(x)$) ($-1*(x)$)--++($-4*(x)$);
\draw[orange] ($1*(y)$)--++($4*(x)$) ($2*(y)$)--++($4*(x)$) ($3*(y)$)--++($4*(x)$) ($4*(y)$)--++($4*(x)$) ($5*(y)$)--++($4*(x)$);
\draw[orange] ($1*(y)-(x)$)--++($-4*(x)$) ($2*(y)-(x)$)--++($-4*(x)$) ($3*(y)-(x)$)--++($-4*(x)$) ($4*(y)-(x)$)--++($-4*(x)$) ($5*(y)-(x)$)--++($-4*(x)$);
\draw[smoothbd] ($0*(x)$)--++($5*(y)$) ($-1*(x)$)--++($5*(y)$);
\draw[orange] ($1*(x)$)--++($5*(y)$) ($-2*(x)$)--++($5*(y)$);
\foreach \x in {-3,-2,-1,0,1,2}{
\foreach \y in {0,1,2,3,4}{
\atoms{vertex,astyle={\qubitb}}{g\x\y/p={$\x*(x)+{\y+0.5}*(y)$}}
\atoms{vertex,astyle={\qubita}}{r\x\y/p={${\x+0.5}*(x)+{\y+0.5}*(y)$}}
}}
\foreach \y in {0,1,2,3}{
\draw[line width=0.3cm,cyan,opacity=0.5,line cap=round] (r-3\y-c)--(g-2\y-c) (r-2\y-c)--(g-1\y-c) (r0\y-c)--(g1\y-c);
\node at ($0.5*(r-3\y-c)+0.5*(g-2\y-c)+(90:0.15)$) {$\scriptstyle{CX}$};
\node at ($0.5*(r-2\y-c)+0.5*(g-1\y-c)+(90:0.15)$) {$\scriptstyle{CX}$};
\node at ($0.5*(r0\y-c)+0.5*(g1\y-c)+(90:0.15)$) {$\scriptstyle{CX}$};
}
\end{tikzpicture}
\end{tabular}
\;.
\end{equation}

\section{Decoding and logic operation}
\label{sec:decoding}
In this section, we discuss how preform decoding and corrections of the proposed fault-tolerant circuit, and derive the logical operation implemented by the lattice surgery in Section~\ref{sec:lattice_surgery}.

\subsection{Decoding in the bulk}
Let us start by describing decoding of the circuit from Section~\ref{sec:circuit} in the bulk.
The most commonly considered types of noise are single-qubit bit-flip and phase-filp, as well as measurement errors
\footnote{
The path-integral approach allows us to directly prove the existance of a fault-tolerant threshold for arbitrary local noise \cite{twisted_double_code}.
Here, we restrict to Pauli and measurement noise for simplicity.
}
.
Bit and phase-flip errors insert a Pauli-$X$ or Pauli-$Z$ operator into the circuit.
These operators are 2-index charged $\delta$ or $\zz_2$-tensors,
\begin{equation}
\begin{tikzpicture}
\atoms{delta,bdastyle=red}{0/}
\draw (0)edge[ind=$a$]++(-90:0.5) (0)edge[ind=$b$]++(90:0.5);
\end{tikzpicture}
=
\delta_{a=b} (-1)^a
=
\bra{b} Z \ket{a}
\;,\quad
\begin{tikzpicture}
\atoms{z2,bdastyle=red}{0/}
\draw (0)edge[ind=$a$]++(-90:0.5) (0)edge[ind=$b$]++(90:0.5);
\end{tikzpicture}
=
\delta_{a=b+1}
=
\bra{b} X \ket{a}
\;.
\end{equation}
These additional tensors can be absorbed into the adjacent $\zz_2$ or $\delta$-tensors of the path integral, for example
\begin{equation}
\begin{tikzpicture}
\atoms{delta}{{0/lab={t=$\ldots$,p=-90:0.25}}, {1/p={0,0.6},bdastyle=red}}
\draw (0)--(1) (0)--++(0:0.5) (0)--++(180:0.5) (1)--++(90:0.5);
\end{tikzpicture}
=
\begin{tikzpicture}
\atoms{delta}{{0/lab={t=$\ldots$,p=-90:0.25},bdastyle=red}}
\draw (0)--++(0:0.5) (0)--++(180:0.5) (0)--++(90:0.5);
\end{tikzpicture}
\;.
\end{equation}
So each Pauli-$X$ or $Z$ error corresponds to anyon worldlines at one or multiple edges or faces in the spacetime lattice.
Each measurement error corresponds to the same anyon worldlines as the $-1$ outcome of the measurement.
So each spacetime configuration of Pauli-$X$, $Z$, and measurement errors corresponds to a configuration of $e$ and $m$ anyon worldline segments, which we will call the \emph{error worldline configuration}.
As we have seen in Section~\ref{sec:circuit}, also every configuration of measurement outcomes gives rise to a configuration of anyon worldline segments, which we will call the \emph{measurement worldline configuration}.
The path integral is zero if the overall configuration of anyon worldlines is not a closed-loop pattern, as discussed around Eq.~\eqref{eq:cocycle_constraint}.
So the error and measurement worldline configurations together will form a closed-loop pattern with probability $1$.
The measurement results thus tell us the configuration of endpoints of the error worldline configuration, which is the same as for the measurement worldline configuration.
So we can estimate the errors worldline configuration by connecting (matching) these endpoints in pairs.
Luckily, it suffices to estimate the homology class
\footnote{This is a ``homology class'' with respect to a fixed endpoint configuration.}
of the error worldline configuration, since the path integral is invariant under local deformations of the anyon worldlines as discussed around Eq.~\eqref{eq:anyon_worldline_invariance}.
Assuming weak enough noise, a match of low weight will succeed with high probability, since the size of the smallest homologically non-trivial loop is of order $L$.
When matching, we allow to match endpoints to the temporal state boundary corresponding to the final state of the qubits.
We do not allow matching to the initial state boundary since we assume that the initial state is a ground state.
The endpoints of $e$ and $m$ anyon worldlines at the final state boundary correspond to $e$ and $m$ anyons in the final state.
The corrections are given by removing these $e$ or $m$ anyons by $Z$ and $X$ string operators.
In the spacetime picture, we insert additional anyon worldlines near the final temporal state boundary to close off the measured anyon worldlines.
We need to choose these worldlines in a way such that the overall worldline configuration is homologically trivial.
Then due to the homological invariance of the anyon worldlines shown in Eq.~\eqref{eq:anyon_worldline_invariance}, the logical operation performed the same as for the $+1$ post-selected circuit, independently of the configuration of errors and measurement outcomes.

All in all, the decoding procedure is not very different from the one proposed in Ref.~\cite{Dennis2001} for the stabilizer toric code.
The only difference is in the faces and edges on which potential measurement worldlines can be supported.
For the stabilizer toric code, all measurement worldline segments point in the $t$ direction, so without errors the worldlines are just straight lines in $t$ direction.
In particular, if we start in the ground state with no errors, all measurement outcomes will be $+1$.
For measurement-based topological quantum computation, the measurement worldlines can be supported on any edge of the underlying cubic lattice.
The measurement worldline configuration is a completely arbitrary closed-loop pattern, even if there are no errors and we start from the ground state.
The subsystem toric code (with a schedule where we alternatingly measure $X$ and $Z$ gauge checks) behaves like the toric code, just that every $t$-directed edge is split into two alternative edges and small loops of edge pairs are possible.
The (CSS) honeycomb Floquet code behaves most similarly to measurement-based topolgical quantum computation:
The worldline segments can go in any direction, but they are diagonal to the time direction.
Finally, for the proposed x+y Floquet code, the worldline segments are constrained to $(\overline x,t)$-planes, but can go in any direction (diagonal to time) within each plane.
The following picture illustrates the different behaviors:
\begin{equation}
\begin{tabular}{cc}
Stabilizer & Measurement-based\\
\begin{tikzpicture}[scale=0.4]
\atoms{void}{x/p={2,0}, y/p={0,2}, z/p={1.6,0.8}}
\draw (0,0)--++($3*(x)$)--++($3*(y)$)--++($-3*(x)$)--cycle ($3*(x)$)--++($2*(z)$)--++($3*(y)$) ($3*(x)+3*(y)$)--++($2*(z)$) ($3*(y)$)--++($2*(z)$)--++($3*(x)$);
\draw[back] (0,0)--++($2*(z)$)--++($3*(x)$) ($2*(z)$)--++($3*(y)$);
\foreach \x in {0,1,2,3}{
\foreach \z in {0,1,2}{
\draw[opacity=0.2,line width=0.2cm] ($\x*(x)+\z*(z)$)--++($3*(y)$);
}}
\foreach \x/\z in {1/0, 1/1, 3/2}{
\draw[red,line width=0.1cm] ($\x*(x)+\z*(z)$)--++($3*(y)$);
}
\end{tikzpicture}
&
\begin{tikzpicture}[scale=0.4]
\atoms{void}{x/p={2,0}, y/p={0,2}, z/p={1.6,0.8}}
\draw (0,0)--++($3*(x)$)--++($3*(y)$)--++($-3*(x)$)--cycle ($3*(x)$)--++($2*(z)$)--++($3*(y)$) ($3*(x)+3*(y)$)--++($2*(z)$) ($3*(y)$)--++($2*(z)$)--++($3*(x)$);
\draw[back] (0,0)--++($2*(z)$)--++($3*(x)$) ($2*(z)$)--++($3*(y)$);
\foreach \x/\y in {0/1,0/2,1/0,1/1,1/2,1/3,2/0,2/1,2/2,2/3,3/1,3/2}{
\draw[gray] ($\x*(x)+\y*(y)$)--++($2*(z)$);
}
\foreach \x/\z in {0/1,1/0,1/1,1/2,2/0,2/1,2/2,3/1}{
\draw[gray] ($\x*(x)+\z*(z)$)--++($3*(y)$);
}
\foreach \y/\z in {0/1,1/0,1/1,1/2,2/0,2/1,2/2,3/1}{
\draw[gray] ($\y*(y)+\z*(z)$)--++($3*(x)$);
}
\draw[red,line width=0.1cm] ($1*(x)+1*(z)$)--++(y)--++(x)--++(z)--++(y)--++($-1*(x)$)--++(y) ($2*(y)$)--++(x)--++(z)--++($-1*(x)$)--cycle;
\end{tikzpicture}\\
\rule{0pt}{5ex}
Subsystem & Honeycomb Floquet\\
\begin{tikzpicture}[scale=0.4]
\atoms{void}{x/p={2,0}, y/p={0,2}, z/p={1.6,0.8}}
\draw (0,0)--++($3*(x)$)--++($3*(y)$)--++($-3*(x)$)--cycle ($3*(x)$)--++($2*(z)$)--++($3*(y)$) ($3*(x)+3*(y)$)--++($2*(z)$) ($3*(y)$)--++($2*(z)$)--++($3*(x)$);
\draw[back] (0,0)--++($2*(z)$)--++($3*(x)$) ($2*(z)$)--++($3*(y)$);
\foreach \x in {0,1,2,3}{
\foreach \z in {0,1,2}{
\draw[opacity=0.2,line width=0.2cm] ($\x*(x)+\z*(z)$)--++($3*(y)$);
}}
\draw[red,line width=0.1cm] ($1*(x)+0*(z)$)to[bend left]++(y)to[bend right]++(y)to[bend right]++(y);
\draw[red,line width=0.1cm] ($1*(x)+2*(z)$)to[bend left]++(y)to[bend left]++(y)to[bend left]++(y);
\draw[red,line width=0.1cm] ($3*(x)+1*(z)+(y)$)to[bend left]++(y) ($3*(x)+1*(z)+(y)$)to[bend right]++(y);
\end{tikzpicture}
&
\begin{tikzpicture}[scale=0.4]
\atoms{void}{x/p={2,0}, y/p={0,2}, z/p={1.6,0.8}}
\draw (0,0)--++($3*(x)$)--++($3*(y)$)--++($-3*(x)$)--cycle ($3*(x)$)--++($2*(z)$)--++($3*(y)$) ($3*(x)+3*(y)$)--++($2*(z)$) ($3*(y)$)--++($2*(z)$)--++($3*(x)$);
\draw[back] (0,0)--++($2*(z)$)--++($3*(x)$) ($2*(z)$)--++($3*(y)$);
\foreach \x/\y in {0/1,0/2,1/0,1/1,1/2,1/3,2/0,2/1,2/2,2/3,3/1,3/2}{
\draw[gray] ($\x*(x)+\y*(y)$)--++($2*(z)$);
}
\foreach \x/\z in {0/1,1/0,1/1,1/2,2/0,2/1,2/2,3/1}{
\draw[gray] ($\x*(x)+\z*(z)$)--++($3*(y)$);
}
\foreach \y/\z in {0/1,1/0,1/1,1/2,2/0,2/1,2/2,3/1}{
\draw[gray] ($\y*(y)+\z*(z)$)--++($3*(x)$);
}
\draw[red,line width=0.1cm] ($1*(x)+1*(z)$)--++($0.5*(y)-0.5*(z)$)--++($0.5*(y)+0.5*(z)$)--++($1*(y)+1*(x)$)--++($1*(y)+1*(z)$) ($(x)+3*(y)$)--++($-0.5*(x)-0.5*(y)$)--++($0.5*(x)-0.5*(y)$)--++($(x)+(y)$) ($3*(x)+1*(y)+1*(z)$)--++($-0.5*(z)+0.5*(y)$)--++($0.5*(z)+0.5*(y)$)--++($0.5*(z)-0.5*(y)$)--cycle;
\end{tikzpicture}
\\
\rule{0pt}{5ex}
x+y Floquet&\\
\begin{tikzpicture}[scale=0.4]
\atoms{void}{x/p={2,0}, y/p={0,2}, z/p={1.6,0.8}}
\draw (0,0)--++($3*(x)$)--++($3*(y)$)--++($-3*(x)$)--cycle ($3*(x)$)--++($2*(z)$)--++($3*(y)$) ($3*(x)+3*(y)$)--++($2*(z)$) ($3*(y)$)--++($2*(z)$)--++($3*(x)$);
\draw[back] (0,0)--++($2*(z)$)--++($3*(x)$) ($2*(z)$)--++($3*(y)$);
\fill[opacity=0.3] ($2*(z)$)rectangle++($3*(x)+3*(y)$);
\draw[red,line width=0.1cm] ($1*(x)+1*(y)+2*(z)$)--++($0.5*(x)+0.5*(y)$)--++($-0.5*(x)+0.5*(y)$)--++($-0.5*(x)+-0.5*(y)$)--cycle;
\fill[opacity=0.3] ($1*(z)$)rectangle++($3*(x)+3*(y)$);
\draw[red,line width=0.1cm] ($1*(y)+1*(z)$)--++($(x)+(y)$)--++($-0.5*(x)+0.5*(y)$)--++($0.5*(x)+0.5*(y)$);
\fill[opacity=0.3] ($0*(z)$)rectangle++($3*(x)+3*(y)$);
\draw[red,line width=0.1cm] ($1*(x)+0*(z)$)--++($(x)+(y)$)--++($-0.5*(x)+0.5*(y)$)--++($0.5*(x)+0.5*(y)$)--++($1*(x)-1*(y)$);
\end{tikzpicture}
&
\end{tabular}
\end{equation}

\subsection{Decoding of the logic operation}
In this section, we discuss two points that we need to take into account when decoding a fault-tolerant logical operation such as the lattice-surgery protocol from Section~\ref{sec:lattice_surgery}, instead of merely storing qubits on a spatial torus.
The first point is due to the presence of physical spacetime boundaries in the protocol:
In addition to the temporal state boundary at the time of decoding, we are allowed to match $e$-anyon endpoints to the rough physical boundary, and an $m$-anyon endpoints to the smooth physical boundary.

The second point is due to the non-trivial spacetime topology:
At the time of decoding, it may not be possible to close off the anyon worldlines in a homologically trivial way.
Instead, we choose one representative of each spacetime homology class of anyon worldlines.
For example, in our lattice-surgery protocol, there is one non-trivial homology class represented by an $m$ anyon worldline connecting the bottom and the top part of the bridge between the two individual spacetime blocks:
\begin{equation}
\label{eq:lattice_surgery_outcome_representative}
M_1=
\begin{tikzpicture}[scale=0.5]
\atoms{void}{x/p={2,0}, y/p={0,2}, z/p={1.6,0.8}}
\fill[roughbdfillback] (z)--++(x)--++($0.5*(y)$)--++($0.4*(x)$)--++($-0.5*(y)$)--++(x)--++($2*(y)$)--++($-1*(x)$)--++($-0.5*(y)$)--++($-0.4*(x)$)--++($0.5*(y)$)--++($-1*(x)$)--cycle;
\draw[red, line width=0.1cm] ($1.2*(x)+0.5*(y)+0.5*(z)$)--++(y);
\draw[very thick] ($(x)+(z)$)--++($0.5*(y)$)--++($0.4*(x)$)--++($-0.5*(y)$);
\draw[very thick] ($2*(y)+(x)+(z)$)--++($-0.5*(y)$)--++($0.4*(x)$)--++($0.5*(y)$);
\fill[red,opacity=0.6] (0,0)--++(x)--++(z)--++($-1*(x)$)--cycle;
\fill[red,opacity=0.6] ($1.4*(x)$)--++(x)--++(z)--++($-1*(x)$)--cycle;
\fill[opacity=0.6] (0,0)--++(z)--++($2*(y)$)--++($-1*(z)$)--cycle;
\fill[opacity=0.6] ($2.4*(x)$)--++(z)--++($2*(y)$)--++($-1*(z)$)--cycle;
\fill[opacity=0.6] (x)--++(z)--++($0.5*(y)$)--++($-1*(z)$)--cycle;
\fill[opacity=0.6] ($1.4*(x)$)--++(z)--++($0.5*(y)$)--++($-1*(z)$)--cycle;
\fill[opacity=0.6] ($(x)+0.5*(y)$)--++(z)--++($0.4*(x)$)--++($-1*(z)$)--cycle;
\fill[opacity=0.6] ($2*(y)+(x)$)--++(z)--++($-0.5*(y)$)--++($-1*(z)$)--cycle;
\fill[opacity=0.6] ($2*(y)+1.4*(x)$)--++(z)--++($-0.5*(y)$)--++($-1*(z)$)--cycle;
\fill[opacity=0.6] ($(x)+1.5*(y)$)--++(z)--++($0.4*(x)$)--++($-1*(z)$)--cycle;
\fill[red,opacity=0.6] ($2*(y)$)--++(x)--++(z)--++($-1*(x)$)--cycle;
\fill[red,opacity=0.6] ($2*(y)+1.4*(x)$)--++(x)--++(z)--++($-1*(x)$)--cycle;
\fill[roughbdfill] (0,0)--++(x)--++($0.5*(y)$)--++($0.4*(x)$)--++($-0.5*(y)$)--++(x)--++($2*(y)$)--++($-1*(x)$)--++($-0.5*(y)$)--++($-0.4*(x)$)--++($0.5*(y)$)--++($-1*(x)$)--cycle;
\draw[very thick] (x)--++($0.5*(y)$)--++($0.4*(x)$)--++($-0.5*(y)$);
\draw[very thick] ($2*(y)+(x)$)--++($-0.5*(y)$)--++($0.4*(x)$)--++($0.5*(y)$);
\end{tikzpicture}
\;.
\end{equation}
Now, we close off the measured anyon worldlines such that they are either in the trivial homology class $M_0$ or in $M_1$ shown above.
The homology class $M_0$ or $M_1$ is used as a classical logical measurement outcome
\footnote{
Note that the set of logical measurement outcomes corresponds to the homology classes where anyon worldlines are allowed to terminate on the final state boundary.
However, the resulting logical quantum instrument depends on a representative homology class where anyon worldlines are not allowed to terminate.
For example, instead of Eq.~\eqref{eq:lattice_surgery_outcome_representative}, we could have instead chosen $M_1$ as a worldline connecting the bottom part of the bridge with the spatial smooth boundary on the right.
While this different choice of $M_1$ would correspond to the same logical measurement result, the correction differs by a $Z$ string across right part of the final state boundary, which is a logical $Z$ operator on the second qubit.
}
.

\subsection{Deriving the logic operation}
In this section, we concretely derive the logical operation performed by the lattice surgery protocol from Section~\ref{sec:lattice_surgery}.
The circuit with arbitrary measurement outcomes and correction defines a linear operator acting on the ground-state space of two spatial rectangular blocks of qubits.
If the logical measurement result is $M_0$, then this linear operator is the same as the path integral without anyon worldlines, where all measurements are post-selected to $+1$.
If the logical measurement result is $M_1$, then this linear operator is the same as the path integral with worldline configuration $M_1$ shown in Eq.~\eqref{eq:lattice_surgery_outcome_representative}.
This is by construction of the correction and due to homological invariance of the anyon worldlines discussed around Eq.~\eqref{eq:anyon_worldline_invariance}.

In order to determine the logical quantum operation corresponding to the logical measurement outcome $M_m$, we just need to evaluate the corresponding path integral with anyon pattern given by $M_m$.
This can be done using the closed-membrane picture.
As shown in Eq.~\eqref{eq:surface_code_cohomology_classes}, a natural basis for the ground-state space of the toric code on a rectangular spatial block is labeled by the two homology classes $B_0$ and $B_1$.
To evaluate the path integral for given initial homology classes $(B_a,B_b)$ and final homology classes $(B_c,B_d)$, we first determine the homology classes of spacetime closed-membrane patterns that are compatible with the initial and final homology classes, as well as with $M_m$.
Then we simply sum over their weights.
We can reduce the sum over all closed-membrane patterns to the sum over homology classes, since the weights of all patterns in the same homology class are equal.

For our lattice-surgery protocol, there are two generating bulk homology classes:
\begin{equation}
\label{eq:lattice_surgery_bulk_homology}
I_{10}=
\begin{tikzpicture}[scale=0.5]
\atoms{void}{x/p={2,0}, y/p={0,2}, z/p={1.6,0.8}}
\fill[roughbdfillback] (z)--++(x)--++($0.5*(y)$)--++($0.4*(x)$)--++($-0.5*(y)$)--++(x)--++($2*(y)$)--++($-1*(x)$)--++($-0.5*(y)$)--++($-0.4*(x)$)--++($0.5*(y)$)--++($-1*(x)$)--cycle;
\draw[red, line width=0.1cm] ($1.2*(x)+0.5*(y)+0.5*(z)$)--++(y);
\draw[very thick] ($(x)+(z)$)--++($0.5*(y)$)--++($0.4*(x)$)--++($-0.5*(y)$);
\draw[very thick] ($2*(y)+(x)+(z)$)--++($-0.5*(y)$)--++($0.4*(x)$)--++($0.5*(y)$);
\fill[red,opacity=0.6] (0,0)--++(x)--++(z)--++($-1*(x)$)--cycle;
\fill[red,opacity=0.6] ($1.4*(x)$)--++(x)--++(z)--++($-1*(x)$)--cycle;
\fill[opacity=0.6] (0,0)--++(z)--++($2*(y)$)--++($-1*(z)$)--cycle;
\fill[membranecol, opacity=0.6] ($0.5*(z)$)--++(x)--++($2*(y)$)--++($-1*(x)$)--cycle;
\fill[opacity=0.6] ($2.4*(x)$)--++(z)--++($2*(y)$)--++($-1*(z)$)--cycle;
\fill[opacity=0.6] (x)--++(z)--++($0.5*(y)$)--++($-1*(z)$)--cycle;
\fill[opacity=0.6] ($1.4*(x)$)--++(z)--++($0.5*(y)$)--++($-1*(z)$)--cycle;
\fill[opacity=0.6] ($(x)+0.5*(y)$)--++(z)--++($0.4*(x)$)--++($-1*(z)$)--cycle;
\fill[membranecol, opacity=0.6] ($0.5*(z)+(x)+0.5*(y)$)--++($0.2*(x)$)--++($1*(y)$)--++($-0.2*(x)$)--cycle;
\fill[opacity=0.6] ($2*(y)+(x)$)--++(z)--++($-0.5*(y)$)--++($-1*(z)$)--cycle;
\fill[opacity=0.6] ($2*(y)+1.4*(x)$)--++(z)--++($-0.5*(y)$)--++($-1*(z)$)--cycle;
\fill[opacity=0.6] ($(x)+1.5*(y)$)--++(z)--++($0.4*(x)$)--++($-1*(z)$)--cycle;
\fill[red,opacity=0.6] ($2*(y)$)--++(x)--++(z)--++($-1*(x)$)--cycle;
\fill[red,opacity=0.6] ($2*(y)+1.4*(x)$)--++(x)--++(z)--++($-1*(x)$)--cycle;
\draw[membranecol, line width=0.05cm] ($0.5*(z)$)--++(x)--++($0.5*(y)$)--++($0.2*(x)$)  ($0.5*(z)+2*(y)$)--++(x)--++($-0.5*(y)$)--++($0.2*(x)$) ($0.5*(z)$)--++($2*(y)$);
\fill[roughbdfill] (0,0)--++(x)--++($0.5*(y)$)--++($0.4*(x)$)--++($-0.5*(y)$)--++(x)--++($2*(y)$)--++($-1*(x)$)--++($-0.5*(y)$)--++($-0.4*(x)$)--++($0.5*(y)$)--++($-1*(x)$)--cycle;
\draw[very thick] (x)--++($0.5*(y)$)--++($0.4*(x)$)--++($-0.5*(y)$);
\draw[very thick] ($2*(y)+(x)$)--++($-0.5*(y)$)--++($0.4*(x)$)--++($0.5*(y)$);
\end{tikzpicture}
\;,
I_{01}=
\begin{tikzpicture}[scale=0.5]
\atoms{void}{x/p={2,0}, y/p={0,2}, z/p={1.6,0.8}}
\fill[roughbdfillback] (z)--++(x)--++($0.5*(y)$)--++($0.4*(x)$)--++($-0.5*(y)$)--++(x)--++($2*(y)$)--++($-1*(x)$)--++($-0.5*(y)$)--++($-0.4*(x)$)--++($0.5*(y)$)--++($-1*(x)$)--cycle;
\draw[red, line width=0.1cm] ($1.2*(x)+0.5*(y)+0.5*(z)$)--++(y);
\draw[very thick] ($(x)+(z)$)--++($0.5*(y)$)--++($0.4*(x)$)--++($-0.5*(y)$);
\draw[very thick] ($2*(y)+(x)+(z)$)--++($-0.5*(y)$)--++($0.4*(x)$)--++($0.5*(y)$);
\fill[red,opacity=0.6] (0,0)--++(x)--++(z)--++($-1*(x)$)--cycle;
\fill[red,opacity=0.6] ($1.4*(x)$)--++(x)--++(z)--++($-1*(x)$)--cycle;
\fill[opacity=0.6] (0,0)--++(z)--++($2*(y)$)--++($-1*(z)$)--cycle;
\fill[membranecol, opacity=0.6] ($0.5*(z)+2.4*(x)$)--++($-1*(x)$)--++($2*(y)$)--++($1*(x)$)--cycle;
\fill[opacity=0.6] ($2.4*(x)$)--++(z)--++($2*(y)$)--++($-1*(z)$)--cycle;
\fill[opacity=0.6] (x)--++(z)--++($0.5*(y)$)--++($-1*(z)$)--cycle;
\fill[opacity=0.6] ($1.4*(x)$)--++(z)--++($0.5*(y)$)--++($-1*(z)$)--cycle;
\fill[opacity=0.6] ($(x)+0.5*(y)$)--++(z)--++($0.4*(x)$)--++($-1*(z)$)--cycle;
\fill[membranecol, opacity=0.6] ($0.5*(z)+1.4*(x)+0.5*(y)$)--++($-0.2*(x)$)--++($1*(y)$)--++($0.2*(x)$)--cycle;
\fill[opacity=0.6] ($2*(y)+(x)$)--++(z)--++($-0.5*(y)$)--++($-1*(z)$)--cycle;
\fill[opacity=0.6] ($2*(y)+1.4*(x)$)--++(z)--++($-0.5*(y)$)--++($-1*(z)$)--cycle;
\fill[opacity=0.6] ($(x)+1.5*(y)$)--++(z)--++($0.4*(x)$)--++($-1*(z)$)--cycle;
\fill[red,opacity=0.6] ($2*(y)$)--++(x)--++(z)--++($-1*(x)$)--cycle;
\fill[red,opacity=0.6] ($2*(y)+1.4*(x)$)--++(x)--++(z)--++($-1*(x)$)--cycle;
\draw[membranecol, line width=0.05cm] ($0.5*(z)+2.4*(x)$)--++($-1*(x)$)--++($0.5*(y)$)--++($-0.2*(x)$)  ($0.5*(z)+2.4*(x)+2*(y)$)--++($-1*(x)$)--++($-0.5*(y)$)--++($-0.2*(x)$) ($0.5*(z)+2.4*(x)$)--++($2*(y)$);
\fill[roughbdfill] (0,0)--++(x)--++($0.5*(y)$)--++($0.4*(x)$)--++($-0.5*(y)$)--++(x)--++($2*(y)$)--++($-1*(x)$)--++($-0.5*(y)$)--++($-0.4*(x)$)--++($0.5*(y)$)--++($-1*(x)$)--cycle;
\draw[very thick] (x)--++($0.5*(y)$)--++($0.4*(x)$)--++($-0.5*(y)$);
\draw[very thick] ($2*(y)+(x)$)--++($-0.5*(y)$)--++($0.4*(x)$)--++($0.5*(y)$);
\end{tikzpicture}
\;.
\end{equation}
Both generators are compatible with the non-trivial measurement outcome $m=1$, but their sum corresponds to the trivial outcome $m=0$.
In total there are four homology classes obtained from adding the generating membrane patterns:
\begin{equation}
I_{ij}=i\cdot I_{10}+j\cdot I_{01}\;,\quad i,j\in \{0,1\}\;.
\end{equation}
The weight evaluates to $1$ for all bulk homology classes
\footnote{
Slightly abusing notation, we are using $Z$ below to denote the weight of a fixed variable configuration representing the homology class $I_{ij}$ instead of the path integral evaluation.
},
\begin{equation}
Z(I_{ij})=1\quad\forall i,j\;.
\end{equation}
$I_{ij}$ restricts to the spatial homology classes $(B_i,B_j)$ for the two input rectangular blocks, and the same homology classes for the output.
Furthermore, $I_{ij}$ is compatible with the measurement homology class $M_{i+j}$.
So if $a=c$, $b=d$, and $a+b=m$, there is a single compatible bulk homology class $I_{ab}$, and the amplitude is $1$.
Otherwise, there is no compatible bulk homology class, and we sum over the empty set yielding $0$.
So the amplitudes of the logic operation with measurement result $M_m$ are
\begin{equation}
\bra{B_c,B_d}Z[M_m]\ket{B_a,B_b}= \delta_{a=c} \delta_{b=d} \delta_{a+b=m}\;.
\end{equation}
So we indeed find for the logical action:
\begin{equation}
Z[M_m] = \frac12(1+(-1)^m ZZ)\;.
\end{equation}

\section{Conclusion and outlook}
In this work, we have demonstrated how to construct fault-tolerant circuits for topological quantum computation in the path-integral approach.
We have shown how to obtain circuits in the bulk as well as for boundaries and corners, and for storage as well as performing fault-tolerant logic gates.
We have focused on a specific example, the x+y Floquet code, a circuit obtained from traversing the cubic-lattice path integral in the $x+y$-direction.

It is not difficult to adapt the methods presented in this work to construct other fault-tolerant logic gates.
For example, starting or ending the circuit on a spatial rectangular block by a temporal smooth boundary corresponds to a single-qubit $\ket+$ state preparation or $X$ measurement.
It is also not hard to find the path-integral representation of other sorts of topological defects, like the duality domain wall between the toric code and itself, or its termination at a twist defect \cite{Bombin2010}.
Using these, we can for example construct protocols for our $x+y$ circuit that perform a logical Hadamard via a temporal duality domain wall and a $\frac\pi2$ rotation, or quantum computation by braiding twist defects.

Unfortunately, purely-topological fault-tolerant protocols that only involve the toric-code topological phase do not seem to provide a universal gate set for quantum computation.
However, a great feature of the path-integral approach is that it is not based on Pauli stabilizers or Clifford operations.
This makes it possible to construct fault-tolerant circuits for more exotic topological phases, as was demonstrated in Ref.~\cite{twisted_double_code}.
This applies to abelian topological phases, but also to specific non-Abelian phases through just-in-time decoding \cite{Bombin2018,Brown2019,twisted_double_code}.
This non-Abelian topological phase allows us to complete a universal gate set, for example by a $T$ gate \cite{Bombin2018} or a $CCZ$ gate \cite{Brown2019}.
For more background on fault-tolerant protocols for non-Clifford gates from non-Abelian topological phases, we refer the reader to upcoming work by Davydova et al.~\cite{Davydova2025}.
In this endeavor of performing topological quantum computation purely topologically, the path-integral approach turns out to be very useful.
First of all, a spacetime view that is manifest at all levels in the path-integral approach, is essential for understanding these protocols.
Second, the path-integral approach makes it easy to construct a great variety of low-overhead microscopic protocols, which experimentalists can choose from.
For example, in Ref.~\cite{twisted_double_code}, a schedule for the non-Abelian phase was constructed, consisting of three copies of the ordinary stabilizer toric code protocol together with two layers of $CCZ$ gates acting on all qubits in parallel.

\subsubsection*{Acknowledgments}
I would like to thank Ben Brown, Julio Magdalena de la Fuente, Margarita Davydova, Jens Eisert, and Peter-Jan Derks for discussions and feedback on the manuscript.
This work was supported by the U. S. Army Research Laboratory and the U. S. Army Research Office under contract/grant number W911NF2310255, and by the U.S. Department of Energy, Office of Science, National Quantum Information Science Research Centers, and the Co-design Center for Quantum Advantage (C2QA) under contract number DE-SC0012704.

%apsrev4-2.bst 2019-01-14 (MD) hand-edited version of apsrev4-1.bst
%Control: key (0)
%Control: author (8) initials jnrlst
%Control: editor formatted (1) identically to author
%Control: production of article title (0) allowed
%Control: page (0) single
%Control: year (1) truncated
%Control: production of eprint (0) enabled
%

%\bibliography{110_dynamic_code_refs}{}

\end{document}